\documentclass[%
nofootinbib,
twocolumn,
amsmath,amssymb,
aps,
prd,
floatfix,
a4paper
]{revtex4-2}

\usepackage[colorlinks=true,linkcolor=blue,citecolor=blue,linktocpage=false]{hyperref}

\usepackage{newtxtext,newtxmath}
\usepackage{graphicx}% Include figure files
\usepackage{dcolumn}% Align table columns on decimal point
\usepackage{bm}% bold math

\newcommand{\VEC}{\bm}
\newcommand{\MAT}{\bm}
\newcommand{\hMpc}{\,h^{-1}\,{\rm Mpc}}

\newcommand{\PP}{\bm{\Psi}}
\newcommand{\YY}{\overline{\bm{\Psi}}}
\newcommand{\SL}{\overline{\bm{s}}}

\begin{document}

\preprint{}

\title{Developing a Theoretical Model for the Resummation of Infrared Effects \\ in the Post-Reconstruction Power Spectrum}

\author{Naonori Sugiyama}
\email{nao.s.sugiyama@gmail.com}
\affiliation{National Astronomical Observatory of Japan, Mitaka, Tokyo 181-8588, Japan}%Lines break automatically or can be forced with \\
%\author{Hee-Jong Seo}
%\affiliation{Department of Physics and Astronomy, Ohio University, Clippinger Labs, Athens, OH 45701 }
%\affiliation{Physics Division, Lawrence Berkeley National Laboratory, 1 Cyclotron Road, Berkeley, CA 94720, USA}
%\affiliation{Berkeley Center for Cosmological Physics, Department of Physics, University of California, Berkeley, CA 94720, USA}
%\author{Shun Saito}
%\affiliation{Institute for Multi-messenger Astrophysics and Cosmology, Department of Physics,\\
%Missouri University of Science and Technology, 1315 N. Pine St., Rolla MO 65409, USA}
%\affiliation{Kavli Institute for the Physics and Mathematics of the Universe (WPI), \\
%Todai Institutes for Advanced Study, The University of Tokyo, Chiba 277-8582, Japan
%}

\date{\today}% It is always \today, today,

\begin{abstract}

Since galaxy distribution reconstruction effectively reduces non-Gaussian terms in the power spectrum covariance matrix, it has attracted interest not only for Baryon Acoustic Oscillation (BAO) signals but also for various cosmological signal analyses. To this end, this paper presents a novel theoretical model that addresses infrared (IR) effects in the post-reconstruction galaxy power spectrum, including 1-loop corrections. In particular, we discuss the importance of incorporating non-perturbative effects arising from IR contributions into the displacement vector $\VEC{s}$ used for reconstruction. Consequently, post-reconstruction nonlinear damping of BAO can be described by a single two-dimensional Gaussian function. This is a phenomenon not observed when $\VEC{s}$ is considered to at a linear order in the Zel'dovich approximation. Furthermore, we confirm that the cross-power spectrum of the pre- and post-reconstruction density fluctuations lacks IR effect cancellations, and shows an exponential decay in both the cross-power spectrum and the associated shot-noise term.

\end{abstract}

\maketitle

\section{Introduction}

The reconstruction of galaxy distributions~\cite{Eisenstein:2006nk} has traditionally been a focus of research to amplify baryon acoustic oscillation (BAO) signals~\cite{Sunyaev:1970eu,Peebles:1970ag}. However, recent findings indicate that these reconstruction techniques can effectively reduce non-Gaussian terms in the covariance matrix of cosmological $n$-point statistics. As a result, their application has expanded beyond BAO signals to encompass a variety of cosmological signals. In particular, there is a growing interest in improving constraints on cosmological parameters through post-reconstruction two-point statistics~\cite{Hikage:2020fte,Wang:2022nlx}. In addition, applications in three-point statistics have been reported to strengthen constraints on primordial non-Gaussianity~\cite{Shirasaki:2020vkk}.

Despite the potential of reconstruction techniques, their practical applications, particularly beyond BAO signal analysis, are still emerging. This slow advancement is due to a limited understanding of the nonlinear characteristics of reconstructed density fluctuations. The examination of pre-reconstruction two-point correlation functions and power spectra primarily employs advanced theoretical models, which extend beyond Standard Perturbation Theory (SPT) (e.g.,~\cite{Bernardeau:2001qr}), to precisely extract small-scale information. For instance, sophisticated theoretical models such as the Convolution Lagrangian Perturbation Theory (CLPT)~\cite{Carlson:2012bu,Wang:2013hwa}, the Renormalized Perturbation Theory (RPT)~\cite[][]{Crocce:2005xy}, the TNS model~\cite{Taruya:2010mx}, and the Effective Field Theory of Large-scale Structure (EFT of LSS)~\cite{Baumann:2010tm,Carrasco:2012cv} have been applied to analyze actual galaxy survey datasets, such as the Baryon Oscillation Spectroscopic Survey Data Release 12 (BOSS DR12) galaxy data~\citep{Eisenstein:2011sa,Bolton:2012hz,Dawson:2012va,Alam:2015mbd}. On the other hand, theoretical research into post-reconstruction density fluctuations has been limited to 1-loop calculations in SPT~\cite{Schmittfull:2015mja,Hikage:2017tmm,Hikage:2019ihj} and methods based on the Zel'dovich approximation~\cite{Padmanabhan:2008dd,Seo:2015eyw,White:2015eaa,Chen:2019lpf}. See also \cite{Ota:2021caz,Ota:2022him} for perturbation theory approaches to iterative reconstruction methods~\cite{Zhu:2016sjc,Zhu:2017vtj,Yu:2017tpa,Wang:2017jeq,Schmittfull:2017uhh,Hada:2018fde,Hada:2018ziy,Seo:2021nev}.

In this study, we primarily investigate the infrared (IR) effect on the power spectrum of galaxies. This effect possesses a unique property that allows for non-perturbative analysis, making it crucial for advancing theories beyond SPT. For the pre-reconstruction scenario, several IR effects are found, and we summarize them below. When taking the IR limit and considering only the nonlinearities arising from IR effects, all IR nonlinear effects are negated in the calculations of the power spectrum, leaving only the linear power spectrum~\citep{Jain:1995kx,Scoccimarro:1995if,Kehagias:2013yd,Peloso:2013zw,Sugiyama:2013pwa,Sugiyama:2013gza,Blas:2013bpa,Blas:2015qsi,Lewandowski:2017kes}. This IR cancellation holds true when evaluating all density fluctuations at equal times. However, assessing the correlation of fluctuations at distinct times yields the consistency relation for the LSS, which uniquely connects higher-order statistics to their lower-order counterparts~\citep{Peloso:2013zw,Kehagias:2013yd,Creminelli:2013mca}. Moreover, proper consideration of the IR effect can shed light on the nonlinear decay of the BAO signal~\citep{Zeldovich:1969sb,Eisenstein:2006nj,Crocce:2007dt,Matsubara:2007wj,Sugiyama:2013gza,Senatore:2014via,Baldauf:2015xfa,Blas:2016sfa,Senatore:2017pbn,Ivanov:2018gjr,Lewandowski:2018ywf,Sugiyama:2020uil}.

This paper aims to comprehensively elucidate the impact of non-perturbative IR effects on the BAO signal and the overall power spectrum shape after reconstruction. In this context, we present a new IR-resummed model of the post-reconstruction power spectrum, including 1-loop corrections. Concurrently, we explore its relation to established theories that address partial IR effects after reconstruction, such as the 1-loop correction in SPT and the Zel'dovich approximation. We further demonstrate the fundamental significance of the IR effect in the cross-power spectrum between the pre- and post-reconstruction density fluctuations. The proposed model should be directly applicable to observational galaxy data as it incorporates Redshift Space Distortions (RSDs) (e.g.,~\cite{Kaiser:1987qv}) and galaxy bias effects (e.g.,~\cite{Desjacques:2016bnm}).

The outline of this paper is as follows. Section~\ref{Sec:PreRecon} reviews the non-perturbative treatment of infrared (IR) effects on pre-reconstruction density fluctuations. Section~\ref{Sec:PostRecon} explores the IR effects on post-reconstruction density fluctuations and presents a model for their power spectrum. Section~\ref{Sec:Cross} calculates the cross-power spectrum of density fluctuations before and after reconstruction, highlighting the importance of nonlinear IR effects. Section~\ref{Sec:Conclusions} concludes with the findings and implications in this paper. Appendix~\ref{Sec:DHOST} introduces a kind of modified gravity theory as an example of breaking the IR cancellation. Appendixes~\ref{Sec:SPT} and \ref{Sec:SPT_REC} summarize the behavior in the IR limit for solutions up to third-order fluctuations in SPT before and after reconstruction, respectively.

Our numerical calculations use a flat $\Lambda$CDM model as the fiducial cosmological model with the following parameters: matter density $\Omega_{\rm m0}=0.316$, Hubble constant $h\equiv H_0/(100\,{\rm km\, s^{-1}\, Mpc^{-1}})=0.671$, baryon density $\Omega_{\rm b0}h^2=0.022$, $\sigma_8=0.86$, and spectral tilt $n_{\rm s}=0.962$, which are close to the best-fit values given by Planck2018~\citep{Aghanim:2018eyx}. 

\section{Pre-Reconstruction Case}
\label{Sec:PreRecon}

In preparation for the post-reconstruction case, we review the derivation for the IR-resummed power spectrum model in the pre-reconstruction case. For the behavior of the solution in the IR limit, see \cite{Sugiyama:2013pwa,Sugiyama:2013gza,Sugiyama:2020uil}, and for the details of the derivation of the IR-resummed model, see \cite{Sugiyama:2013gza,Baldauf:2015xfa,Vlah:2015zda,Senatore:2014via,Blas:2016sfa,Senatore:2017pbn,Ivanov:2018gjr,Lewandowski:2018ywf}.

\subsection{Infrared (IR) effects on dark matter density fluctuations}
\label{Sec:IR_kernel}

Consider a density field function $\rho(\VEC{x})$ defined over three-dimensional spatial coordinates $\VEC{x}$. Let $\delta(\VEC{x})$ represent the deviation from the mean density field $\bar{\rho}$. This relation can be expressed as $\rho(\VEC{x}) = \bar{\rho}\left( 1 + \delta(\VEC{x}) \right)$. The Fourier transform of $\delta(\VEC{x})$ can be represented as $\widetilde{\delta}(\VEC{k}) = \int d^3x e^{-i\VEC{x}\cdot \VEC{k}} \delta(\VEC{x})$, where the tilde denotes a Fourier-transformed quantity.
 
Let us begin by considering the density fluctuations of dark matter, denoted $\delta_{\rm m}(\VEC{x})$, wherein the subscript ``m'' represents ``matter''. In perturbation theory, an $n$th-order dark matter density fluctuation in Fourier space, denoted $\widetilde{\delta}^{\,[n]}_{{\rm m}}(\VEC{k})$, can be represented by the convolution integral of $n$ wave vectors ($\VEC{p}_1,\dots,\VEC{p}_n$) with an associated $n$th-order kernel function $F_n(\VEC{p}_1,\dots,\VEC{p}_n)$. Hence, the $n$th-order density fluctuation is expressed as
\begin{eqnarray}
    \widetilde{\delta}^{\,[n]}_{{\rm m}}(\VEC{k}) &=& \int \frac{d^3p_1}{(2\pi)^3}\cdots\frac{d^3p_n}{(2\pi)^3}
    (2\pi)^3\delta_{\rm D}(\VEC{k}-\VEC{p}_{[1,n]}) \nonumber\\ 
    &\times& F_n(\VEC{p}_1,\dots,\VEC{p}_n) 
    \widetilde{\delta}_{\rm m}^{\,[1]}(\VEC{p}_1)\cdots\widetilde{\delta}_{\rm m}^{\,[1]}(\VEC{p}_n)\;,
    \label{Eq:delta_n}
\end{eqnarray}
where $\VEC{p}_{[1,n]}=\VEC{p}_1+\cdots+\VEC{p}_n$, $\widetilde{\delta}^{\,[1]}_{\rm m}$ is the linear (first-order) dark matter density fluctuation in Fourier space, and $\delta_{\rm D}$ is the three-dimensional delta function. 

\subsubsection{Second-order fluctuations}

For example, the second-order kernel function is given by
\begin{eqnarray}
    F_2(\VEC{p}_1,\VEC{p}_2)
    = \frac{17}{21} + \frac{1}{2}\mu_{12}\left( \frac{p_1}{p_2} + \frac{p_2}{p_1} \right)
    + \frac{2}{7} \left( \mu_{12}^2 - \frac{1}{3} \right)\;,
    \label{Eq:F2}
\end{eqnarray}
where $p_n = |\VEC{p}_n|$ and $\mu_{12} = \hat{p}_1\cdot\hat{p}_2$ with $\hat{p}_n = \VEC{p}_n/p_n$. Substituting Eq.~(\ref{Eq:F2}) into Eq.~(\ref{Eq:delta_n}) and performing the inverse Fourier transform, we obtain
\begin{eqnarray}
    \delta_{\rm m}^{[2]}(\VEC{x})
    &=& \frac{17}{21} \big[ \delta_{\rm m}^{[1]}(\VEC{x}) \big]^2
    - \PP^{[1]}(\VEC{x}) \cdot \nabla \delta_{\rm m}^{[1]}(\VEC{x}) \nonumber \\
    &+&
    \frac{2}{7} \left[\left( \frac{\partial_i\partial_j}{\partial^2}-\frac{1}{3}\delta_{ij} \right) \delta_{\rm m}^{[1]}(\VEC{x}) \right] ^2\;.
    \label{Eq:delta_2}
\end{eqnarray}
On the right-hand side of Eq.~(\ref{Eq:delta_2}), the three terms from left to right are, respectively, named the ``nonlinear growth term,'' ``shift term,'' and ``tidal term''~\cite{Schmittfull:2014tca,Schmittfull:2015mja}. The linear displacement vector, associated with the shift term, is defined by the linear dark matter density fluctuation as
\begin{eqnarray}
    \PP^{[1]}(\VEC{x}) = \int \frac{d^3k}{(2\pi)^3}e^{i\VEC{k}\cdot\VEC{x}} \frac{i\VEC{k}}{k^2} \widetilde{\delta}_{\rm m}^{\, [1]}(\VEC{k})\;.
\end{eqnarray}

In the IR limit, where one of the wave vectors in $F_2(\VEC{p}_1,\VEC{p}_2)$ is much smaller than the other (specifically, when $p_2\to0$), the following relation holds:
\begin{eqnarray}
    F_2(\VEC{p}_1,\VEC{p}_2)  \xrightarrow[p_2\to0]{}
    \frac{1}{2}\left( \frac{\VEC{p}_1\cdot\VEC{p}_2}{p_2^2} \right)\;.
    \label{Eq:F2_IR}
\end{eqnarray}
When substituting Eq.~(\ref{Eq:F2_IR}) into Eq.~(\ref{Eq:delta_n}), the following approximation is considered simultaneously:
\begin{eqnarray}
    \delta_{\rm D}(\VEC{k}-\VEC{p}_1-\VEC{p}_2) \xrightarrow[p_2\to0]{}\delta_{\rm D}(\VEC{k}-\VEC{p}_1)\;.
    \label{Eq:delta_D_2}
\end{eqnarray}
Consequently, in the IR limit, we derive
\begin{eqnarray}
    \delta_{\rm m}^{[2]}(\VEC{x}) \xrightarrow[{\rm IR\, limit}]{} - \YY^{[1]} \cdot \nabla \delta_{\rm m}^{[1]}(\VEC{x})\;,
    \label{Eq:delta_2_IR}
\end{eqnarray}
where
\begin{eqnarray}
    \YY^{[1]} = \PP^{[1]}(\VEC{x}=\VEC{0})\;.
    \label{Eq:YY_1}
\end{eqnarray}
In deriving Eq.~(\ref{Eq:delta_2_IR}), we also accounted for the condition $p_1\to0$ and introduced a prefactor of $2$. Comparison with Eq.~(\ref{Eq:delta_2}) reveals that the IR limit solution involves isolating only the shift term, and furthermore, the displacement vector contained in the shift term is evaluated at the origin $\VEC{x}=\VEC{0}$.

Decomposing the second-order fluctuation solution into terms manifesting in the IR limit and the others, the solution can be reformulated as
\begin{eqnarray}
    \delta_{\rm m}^{[2]} = \delta^{[2]}_{\rm (S)m}(\VEC{x}) - \YY^{[1]} \cdot \nabla \delta_{\rm m}^{[1]}(\VEC{x})\;,
    \label{Eq:delta_2_Sp}
\end{eqnarray}
where the subscript $(\rm S)$ denotes the contribution from short-wavelength modes, and $\delta_{\rm (S)m}^{[2]}(\VEC{x})$ is explicitly given by
\begin{eqnarray}
    \delta_{\rm (S)m}^{[2]}(\VEC{x})
    &=& \frac{17}{21} \left[ \delta_{\rm m}^{[1]}(\VEC{x}) \right]^2
    - \PP_{\rm (\rm S)}^{[1]}(\VEC{x}) \cdot \nabla \delta_{\rm m}^{[1]}(\VEC{x}) \nonumber \\
    &+&
    \frac{2}{7} \left[\left( \frac{\partial_i\partial_j}{\partial^2}-\frac{1}{3}\delta_{ij} \right) \delta_{\rm m}^{[1]}(\VEC{x}) \right] ^2\;.
    \label{Eq:delta_2_S}
\end{eqnarray}
In the equation above, we defined the displacement vector contributing from short-wavelength modes as
\begin{eqnarray}
    \PP_{\rm (S)}^{[1]}(\VEC{x}) =  \PP^{[1]}(\VEC{x}) - \YY^{[1]}\;,
\end{eqnarray}
where $\PP_{\rm (S)}^{[1]}(\VEC{x}=\VEC{0})=\VEC{0}$.

\subsubsection{Non-perturbative treatment}
\label{Sec:General}

In the IR limit of the $n$th-order kernel function $F_n(\VEC{p}_1,\dots,\VEC{p}_n)$, one of the $n$ wave numbers contributing to the scale of interest $k=|\VEC{p}_{[1,n]}|$ possesses a magnitude substantially smaller than the other $(n-1)$ wave numbers. For a single dark matter fluid in General Relativity (GR), the recursion relation allows the determination of solutions for $F_n(\VEC{p}_1,\dots,\VEC{p}_n)$ up to an infinite order (e.g., see~\cite{Bernardeau:2001qr}). Using mathematical induction on this recursion relation, the relation below can be proven~\cite{Sugiyama:2013pwa}:
\begin{eqnarray}
    \hspace{-1.0cm} F_n(\VEC{p}_1,\dots,\VEC{p}_{n-1},\VEC{p}) 
    && \xrightarrow[p\to0]{}
    \frac{1}{n}\left( \frac{\VEC{p}_{[1,n-1]}\cdot\VEC{p}}{p^2} \right) \nonumber \\
    \hspace{-1.0cm}&& \hspace{0.5cm} \times \hspace{0.2cm}F_{n-1}(\VEC{p}_1,\dots,\VEC{p}_{n-1})\;.
    \label{Eq:F_n_IR}
\end{eqnarray}
More generally, for $r$ wave numbers approaching zero, 
\begin{eqnarray}
    &&\hspace{-0.4cm}F_n(\VEC{p}_1,\dots, \VEC{p}_{n-r}, \VEC{p}_{n-r+1}, \dots, \VEC{p}_{n})   
     \xrightarrow[p_{n-r+1},\dots,p_n\to0]{} \nonumber \\
   && \hspace{-0.7cm}\frac{(n-r)!}{n!}
     \prod_{\alpha=1}^{r}\left( \frac{\VEC{p}_{[1,n-r]}\cdot\VEC{p}_{n-r+\alpha}}{p_{n-r+\alpha}^2} \right)
     F_{n-r}(\VEC{p}_1,\dots,\VEC{p}_{n-r})\;.
   \label{Eq:delta_nr}
\end{eqnarray}
In this context, we decompose the linear dark matter density fluctuation into long-wavelength modes, with wave numbers approaching zero, and short-wavelength modes which are independent of the former:
\begin{eqnarray}
    \widetilde{\delta}_{\rm m}^{\,[1]}(\VEC{p}) = \widetilde{\delta}_{\rm m}^{\,[1]}(\VEC{p}\to0) +  \widetilde{\delta}_{\rm (S)m}^{\,[1]}(\VEC{p})\;.
    \label{Eq:linear_S}
\end{eqnarray}
Substituting Eqs.~(\ref{Eq:delta_nr}) and (\ref{Eq:linear_S}) into Eq.~(\ref{Eq:delta_n}) and using the approximation 
\begin{eqnarray}
    \delta_{\rm D}(\VEC{k}-\VEC{p}_{[1,n]})\xrightarrow[p_{n-r+1},\dots,p_n\to0]{} 
    \delta_{\rm D}(\VEC{k}-\VEC{p}_{[1,n-r]})\;,
    \label{Eq:delta_D_n}
\end{eqnarray}
the dark matter density fluctuation can be formally expressed as~\cite{Sugiyama:2013gza}
\begin{eqnarray}
    \delta_{\rm m}(\VEC{x}) = \delta_{\rm (S)m}(\VEC{x}-\YY^{[1]})\;,
    \label{Eq:delta_m_S}
\end{eqnarray}
where the $n$th order of $\delta_{\rm (S)m}(\VEC{x})$ is given by
\begin{eqnarray}
    \hspace{-0.5cm}
    \delta^{[n]}_{{\rm (S)m}}(\VEC{x}) &=& \int \frac{d^3p_1}{(2\pi)^3}\cdots\frac{d^3p_n}{(2\pi)^3}\,
    e^{i(\VEC{p}_1+\cdots+\VEC{p}_n)\cdot\VEC{x}}\nonumber\\ 
    &\times& F_n(\VEC{p}_1,\dots,\VEC{p}_n) 
    \widetilde{\delta}_{\rm (S)m}^{\,[1]}(\VEC{p}_1)\cdots\widetilde{\delta}_{\rm (S)m}^{\,[1]}(\VEC{p}_n)\;.
    \label{Eq:delta_n_S}
\end{eqnarray}
Thus, the IR limit effect in a single dark matter fluid can be interpreted as a result of a coordinate transformation due to the displacement vector at the origin.

The IR limit is an operation that extracts nonlinear contributions from large scales (small wave numbers). In other words, the wave number $k$ of interest is considerably larger than the wave numbers contributed in the IR limit. Therefore, historically, the IR limit solution has also been referred to as the high-$k$ ($k\to\infty$) limit solution (e.g., see \cite{Bernardeau:2008fa}). Conversely, the ultraviolet (UV) limit corresponds to the low-$k$ ($k\to 0$) limit and is treated by, e.g., EFT of LSS~\cite{Baumann:2010tm,Carrasco:2012cv}.

In summary, this paper makes three approximations related to the IR limit:
\begin{enumerate}

    \item As illustrated in Eqs.~(\ref{Eq:F2_IR}) and (\ref{Eq:delta_nr}), for the kernel function $F_n(\VEC{p}_1,\dots,\VEC{p}_n)$ that characterizes nonlinear effects, we employ the approximate kernel function when one or multiple wave numbers approach zero. This means that the dominant effect in the IR limit appears as the coordinate transformation of the short-wavelength density fluctuation $\delta_{\rm (S)}$ due to $\YY$ as in Eq.~(\ref{Eq:delta_m_S}).

    \item As presented in Eqs.~(\ref{Eq:delta_D_2}) and (\ref{Eq:delta_D_n}), within the delta function $\delta_{\rm D}(\VEC{k}-\VEC{p}_{[1,n]})$ that characterizes the mode-coupling integral with multiple wave vectors, we assume no mode coupling occurs between the IR effect and other components when one or more wave numbers approach zero. This implies that the displacement vector is evaluated at the origin in Eq.~(\ref{Eq:YY_1}), and furthermore, indicates no correlation between $\delta_{\rm (S)}$ and $\YY$, as will be discussed in Section~\ref{Sec:IR_Cancellation}.

    \item The nonlinear effects of $\delta_{\rm (S)m}$ are truncated at a finite perturbation order. Any higher-order perturbative effects beyond this are treated as IR effects through $\YY$.
\end{enumerate}

The clearest example in the IR limit occurs when $\delta_{\rm (S)m}(\VEC{x})\sim \delta_{\rm m}^{[1]}(\VEC{x})$, approximating the nonlinear dark matter density fluctuation as
\begin{eqnarray}
    \delta_{\rm m}(\VEC{x}) \xrightarrow[\rm IR\, limit]{} \delta_{\rm m}^{[1]}(\VEC{x}-\YY^{[1]}).
\end{eqnarray}
The conclusions of this paper will be mainly based on this simplest case in Sections~\ref{Sec:PostRecon} and~\ref{Sec:Cross}. When higher-order correction terms up to the 1-loop order are to be considered, they will be explained on there as appropriate.

\subsection{Extensions to RSD and bias effects}
\label{Sec:IR_RSD_Bias}

In cases affected by RSD and bias effects, such as observed galaxies, the use of Lagrangian Perturbation Theory (LPT) is advantageous. This is because LPT naturally accommodates displacement vectors. Within LPT, RSD effects are integrated into the displacement vector. Bias effects, on the other hand, can be treated using the equation~\citep{Matsubara:2008wx}
\begin{eqnarray}
    && 1 + \delta_{\rm g}(\VEC{x},\hat{n})  \nonumber \\
    &=& 
    \int d^3q \left[ 1 + \delta_{\rm bias}(\VEC{q}) \right] 
      \delta_{\rm D}\left( \VEC{x} - \VEC{q} - \PP_{\rm red}(\VEC{q},\hat{n}) \right),
    \label{Eq:delta_g}
\end{eqnarray}
where $\delta_{\rm g}$ represents the galaxy density fluctuation, $\delta_{\rm bias}$ is the density fluctuation including the bias effect in Lagrangian space, $\PP_{\rm red}$ is the displacement vector including the RSD effect, and $\hat{n}$ is a unit vector directed along the line of sight (LOS). The relation between $\PP$ and $\PP_{\rm red}$ is given by
\begin{eqnarray}
    \PP_{\rm red} = \PP + \frac{\dot{\PP}\cdot\hat{n}}{aH}\hat{n},
\end{eqnarray}
where $\dot{\PP}(\VEC{q})$ is the time derivative of $\PP(\VEC{q})$, and $a$ and $H$ are the scale factor and the Hubble parameter, respectively.

The $n$th order of $\PP(\VEC{q})$ in Fourier space is represented as
\begin{eqnarray}
    \widetilde{\PP}^{\,[n]}_{{\rm m}}(\VEC{k}) &=& i\,\int \frac{d^3p_1}{(2\pi)^3}\cdots\frac{d^3p_n}{(2\pi)^3}
    (2\pi)^3\delta_{\rm D}(\VEC{k}-\VEC{p}_{[1,n]}) \nonumber\\ 
    &\times& \VEC{L}_n(\VEC{p}_1,\dots,\VEC{p}_n) 
    \widetilde{\delta}_{\rm m}^{\,[1]}(\VEC{p}_1)\cdots\widetilde{\delta}_{\rm m}^{\,[1]}(\VEC{p}_n),
    \label{Eq:Psi_n}
\end{eqnarray}
where the nonlinear kernel vector functions up to the second order are given by
\begin{eqnarray}
    \VEC{L}_1(\VEC{p}_1) &=& \frac{\VEC{p}_1}{p_1^2}, \nonumber \\
    \VEC{L}_2(\VEC{p}_1,\VEC{p}_2) &=& \frac{3}{14} \frac{\VEC{p}_{[1,2]}}{|\VEC{p}_{[1,2]}|^2} \left( 1- \mu_{12}^2 \right)^2.
    \label{Eq:L2}
\end{eqnarray}
Note that the second-order kernel vector function $\VEC{L}_2(\VEC{p}_1,\VEC{p}_2)$ does not include a shift term but consists of a combination of a growth term and a tidal term, unlike $F_2(\VEC{p}_1,\VEC{p}_2)$.

The displacement vector is decomposed into two parts: the value at the origin $\VEC{q}=\VEC{0}$ and the other values. Thus, we can write
\begin{eqnarray}
    \PP_{\rm red}(\VEC{q},\hat{n}) = \YY_{\rm red}(\hat{n}) + \PP_{\rm (S)red}(\VEC{q},\hat{n}).
    \label{Eq:PPYY}
\end{eqnarray}
Here, $\PP_{\rm (S)red}=\PP_{\rm red}(\VEC{q})-\YY_{\rm red}$ is termed as the short-wavelength mode of $\PP_{\rm red}$. Consequently, $\delta_{\rm g}$ can be formally rewritten as
\begin{eqnarray}
    \delta_{\rm g}(\VEC{x},\hat{n}) = \delta_{\rm (S)g}(\VEC{x}-\YY_{\rm red}(\hat{n}),\hat{n}),
    \label{Eq:delta_IR_4}
\end{eqnarray}
where 
\begin{eqnarray}
    && 1 + \delta_{\rm (S)g}(\VEC{x},\hat{n}) \nonumber \\
    &=&
    \int \hspace{-0.1cm}d^3q \left[ 1 + \delta_{\rm bias}(\VEC{q}) \right] 
    \delta_{\rm D}\left( \VEC{x} - \VEC{q} - \PP_{\rm (S)red}(\VEC{q},\hat{n}) \right),
    \label{Eq:delta_g_S}
\end{eqnarray}
In the above derivation, we used $1 = \int d^3q \delta_{\rm D}(\VEC{x}-\VEC{q}) =\int d^3q \delta_{\rm D}(\VEC{x}-\VEC{q}-\YY_{\rm red}(\hat{n}))$. In Fourier space, it becomes
\begin{eqnarray}
    \widetilde{\delta}_{\rm g}(\VEC{k},\hat{n}) = e^{-i\VEC{k}\cdot\YY_{\rm red}(\hat{n})}\, \widetilde{\delta}_{\rm (S)g}(\VEC{k},\hat{n}).
    \label{Eq:delta_IR_4_Fourier}
\end{eqnarray}

Note that Eq.~(\ref{Eq:delta_IR_4}) is simply a rewrite of Eq.~(\ref{Eq:delta_g}) and, unlike Eq.~(\ref{Eq:delta_m_S}), is not the result of solving the equation to perturbative infinite order in the IR limit. To consider Eq.~(\ref{Eq:delta_IR_4}) as an extension of the solution of Eq.~(\ref{Eq:delta_m_S}) that takes into account the bias effect, the RSD effect, and even nonlinear effects on the displacement vector, one must make an assumption. That is, \textit{in the IR limit, all contributions from IR modes appear as the coordinate transformation through $\YY_{\rm red}$}. This is expected to be true for a single dark matter fluid in GR. However, in Appendix~\ref{Sec:DHOST}, we introduce Degenerate Higher-Order Scalar-Tensor (DHOST) theories (for reviews, see e.g.~\cite{Langlois:2018dxi,Kobayashi:2019hrl}), a type of modified gravity theory, as an example of how a simple rewrite of Eq.~(\ref{Eq:delta_g}) into Eq.~(\ref{Eq:delta_IR_4}) cannot extract only the IR effect through $\YY_{\rm red}$. Also, while this assumption is considered valid for the standard bias model in GR (e.g., see~\cite{Desjacques:2016bnm}), there is a possibility that it may be violated by other bias effects in the context of modified gravity theories~\cite{Sugiyama:2023tes,Sugiyama:2023zvd} and multicomponent fluids~\cite{Yoo:2011tq}.

\subsubsection{Perturbative expansion}

In Fourier space, the $n$th-order term of the galaxy density fluctuation can be expressed as 
\begin{eqnarray}
    \widetilde{\delta}_{\rm g}^{\,[n]}(\VEC{k}) &=& \int \frac{d^3p_1}{(2\pi)^3}\cdots\frac{d^3p_n}{(2\pi)^3}
    (2\pi)^3\delta_{\rm D}(\VEC{k}-\VEC{p}_{[1,n]}) \nonumber\\ 
    &\times& Z_n(\VEC{p}_1,\dots,\VEC{p}_n,\hat{n}) \,
    \widetilde{\delta}_{\rm m}^{\,[1]}(\VEC{p}_1)\cdots\widetilde{\delta}_{\rm m}^{\,[1]}(\VEC{p}_n).
    \label{Eq:delta_n_red}
\end{eqnarray}
Here, $Z_n(\VEC{p}_1,\dots,\VEC{p}_n)$ represents nonlinear kernel functions that incorporate both the bias effect and the RSD effect. 

Approximating both $\delta^{(\rm S)}(\VEC{x})$ and $\YY_{\rm red}$ as linear, we then obtain~\citep{Sugiyama:2020uil}
\begin{eqnarray}
    \delta_{\rm g}(\VEC{x},\hat{n}) &\xrightarrow[{\rm IR\, limit}]{}& \delta_{\rm g}^{[1]}(\VEC{x} - \YY_{\rm red}^{[1]}(\hat{n}),\hat{n}),
    \nonumber \\
    \widetilde{\delta}_{\rm g}(\VEC{k},\hat{n}) &\xrightarrow[{\rm IR\, limit}]{}& 
    e^{-i\VEC{k}\cdot\YY_{\rm red}^{[1]}(\hat{n})}\widetilde{\delta}_{\rm g}^{\,[1]}(\VEC{k},\hat{n}).
    \label{Eq:delta_g_sb}
\end{eqnarray}
The linear galaxy density fluctuation is given by
\begin{eqnarray}
\widetilde{\delta}_{\rm g}^{\,[1]}(\VEC{k},\hat{n}) = Z^{[1]}(\VEC{k},\hat{n}) \widetilde{\delta}^{\,[1]}_{\rm m}(\VEC{k}),
\end{eqnarray}
where $Z_1(\VEC{k},\hat{n})$ is the first-order kernel function~\cite{Kaiser:1987qv}:
\begin{eqnarray}
    Z_1(\VEC{k},\hat{n}) = b_1+f\mu^2.
    \label{Eq:Z1}
\end{eqnarray}
Here, $b_1$ is the linear bias parameter, $f$ represents the linear growth rate function, and $\mu=\hat{k}\cdot\hat{n}$ denotes the cosine of the angle between $\hat{k}$ and $\hat{n}$. 

From Eq.~(\ref{Eq:delta_IR_4_Fourier}), up to the third order in perturbation theory, $\widetilde{\delta}^{\,[n]}_{\rm (S)g}(\VEC{k},\hat{n})$ are expressed as
\begin{eqnarray}
    \widetilde{\delta}^{\,[1]}_{\rm (S)g}(\VEC{k},\hat{n}) &=&\widetilde{\delta}^{\,[1]}_{\rm g}(\VEC{k},\hat{n})
    \nonumber \\
    \widetilde{\delta}^{\,[2]}_{\rm (S)g}(\VEC{k},\hat{n}) &=& 
    \widetilde{\delta}^{\,[2]}_{\rm g}(\VEC{k},\hat{n})
    - \left( -i\VEC{k}\cdot\YY_{\rm red}^{[1]}(\hat{n}) \right) \widetilde{\delta}^{\,[1]}_{\rm g}(\VEC{k},\hat{n})
    \nonumber \\
    \widetilde{\delta}^{\,[3]}_{\rm (S)g}(\VEC{k},\hat{n}) &=& 
    \widetilde{\delta}^{\,[3]}_{\rm g}(\VEC{k},\hat{n})
    - \left( -i\VEC{k}\cdot\YY_{\rm red}^{[2]}(\hat{n}) \right) \widetilde{\delta}^{\,[1]}_{\rm g}(\VEC{k},\hat{n})
    \nonumber \\
    &-&
    \left( -i\VEC{k}\cdot\YY_{\rm red}^{[1]}(\hat{n}) \right) \widetilde{\delta}^{\,[2]}_{\rm g}(\VEC{k},\hat{n})
    \nonumber \\
    &+& \frac{1}{2}\left( -i\VEC{k}\cdot\YY_{\rm red}^{[1]}(\hat{n}) \right)^2\widetilde{\delta}^{\,[1]}_{\rm g}(\VEC{k},\hat{n}).
    \label{Eq:delta_S_n}
\end{eqnarray}
The $n$th-order displacement vector in redshift space is computed by the following linear transformation as~\cite{Matsubara:2008wx}
\begin{eqnarray}
    \PP_{\rm red}^{[n]}(\VEC{q},\hat{n}) = \MAT{R}_n(\hat{n}) \cdot \PP^{[n]}(\VEC{q})\;.
\end{eqnarray}
The transformation matrix is given by
\begin{eqnarray}
    [ \MAT{R}_n ]_{ij} = \MAT{I}_{ij} + n\, f\, \hat{n}_i\hat{n}_j ,
    \label{Eq:R}
\end{eqnarray}
where $\VEC{I}$ is the three-dimensional identity matrix and $i,j=1,2,3$. 

Throughout the remainder of this paper, we always consider both RSD and bias effects. To simplify notation, the subscripts ``g'' and ``red,'' as well as the LOS-dependence $\hat{n}$, are omitted in the subsequent sections. We retain the subscript ``m'' only for quantities related to dark matter: i.e., $\PP_{\rm red}(\VEC{q},\hat{n}) = \PP(\VEC{q})$, $\YY_{\rm red}(\hat{n}) = \YY$, $\delta_{\rm g}(\VEC{x},\hat{n}) = \delta(\VEC{x})$, and $Z_n(\VEC{p}_1,\dots,\VEC{p}_n,\hat{n})=Z_n(\VEC{p}_1,\dots,\VEC{p}_n)$. Furthermore, a simple right arrow will represent the IR limit: $\xrightarrow[\rm IR\, limit]{} = \to$.

\subsection{IR-cancellation of two-point statistics in the IR limit}
\label{Sec:IR_Cancellation}

In this subsection, we provide a brief overview of the IR effect cancellation in the 2PCF and its Fourier-transformed counterpart, the power spectrum. A fundamental assumption is \textit{the complete absence of correlation between $\delta_{\rm (S)}(\VEC{x})$ and $\YY$}, which corresponds to the second assumption in the IR limit discussed at the end of Section~\ref{Sec:IR_kernel}. Owing to the translational symmetry of the ensemble average, all IR effects originating from $\YY$ are canceled out, leaving only the short-wavelength 2PCF:
\begin{eqnarray}
      \left\langle\delta(\VEC{x}) \delta(\VEC{x}') \right\rangle 
      &=&
      \big\langle \delta_{\rm (S)}(\VEC{x} - \YY) \delta_{\rm (S)}(\VEC{x}' - \YY) \big\rangle \nonumber \\
    &\to&
    \big\langle \delta_{\rm (S)}(\VEC{x}) \delta_{\rm (S)}(\VEC{x}')\big\rangle\;.
\end{eqnarray}
Assuming that all nonlinear effects emerge only from the IR effects through $\YY^{[1]}$, then the nonlinear 2PCF reduces to the linear 2PCF:
\begin{eqnarray}
      \left\langle\delta(\VEC{x}) \delta(\VEC{x}') \right\rangle 
      &\to&
      \big\langle \delta^{[1]}(\VEC{x} - \YY^{[1]}) \delta^{[1]}(\VEC{x}' - \YY^{[1]}) \big\rangle \nonumber \\
      &=& 
      \big\langle \delta^{[1]}(\VEC{x}) \delta^{[1]}(\VEC{x}') \big\rangle\;.
\end{eqnarray}
In Fourier space, it becomes
\begin{eqnarray}
    \left\langle \delta(\VEC{k}) \delta(\VEC{k}') \right\rangle
    &\to&
    \big\langle e^{-i\VEC{k}\cdot\YY^{[1]}} e^{-i\VEC{k}'\cdot\YY^{[1]}} \big\rangle
    \big\langle \widetilde{\delta}^{\,[1]}(\VEC{k}) 
    \widetilde{\delta}^{\,[1]}(\VEC{k}') \big\rangle \nonumber \\
    &=& (2\pi)^3\delta_{\rm D}(\VEC{k}+\VEC{k}')
    [Z_1(\VEC{k})]^2\, P^{[11]}_{\rm m}(k)\;,
    \label{Eq:IR_Fourier}
\end{eqnarray}
where $P_{\rm m}^{[11]}(k)$ is the linear matter power spectrum, calculated as
\begin{eqnarray}
    \langle  \widetilde{\delta}_{\rm m}^{\,[1]}(\VEC{k}) \widetilde{\delta}_{\rm m}^{\,[1]}(\VEC{k}') \rangle
    =(2\pi)^3\delta_{\rm D}\left( \VEC{k}+\VEC{k}' \right) P_{\rm m}^{[11]}(k).
\end{eqnarray}

In the following, we provide a detailed analysis of the exponential function $\big\langle e^{-i\VEC{k}\cdot\YY^{[1]}} e^{-i\VEC{k}'\cdot\YY^{[1]}} \big\rangle$ in Eq.~(\ref{Eq:IR_Fourier}). Using cumulants, denoted by $\langle \cdots \rangle_{\rm c}$~\footnote{The cumulants of a statistical quantity $X$, denoted $\langle X^n \rangle_{\rm c}$, are obtained from a power series expansion of the cumulant-generating function $\ln \langle e^{X}\rangle$:
\begin{eqnarray}
    \ln\left(  \langle e^X \rangle \right)
    =  \sum_{n=1}\frac{1}{n!} \langle X^n \rangle_{\rm c}.
\end{eqnarray}
}, this can be reconstructed as
\begin{eqnarray}
    \big\langle e^{-i\VEC{k}\cdot\YY^{[1]}} e^{-i\VEC{k}'\cdot\YY^{[1]}} \big\rangle 
    = {\cal D}(\VEC{k}) {\cal D}(\VEC{k}') {\cal E}(\VEC{k},\VEC{k}'),
   \label{Eq:cumulant}
\end{eqnarray}
where ${\cal D}(\VEC{k})$ and ${\cal E}(\VEC{k},\VEC{k}')$ are defined as
\begin{eqnarray}
    {\cal D}(\VEC{k})
   &=&
   \exp\left( \frac{1}{2}\left\langle \left( -i\VEC{k}\cdot\YY^{[1]} \right)^2\right\rangle_{\rm c} \right), \nonumber \\
    {\cal E}(\VEC{k},\VEC{k}')
   &=&
   \exp\left( \left\langle \left( -i\VEC{k}\cdot\YY^{[1]}\right)\left( -i\VEC{k}'\cdot\YY^{[1]} \right)\right\rangle_{\rm c} \right).
\end{eqnarray}
The function ${\cal D}(\VEC{k})$ is subsequently described as a two-dimensional exponentially decaying function~\citep{Zeldovich:1969sb,Eisenstein:2006nj,Crocce:2007dt,Matsubara:2007wj}:
\begin{eqnarray}
    {\cal D}(\VEC{k}) = \exp\left( -\frac{ k^2(1-\mu^2)\sigma_{\perp}^2 + k^2\mu^2 \sigma^2_{\parallel} }{2} \right),
    \label{Eq:Damping}
\end{eqnarray}
where the radial and transverse components of smoothing factors, $\sigma_{\parallel}^2$ and $\sigma_{\perp}^2$, are given by
\begin{eqnarray}
    \sigma^2_{\perp} &=&
    \frac{1}{3}\int \frac{dp}{2\pi^2} P_{\rm m}^{[11]}(p), \nonumber \\
    \sigma^2_{\parallel} &=& (1+f)^2\, \sigma^2_{\perp}.
    \label{Eq:sigma_sigma}
\end{eqnarray}
In addition, the function ${\cal E}(\VEC{k},\VEC{k}')$ satisfies
\begin{eqnarray}
    {\cal E}(\VEC{k},\VEC{k}) &=& {\cal D}^2(\VEC{k}), \nonumber \\
    {\cal E}(\VEC{k},-\VEC{k}') &=& {\cal E}^{-1}(\VEC{k},\VEC{k}'), \nonumber \\
    {\cal E}(\VEC{k},-\VEC{k}) &=& {\cal D}^{-2}(\VEC{k}),
    \label{Eq:cal_E}
\end{eqnarray}
and generates an exponentially increasing function as the square of the inverse of Eq.~(\ref{Eq:Damping}) when the statistical translational symmetry $\VEC{k}+\VEC{k'}=\VEC{0}$ is satisfied. Finally, as indicated in Eq.~(\ref{Eq:IR_Fourier}), the nonlinear galaxy power spectrum converges to the linear galaxy power spectrum in the IR limit:
\begin{eqnarray}
    P(\VEC{k}) 
    &\to&
    \big\langle e^{-i\VEC{k}\cdot\YY^{[1]}} e^{i\VEC{k}\cdot\YY^{[1]}} \big\rangle 
    [Z_1(\VEC{k})]^2 P_{\rm m}^{[11]}(k) \nonumber \\
    &=&
    {\cal D}^2(\VEC{k}){\cal D}^{-2}(\VEC{k}) [Z_1(\VEC{k})]^2 P_{\rm m}^{[11]}(k)\nonumber \\
    &=&[Z_1(\VEC{k})]^2 P_{\rm m}^{[11]}(k)\;.
    \label{Eq:IRcancel}
\end{eqnarray}

\subsection{Relations to previous works}
\label{Sec:PreviousWorks}

\subsubsection{1-loop corrections in SPT}

Consider the next leading order, i.e., the 1-loop order contribution, in the framework of SPT~\cite{Bernardeau:2001qr}. This is divided into $P^{[22]}(\VEC{k})$, which is the product of the dual second-order density fluctuations, and $P^{[13]}(\VEC{k})$, which is the product of the first- and third-order density fluctuations, where $P^{\rm 1 \mathchar`-loop}(\VEC{k}) = P^{[22]}(\VEC{k}) + P^{[13]}(\VEC{k})$. 

In this context, $P_{22}$ and $P_{13}$ become
\begin{eqnarray}
P^{[22]}(\VEC{k}) &=& 2 \int \frac{d^3p_1}{(2\pi)^3}\int \frac{d^3p_2}{(2\pi)^3} 
(2\pi)^3\delta_{\rm D}(\VEC{k}-\VEC{p}_{[1,2]})
\nonumber \\
&\times&
[Z_2(\VEC{p}_1,\VEC{p}_2)]^2 P_{\rm m}^{[11]}(p_1)P_{\rm m}^{[11]}(p_2), \nonumber \\
P^{[13]}(\VEC{k}) &=& 6 Z_1(\VEC{k})\, P_{\rm m}^{[11]}(k)  \nonumber \\
&\times& \int \frac{d^3p}{(2\pi)^3}Z_3(\VEC{k},\VEC{p},-\VEC{p}) P_{\rm m}^{[11]}(p).
    \label{Eq:P22_P13}
\end{eqnarray}
In the IR limit, both $Z_2(\VEC{k},\VEC{p})$ and $Z_3(\VEC{k},\VEC{p},-\VEC{p})$ are approximated as (see Appendix~\ref{Sec:SPT})
\begin{eqnarray}
    Z_2(\VEC{k},\VEC{p}) &\xrightarrow[p\to0]{}& 
    \frac{1}{2} \left( \frac{\VEC{k}\cdot\MAT{R}_1\cdot\VEC{p}}{p^2} \right) Z_1(\VEC{k})\;, 
    \nonumber \\
    \hspace{-0.5cm} Z_3(\VEC{k},\VEC{p},-\VEC{p})&\xrightarrow[p\to0]{}&
    - \frac{1}{3!} \left( \frac{\VEC{k}\cdot\MAT{R}_1\cdot\VEC{p}}{p^2} \right)^2 Z_1(\VEC{k})\;,
    \label{Eq:Z2Z3_IR}
\end{eqnarray}
where $\MAT{R}_1$ is the transformation matrix given in Eq.~(\ref{Eq:R}). Consequently, $P^{[22]}$ and $P^{[13]}$ in the IR limit are described as
\begin{eqnarray}
    P_{\rm IR}^{[22]}(\VEC{k}) &\to&
    \left( k^2(1-\mu^2)\sigma_{\perp}^2 + k^2\mu^2 \sigma^2_{\parallel} \right) 
    \nonumber \\
    &\times&
    [Z_1(\VEC{k})]^2P_{\rm m}^{[11]}(k) \;,\nonumber \\
    P_{\rm IR}^{[13]}(\VEC{k}) &\to&
    -\left( k^2(1-\mu^2)\sigma_{\perp}^2 + k^2\mu^2 \sigma^2_{\parallel} \right) 
    \nonumber \\
    &\times&
    [Z_1(\VEC{k})]^2P_{\rm m}^{[11]}(k) \;,
    \label{Eq:P22_P13_IR}
\end{eqnarray}
where the subscript ``IR'' means the values evaluated in the IR limit. When summing up $P^{[22]}_{\rm IR}$ and $P_{\rm IR}^{[13]}$, they mutually cancel out, resulting in a net value of zero. These results align with the form obtained when expanding ${\cal D}(\VEC{k}){\cal D}^{-1}(\VEC{k})$ up to the 1-loop order in Eq.~(\ref{Eq:IRcancel}).

\subsubsection{$\Gamma$-expansion}
\label{Sec:GammaExp}

The nonlinear power spectrum can generally be decomposed into two components. One component is proportional to the linear matter power spectrum, and the other is associated with mode-coupling integrals. This relation is presented in~\citep{Crocce:2005xy,Crocce:2007dt}
\begin{eqnarray}
    P(\VEC{k}) = G^2(\VEC{k}) P_{\rm m}^{[11]}(k) + P_{\rm MC}(\VEC{k}).
    \label{Eq:P_G_MC}
\end{eqnarray}
Here, the function $G(\VEC{k})$ appearing in the first term on the right-hand side is referred to as the propagator, and the second term is known as the mode-coupling term.

The $\Gamma$-expansion corresponds to the decomposition of the power spectrum within the framework of mode-coupling integrals (e.g., see~\cite{Bernardeau:2008fa}):
\begin{eqnarray}
    P(\VEC{k}) &=& \sum_{r=1}^{\infty} r! \int \frac{d^3k_1}{(2\pi)^3} \cdots \int \frac{d^3k_r}{(2\pi)^3}
    (2\pi)^3\delta_{\rm D}(\VEC{k} - \VEC{k}_{[1,r]})\nonumber \\
    &\times& 
    \left[ \Gamma^{(r)}(\VEC{k}_1,\cdots,\VEC{k}_r) \right]^2
    P_{\rm m}^{[11]}(k_1) \cdots P_{\rm m}^{[11]}(k_r)\;,
    \label{Eq:Gamma_P}
\end{eqnarray}
where the $r$th-order kernel function $\Gamma^{(r)}$ in the $\Gamma$-expansion can be expressed using the kernel functions $Z_n$ in SPT as
\begin{eqnarray}
    && \Gamma^{(r)}(\VEC{k}_1,\cdots,\VEC{k}_r) \nonumber \\
    &=& \frac{1}{r!}\sum_{s=0}^{\infty} \frac{(r+2s)!}{2^{s}s!}
    \int \frac{d^3p_1}{(2\pi)^3} \cdots \int \frac{d^3p_s}{(2\pi)^3} \nonumber \\
    &\times& 
    Z_{r+2s}\left( \VEC{k}_1,\cdots,\VEC{k}_r,\VEC{p}_1,-\VEC{p}_1,\cdots,\VEC{p}_s,-\VEC{p}_s \right) \nonumber \\
    &\times&  
    P_{\rm m}^{[11]}(p_1) \cdots P_{\rm m}^{[11]}(p_s)\;.
    \label{Eq:Gamma_coeff}
\end{eqnarray}
When compared with Eq.~(\ref{Eq:P_G_MC}), the first-order $\Gamma$-expansion is related to the propagator, and the sum of the degrees of the $\Gamma$-expansion from the second order up to infinity constitutes the mode-coupling term. 

Assuming that all nonlinear effects arise only from the linear IR effect, Eq.~(\ref{Eq:delta_IR_4_Fourier}) shows 
\begin{eqnarray}
    \widetilde{\delta}(\VEC{k}) \to e^{-i\VEC{k}\cdot\YY^{[1]}}Z_1(\VEC{k})\widetilde{\delta}_{\rm m}^{[1]}(\VEC{k})\;,
\end{eqnarray}
and the kernel functions in SPT corresponding to this expression are given by
\begin{eqnarray}
    && Z_n(\VEC{k},\VEC{p}_1,\cdots,\VEC{p}_{n-1}) \xrightarrow[p_1,\cdots,p_{n-1}\to0]{} \nonumber \\
    &&
    \frac{1}{n!}\prod_{i=1}^{n-1} \left( \frac{\VEC{k}\cdot\VEC{R}_1\cdot\VEC{p}_i}{p_i^2} \right) Z_1(\VEC{k})\;.
\end{eqnarray}
Substituting these kernel functions $Z_n$ in this IR limit into Eq.~(\ref{Eq:Gamma_coeff}) leads to the coefficients of the $\Gamma$-expansion in the IR limit:
\begin{eqnarray}
    && \Gamma^{(r)}(\VEC{k}_1,\cdots,\VEC{k}_r) \xrightarrow[k_2,\cdots,k_r\to0;p_1,\cdots,p_s\to0]\nonumber \\
    && 
    {\cal D}(\VEC{k})\frac{1}{r!}\prod_{i=2}^{r}\left( \frac{\VEC{k}_1\cdot\VEC{R}_1\cdot\VEC{k}_i}{k_i^2} \right)Z_1(\VEC{k}_1)\;.
    \label{Eq:Gamma_coeff_IR}
\end{eqnarray}
Substituting this $\Gamma^{(r)}$ into Eq.~(\ref{Eq:Gamma_P}), we finally obtain~\cite{Sugiyama:2013pwa,Sugiyama:2013gza}
\begin{eqnarray}
    P(\VEC{k})&\to&
    {\cal D}^2(\VEC{k})
    \sum_{r=1}^{\infty} \frac{\left( k^2(1-\mu^2)\sigma_{\perp}^2 + k^2\mu^2 \sigma^2_{\parallel} \right)^{r-1} }{(r-1)!}  \nonumber \\
    &\times& 
    [Z_1(\VEC{k})]^2\, P_{\rm m}^{[11]}(k)\;.
    \label{Eq:Gamma}
\end{eqnarray}
To derive the above expression, we assumed 
\begin{eqnarray}
    \delta_{\rm D}(\VEC{k}-\VEC{k}_{[1,r]})\to
    \xrightarrow[k_2,\cdots,k_r\to0]{}
    r\, \delta_{\rm D}(\VEC{k}-\VEC{k}_1)\;,
    \label{Eq:GammaExp_delta_r}
\end{eqnarray}
where the factor $r$ appears because we select one wave vector $\VEC{k}_1$ from $r$ wave vectors. 

Furthermore, we obtain the IR limit expressions of $G^2(\VEC{k})$ and $P_{\rm MC}(\VEC{k})$ from Eq.~(\ref{Eq:Gamma}):
\begin{eqnarray}
    G^2(\VEC{k}) P_{\rm m}^{[11]}(k) 
    &\to& {\cal D}^2(\VEC{k}) [Z_1(\VEC{k})]^2P_{\rm m}^{[11]}(k) \;, \nonumber \\
    P_{\rm MC}(\VEC{k}) 
    &\to&
    {\cal D}^2(\VEC{k}) \left( {\cal E}(\VEC{k},-\VEC{k})-1 \right)[Z_1(\VEC{k})]^2P_{\rm m}^{[11]}(k) 
    \nonumber \\
    &=&
    \big( 1 - {\cal D}^2(\VEC{k}) \big)[Z_1(\VEC{k})]^2P_{\rm m}^{[11]}(k) \;,
    \label{Eq:GMC}
\end{eqnarray}
where ${\cal E}(\VEC{k},-\VEC{k})={\cal D}^{-2}(\VEC{k})$ in Eq.~(\ref{Eq:cal_E}). It is important to note that the function ${\cal E}(\VEC{k},\VEC{k}')$ serves to connect different wave vectors $\VEC{k}$ and $\VEC{k}'$. Consequently, a series expansion of ${\cal E}(\VEC{k},\VEC{k}')-1$ yields a term related to the mode-coupling integrals.

\subsubsection{Propagator}

In this paper, we focus on the impact of the BAO signal due to non-perturbative IR effects. To that end, we provide a detailed explanation of the calculation of the propagator, which directly influences the nonlinear decay of BAO.

The propagator is defined by the correlation between two factors: nonlinear galaxy density fluctuations and linear dark matter density fluctuations.
\begin{eqnarray}
    \langle \delta(\VEC{x}) \delta_{\rm m}^{[1]}(\VEC{x}')\rangle
    = \int \frac{d^3k}{(2\pi)^3} e^{i\VEC{k}\cdot(\VEC{x}-\VEC{x}')} G(\VEC{k}) P_{\rm m}^{[11]}(k).
\end{eqnarray}
Note that this definition assumes the absence of primordial non-Gaussianities. When the nonlinear density fluctuations are decomposed into long-wavelength displacement vectors and short-wavelength density fluctuations, as shown in Eq.~(\ref{Eq:delta_IR_4}), the propagator can be computed as 
\begin{eqnarray}
    && \langle \delta_{\rm (S)}(\VEC{x}-\YY) \delta_{\rm m}^{[1]}(\VEC{x}')\rangle \nonumber \\
    &=& \int \frac{d^3k}{(2\pi)^3} e^{i\VEC{k}\cdot(\VEC{x}-\VEC{x}')}
    \left\langle e^{-i\VEC{k}\cdot\YY} \right\rangle\nonumber \\
    &\times&
    \left[   Z_1(\VEC{k}) P_{\rm m}^{[11]}(k) 
       +
       \sum_{n=1}^{\infty} P_{\rm (S)}^{[1(2n+1)]}(\VEC{k})/(2Z_1(\VEC{k}))\right],
    \label{Eq:prop}
\end{eqnarray}
where $\delta_{\rm m}^{[1]}(\VEC{x})$ corresponds to short-wavelength modes and thus correlates only with $\delta_{\rm (S)}(\VEC{x})$, and $P_{\rm (S)}^{[1(2n+1)]}$ represents the cross-power spectrum of $\delta_{\rm (S)}^{[2n+1]}$ and $\delta_{\rm m}^{[1]}$. Note that only odd orders appear in $P_{\rm (S)}^{[1(2n+1)]}$ since primordial non-Gaussianities are absent. 

By truncating $\delta_{\rm (S)}$ up to the third order and by further approximating $\YY$ to be of linear order, we obtain 
\begin{eqnarray}
    \hspace{-0.4cm} &&G(\VEC{k}) \nonumber \\
   \hspace{-0.4cm}
   &\simeq& {\cal D}(\VEC{k})
    \left[ Z_1(\VEC{k}) 
     + P_{\rm (S)}^{[13]}(\VEC{k})\Big/\left(2Z_1(\VEC{k})P_{\rm m}^{[11]}(k)  \right) \right].
    \label{Eq:prop_1loop}
\end{eqnarray}
Here, $P_{\rm (S)}^{[13]}$ is defined as
\begin{eqnarray}
    P_{\rm (S)}^{[13]}(\VEC{k}) = P^{[13]}(\VEC{k}) - P_{\rm IR}^{[13]}(\VEC{k}),
\end{eqnarray}
where $P_{\rm IR}^{[13]}$ represents the IR limit value of $P^{[13]}$ as given in Eq.~(\ref{Eq:P22_P13_IR}).

In the definition of the propagator (\ref{Eq:prop}), it is important to note that nonlinear density fluctuations and linear density fluctuations possess different IR effects. In particular, there is no IR cancellation based on statistical translational symmetry in the calculation of the propagator. The exponential decay of the propagator is due to the absence of this IR cancellation. Similarly, other higher-order coefficients in the $\Gamma$-expansion decay exponentially, as shown in Eq.~(\ref{Eq:Gamma}). 

This paper repeatedly emphasizes that an exponential decay effect is consistently observed when calculating cross-correlations between fluctuations with distinct IR effects. Notably, Section~\ref{Sec:Cross} will demonstrate the exponential decay of the cross-power spectrum for both pre- and post-reconstruction density fluctuations~\footnote{As another example, \citet{Chisari:2019tig} showed in the Zel'dovich approximation that the cross-power spectrum of density fluctuations evaluated at two different times decays exponentially due to IR cancellation breaking.}.

\subsection{IR-resummed model for \\ the pre-reconstruction power spectrum}
\label{Sec:IRresummed_pre}

In the previous section, we discussed the IR cancellation under the assumption that $\delta^{[1]}(\VEC{x})$ and $\YY^{[1]}$ are uncorrelated. This lack of correlation is due to the three assumptions mentioned at the end of Section~\ref{Sec:General}, especially the one stating that there is no mode coupling between the IR effect and the short-wavelength density fluctuation. In practice, however, there is mode coupling that prevents complete IR cancellation. By addressing this imperfection in IR cancellation and isolating only the effects of BAO, we can construct a theoretical template model that effectively handles the nonlinear damping effects of BAO. 

\subsubsection{Linear level}

We begin by relaxing the approximation of the IR limit concerning the behavior of the power spectrum in the context of the $\Gamma$-expansion discussed in Section~\ref{Sec:GammaExp}. Specifically, we do not use the approximation that decouples the coupling between different wave vectors, as in Eq.~(\ref{Eq:GammaExp_delta_r}), while maintaining the IR limit approximation of the kernel functions of the $\Gamma$-expansion given in Eq.~(\ref{Eq:Gamma_coeff_IR}). We then obtain
\begin{eqnarray}
    && P(\VEC{k}) \nonumber \\
    &\to& {\cal D}^2(\VEC{k})\sum_{r=1}^{\infty} \frac{1}{(r-1)!} \int \frac{d^3k_1}{(2\pi)^3} \cdots \int \frac{d^3k_r}{(2\pi)^3}\nonumber \\
    &\times& (2\pi)^3\delta_{\rm D}(\VEC{k} - \VEC{k}_{[1,r]})
    \prod_{i=2}^{r}\left( \frac{\VEC{k}_1\cdot\VEC{R}_1\cdot\VEC{k}_i}{k_i^2} \right)^2\nonumber \\
    &\times& \left[ Z_1(\VEC{k}_1) \right]^2P_{\rm m}^{[11]}(k_1) \cdots P_{\rm m}^{[11]}(k_r)\;.
    \label{Eq:Gamma_P_mode}
\end{eqnarray}

To account for the damping effect of the BAO signal, we decompose the linear matter power spectrum $P_{\rm m}^{[11]}(k_1)$ in Eq.~(\ref{Eq:Gamma_P_mode}) into two components: the ``wiggle'' component containing only the BAO signal and the ``no-wiggle'' component without BAO~\cite{Eisenstein:1997ik}. Thus, it is expressed as $P_{\rm m}^{[11]}(k_1) = P_{\rm w}(k_1) + P_{\rm nw}(k_1)$, where the subscripts ``w'' and ``nw'' denote ``wiggle'' and ``no-wiggle,'' respectively. Note that this decomposition is only applied to $P_{\rm m}^{[11]}(k_1)$, and not to other linear power spectra $P_{\rm m}^{[11]}(k_2)\dots P_{\rm m}^{[11]}(k_r)$. According to this decomposition, the nonlinear power spectrum that includes the wiggle part and the no-wiggle part is denoted as $P(\VEC{k})|_{\rm w}$ and $P(\VEC{k})|_{\rm nw}$, respectively:
\begin{eqnarray}
    P(\VEC{k}) =  P(\VEC{k})|_{\rm w} +  P(\VEC{k})|_{\rm nw}\;.
    \label{Eq:P_NL_w_nw}
\end{eqnarray}

For the $P(\VEC{k})|_{\rm nw}$ term, we adopt the approximation in Eq.~(\ref{Eq:GammaExp_delta_r}) and achieve full IR cancellation, leaving only the linear no-wiggle power spectrum:
\begin{eqnarray}
    P(\VEC{k})|_{\rm nw} \to [Z_1(\VEC{k})]^2 P_{\rm nw}(k)\;.
    \label{Eq:P_NL_nw}
\end{eqnarray}

For the $P(\VEC{k})|_{\rm w}$ term, Eq.~(\ref{Eq:Gamma_P_mode}) is transformed using $\delta_{\rm D}(\VEC{k}-\VEC{k}_{[1,r]})=\int d^3r e^{-i\VEC{r}\cdot(\VEC{k}-\VEC{k}_{[1,r]})}$ to become
\begin{eqnarray}
    P(\VEC{k})|_{\rm w} 
    \to {\cal D}^2(\VEC{k})
    \int d^3r e^{-i\VEC{k}\cdot\VEC{r}} \xi_{\rm w}(\VEC{r}) {\cal A}^2(\VEC{k},\VEC{r})\;,
\end{eqnarray}
where
\begin{eqnarray}
    \xi_{\rm w}(\VEC{r}) = \int \frac{d^3k_1}{(2\pi)^3} e^{i\VEC{k}_1\cdot\VEC{r}} \left[ Z_1(\VEC{k}_1) \right]^2 P_{\rm w}(k_1)\;,
    \label{Eq:xi_w}
\end{eqnarray}
and
\begin{eqnarray}
    && {\cal A}^2(\VEC{k},\VEC{r}) \nonumber \\
    &=& 
    \exp
    \left\{ \int \frac{d^3p}{(2\pi)^3} e^{i\VEC{p}\cdot\VEC{r}} \left( \frac{\VEC{k}\cdot\VEC{R}_1\cdot\VEC{p}}{p^2} \right)^2 P_{\rm m}^{[11]}(p) \right\}\;.
\end{eqnarray}
The wave vector integral in the ${\cal A}$ function is calculated as 
\begin{eqnarray}
    && \int \frac{d^3p}{(2\pi)^3} e^{i\VEC{p}\cdot\VEC{r}} \left(\frac{\hat{p}_i\hat{p}_j}{p^2}  \right) P_{\rm m}^{[11]}(p)  \nonumber \\
    &=& \delta_{ij} \sigma^2_0(r) + 2 \left( \frac{3\hat{r}_i\hat{r}_j-\delta_{ij}}{2} \right) \sigma^2_2(r)\;,
\end{eqnarray}
where
\begin{eqnarray}
    \sigma^2_{\ell}(r) = \frac{1}{3}i^{\ell} \int \frac{dp}{2\pi^2} j_{\ell}(pr) P_{\rm m}^{[11]}(p) \;.
\end{eqnarray}
To simplify the calculation, we only consider the isotropic component of the ${\cal A}$ function and ignore the $\sigma_2$ term. Furthermore, the 2PCF of the wiggle part in Eq.~(\ref{Eq:xi_w}) has a peak around $r_{\rm BAO}\sim110\hMpc$ and is zero at other scales, so we fix the scale dependence that appears in the $\sigma_0(r)$ function to $r=r_{\rm BAO}$. With these approximations, the ${\cal A}$ function becomes
\begin{eqnarray}
    \hspace{-0.7cm}
    {\cal A}(\VEC{k}) \approx \exp\left( \frac{ k^2(1-\mu^2)\sigma_{0}^2(r_{\rm BAO}) + k^2\mu^2 \sigma^2_{0,\,\parallel}(r_{\rm BAO}) }{2} \right),
    \label{Eq:A}
\end{eqnarray}
where $\sigma^2_{0,\,\parallel}=(1+f)^2\sigma_0^2$. Note that the $\VEC{r}$-dependence has been omitted here. Then, we finally obtain 
\begin{eqnarray}
    P(\VEC{k})|_{\rm w} \to {\cal D}_{\rm BAO}^2(\VEC{k}) \left[ Z_1(\VEC{k}) \right]^2 P_{\rm w}(k)\;.
    \label{Eq:P_NL_w}
\end{eqnarray}
The function that expresses the nonlinear damping of the BAO signal is defined as follows:
\begin{eqnarray}
    {\cal D}(\VEC{k}){\cal A}(\VEC{k}) = {\cal D}_{\rm BAO}(\VEC{k})\;,
\end{eqnarray}
and its specific form is given by
\begin{eqnarray}
    \hspace{-0.7cm}
    {\cal D}_{\rm BAO}(\VEC{k}) = \exp\left( -\frac{ k^2(1-\mu^2)\sigma_{{\rm BAO},\,\perp}^2 + k^2\mu^2 \sigma^2_{ {\rm BAO},\,\parallel} }{2} \right),
    \label{Eq:Damping_BAO}
\end{eqnarray}
where the radial and transverse components of smoothing factors are calculated as
\begin{eqnarray}
    \sigma^2_{ {\rm BAO},\,\perp} &=& 
    \frac{1}{3}\int \frac{dp}{2\pi^2} \left( 1 - j_0(pr_{\rm BAO}) \right)P_{\rm m}^{[11]}(p), \nonumber \\
    \sigma^2_{{\rm BAO},\,\parallel} &=& (1+f)^2\, \sigma^2_{{\rm BAO},\,\perp}.
    \label{Eq:sigma_sigma_BAO}
\end{eqnarray}

Substituting Eqs.~(\ref{Eq:P_NL_nw}) and (\ref{Eq:P_NL_w}) into Eq.~(\ref{Eq:P_NL_w_nw}), we obtain the IR-resummed power spectrum model at the linear level:
\begin{eqnarray}
    P(\VEC{k}) = \big[ Z_1(\VEC{k}) \big]^2\left[  {\cal D}_{\rm BAO}^2(\VEC{k}) P_{\rm w}(k) + P_{\rm nw}(k)\right]\;.
    \label{Eq:P_IR_BAO}
\end{eqnarray}
This form is consistent with the template model proposed by~\cite{Eisenstein:2006nj}, which has been widely used for BAO analyses. 

When the decomposition of the nonlinear power spectrum into the wiggle and no-wiggle parts is applied to the propagator and mode-coupling term in Eq.~(\ref{Eq:P_G_MC}), we obtain
\begin{eqnarray}
    \hspace{-0.7cm}
    G^2(\VEC{k}) P_{\rm m}^{[11]}(k) \to {\cal D}^2(\VEC{k}) \left[ Z_1(\VEC{k}) \right]^2\left[ P_{\rm w}(k) + P_{\rm nw}(k) \right]\;,
\end{eqnarray}
and
\begin{eqnarray}
    P_{\rm MC}(\VEC{k}) &\to& {\cal D}^2(\VEC{k})\left[ {\cal D}^{-2}(\VEC{k}) -1\right] \left[ Z_1(\VEC{k}) \right]^2 P_{\rm nw}(k) \nonumber \\
    &+& {\cal D}^2(\VEC{k}) \left[ {\cal A}^2(\VEC{k}) -1\right]  \left[ Z_1(\VEC{k}) \right]^2 P_{\rm w}(k) \nonumber \\
    &=&\big[ Z_1(\VEC{k}) \big]^2\left[  {\cal D}_{\rm BAO}^2(\VEC{k}) P_{\rm w}(k) + P_{\rm nw}(k)\right] \nonumber \\
    &-&{\cal D}^2(\VEC{k}) \left[ Z_1(\VEC{k}) \right]^2\left[ P_{\rm w}(k) + P_{\rm nw}(k) \right]\;.
    \label{Eq:P_MC_lin_BAO}
\end{eqnarray}
These expressions show that the nonlinear damping effect of BAO arises from the mode-coupling term, and the contribution from the propagator term cancels out with the second term of the mode-coupling term.

\subsubsection{1-loop level}

We turn to the 1-loop order correction terms. In the IR limit, the nonlinear galaxy power spectrum is described as
\begin{eqnarray}
    P(\VEC{k}) \to {\cal D}^2(\VEC{k}) {\cal D}^{-2}(\VEC{k})
    P_{\rm (S)}(\VEC{k})\;,
\end{eqnarray}
where $P_{\rm (S)}$ is consists only of the short-wavelength density fluctuations, $\widetilde{\delta}_{\rm (S)}$. At the 1-loop level, we derive the short-wavelength power spectra from Eq.~(\ref{Eq:delta_S_n}) as
\begin{eqnarray}
    \hspace{-0.5cm}P_{\rm (S)}^{[22]}(\VEC{k}) &=& P^{[22]}(\VEC{k}) - P_{\rm IR}^{[22]}(\VEC{k}) \;, \nonumber \\
    \hspace{-0.5cm}&=& P^{[22]}(\VEC{k}) - \ln {\cal D}^{-2}(\VEC{k}) \left[ Z_1(\VEC{k}) \right]^2 P_{\rm m}^{[11]}(k)\;, \nonumber \\
    \hspace{-0.5cm}P_{\rm (S)}^{[13]}(\VEC{k}) &=& P^{[13]}(\VEC{k}) - P_{\rm IR}^{[13]}(\VEC{k}) \;, \nonumber \\
    \hspace{-0.5cm}&=& P^{[13]}(\VEC{k}) - \ln {\cal D}^2(\VEC{k}) \left[ Z_1(\VEC{k}) \right]^2 P_{\rm m}^{[11]}(k)\;,
\end{eqnarray}
where we used the uncorrelation between $\YY$ and $\delta_{\rm (S)}$. Compared to the propagator including the 1-loop correction term in Eq.~(\ref{Eq:prop_1loop}), the propagator and the mode-coupling term at the 1-loop level are shown as
\begin{eqnarray}
    && G^2(\VEC{k}) P_{\rm m}^{[11]}(k) \nonumber \\
    &=&
    {\cal D}^2(\VEC{k}) \left[ \left[ Z_1(\VEC{k})  \right]^2 P_{\rm m}^{[11]}(k) + P_{\rm (S)}^{[13]}(\VEC{k})  \right]\;,
    \label{Eq:P_G_1loop_IR}
\end{eqnarray}
and
\begin{eqnarray}
   && P_{\rm MC}(\VEC{k}) \nonumber \\
   &=&{\cal D}^2(\VEC{k}) \left( {\cal D}^{-2}(\VEC{k}) - 1 \right)
   \left[ \left[ Z_1(\VEC{k})  \right]^2 P_{\rm m}^{[11]}(k) + P_{\rm (S)}^{[13]}(\VEC{k})  \right] \nonumber \\
    &+& P_{\rm (S)}^{[22]}(\VEC{k}) \;.
    \label{Eq:P_MC_1loop_IR}
\end{eqnarray}
When Eq.~(\ref{Eq:P_G_1loop_IR}) and Eq.~(\ref{Eq:P_MC_1loop_IR}) are added, they yield the 1-loop power spectrum in SPT.

To describe the nonlinear damping of the BAO signal, we recalculate the mode-coupling term in Eq.~(\ref{Eq:P_MC_1loop_IR}). First, given that $\left[ Z_1(\VEC{k})  \right]^2 P_{\rm m}^{[11]}(k) + P_{\rm (S)}^{[13]}(\VEC{k}) \propto P_{\rm m}^{[11]}(k)$, we can directly apply the computational techniques used to derive the linear IR resummation model in Eq.~(\ref{Eq:P_MC_lin_BAO}) to the first term on the right-hand side of Eq.~(\ref{Eq:P_MC_1loop_IR}). Therefore,
\begin{eqnarray}
   && {\cal D}^2(\VEC{k}) \left( {\cal D}^{-2}(\VEC{k}) - 1 \right)
   \left[ \left[ Z_1(\VEC{k})  \right]^2 P_{\rm m}^{[11]}(k) + P_{\rm (S)}^{[13]}(\VEC{k})  \right] \nonumber \\
   && \xrightarrow[ {\rm IR-resummed}]{} \nonumber \\
   && {\cal D}^2(\VEC{k}) \left( {\cal D}^{-2}(\VEC{k}) - 1 \right) \nonumber \\
    &&\times \left[ \left[ Z_1(\VEC{k})  \right]^2 \left( 1- \ln {\cal D}^2(\VEC{k}) \right) P_{\rm nw}(k) + P_{\rm nw}^{[13]}(\VEC{k})  \right] \nonumber \\
    &+& {\cal D}^2(\VEC{k}) \left( {\cal A}^{2}(\VEC{k}) - 1 \right) \nonumber \\
    &&\times \left[ \left[ Z_1(\VEC{k})  \right]^2 \left( 1- \ln {\cal D}^2(\VEC{k}) \right) P_{\rm w}(k) + P_{\rm w}^{[13]}(\VEC{k})  \right]\;,
    \label{Eq:P_MC_BAO_1}
\end{eqnarray}
where
\begin{eqnarray}
    \hspace{-0.5cm}P_{\rm nw}^{[13]}(\VEC{k}) &=& \left( P_{\rm nw}(k)/P_{\rm m}^{[11]}(k) \right) P^{[13]}(\VEC{k})\;, \nonumber \\
    \hspace{-0.5cm}P_{\rm w}^{[13]}(\VEC{k}) &=& \left( P_{\rm w}(k)/P_{\rm m}^{[11]}(k) \right) P^{[13]}(\VEC{k})\;.
\end{eqnarray}

Furthermore, in the IR limit, $P_{ {\rm (S)}}^{[22]}$ can be explicitly written as
\begin{eqnarray}
    \hspace{-0.5cm}P_{\rm (S)}^{[22]}(\VEC{k}) &=&
    {\cal D}^2(\VEC{k}) {\cal D}^{-2}(\VEC{k}) \nonumber \\
    \hspace{-0.5cm}&\times& \left[ P_{22}(\VEC{k}) - \ln {\cal D}^{-2}(\VEC{k}) [Z_1(\VEC{k})]^2 P_{\rm m}^{[11]}(k) \right]\;.
    \label{Eq:P_22_S}
\end{eqnarray}
In this context, since $P_{22}(\VEC{k})$ appropriately includes the wiggle part within the mode-coupling integral, no further manipulation is required. For the second term on the right-hand side of Eq.~(\ref{Eq:P_22_S}), when considering the wiggle part, ${\cal D}^{-2}$ originating from the mode-coupling integral should be replaced by ${\cal A}^2$. Conversely, for the no-wiggle part, ${\cal D}^{-2}$ remains unchanged. Thus, we obtain
\begin{eqnarray}
    && P_{\rm (S)}^{[22]}(\VEC{k}) 
    \xrightarrow[ {\rm IR-resummed}]{} \nonumber \\
   && P_{22}(\VEC{k})  - \ln {\cal D}^{-2}(\VEC{k}) [Z_1(\VEC{k})]^2 P_{\rm nw}(k) \nonumber \\
   && - {\cal D}^2(\VEC{k}){\cal A}^2(\VEC{k}) \left( \ln {\cal A}^{2}(\VEC{k}) \right)[Z_1(\VEC{k})]^2 P_{\rm w}(k) \;.
    \label{Eq:P_MC_BAO_2}
\end{eqnarray}

Substituting Eqs.~(\ref{Eq:P_MC_BAO_1}) and (\ref{Eq:P_MC_BAO_2}) into Eq.~(\ref{Eq:P_MC_1loop_IR}) and then combining with Eq.~(\ref{Eq:P_G_1loop_IR}), we obtain the IR-resummed power spectrum model at the 1-loop level:
\begin{eqnarray}
    \hspace{-0.5cm} P(\VEC{k}) &=& 
    {\cal D}_{\rm BAO}^2(\VEC{k}) \nonumber \\
    \hspace{-0.5cm}&\times&\left[ \left[ Z_1(\VEC{k})  \right]^2
        \left( 1 - \ln {\cal D}^2_{\rm BAO}(\VEC{k}) \right)P_{\rm w}(k) + 
    P_{\rm w}^{[13]}(\VEC{k}) \right]  \nonumber \\
    \hspace{-0.5cm}&+& \left[ Z_1(\VEC{k})  \right]^2 P_{\rm nw}(k) 
    + P^{[13]}_{\rm nw}(\VEC{k}) + P^{[22]}(\VEC{k}) \;.
    \label{Eq:P_IR_BAO_1loop}
\end{eqnarray}

\begin{figure}[t]
    \centering
    \includegraphics[width=\columnwidth]{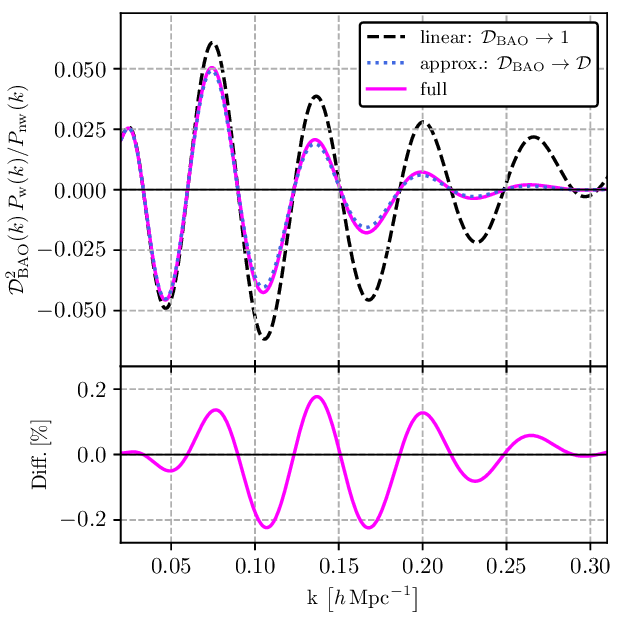}
    \caption{Ratio of the wiggle part to the no-wiggle part in the linear IR-resummed model given by Eq.~(\ref{Eq:P_IR_BAO}) at $z=0$ in real space. The upper panel shows the result using ${\cal D}_{\rm BAO}$ (magenta solid line), the linear theory (black dashed line), and the approximation where ${\cal D}_{\rm BAO}$ is replaced by ${\cal D}$ (blue dotted line). The lower panel plots the difference, ${\rm Diff.}= \left[ {\cal D}^2_{\rm BAO}- {\cal D}^2\right] P_{\rm w}/P_{\rm nw}$. }
    \label{fig:pre_recon}
\end{figure}

\subsection{BAO signal in the mode coupling term}

The IR-resummed power spectrum model, which is obtained by decomposing the linear power spectrum into the wiggle and no-wiggle parts, as in Eqs.~(\ref{Eq:P_IR_BAO}) and E(\ref{Eq:P_IR_BAO_1loop}), was derived by~\citet{Sugiyama:2013gza} based on the results of numerical experiments showing that the BAO signal in the mode-coupling term was negligibly small. The model derived there was one in which all ${\cal D}_{\rm BAO}(\VEC{k})$ appearing in Eqs.~(\ref{Eq:P_IR_BAO}) and (\ref{Eq:P_IR_BAO_1loop}) was replaced with ${\cal D}(\VEC{k})$. 

Subsequent studies~\cite{Baldauf:2015xfa,Vlah:2015zda,Senatore:2014via,Blas:2016sfa,Senatore:2017pbn,Ivanov:2018gjr,Lewandowski:2018ywf} introduced considerations regarding the wiggle part included in the mode-coupling term. It was understood that the wiggle part within the propagator term cancels out with a part of the wiggle part in the mode-coupling term, ultimately leading to the conclusion that the nonlinear damping of the BAO signal is explained by the wiggle part in the mode-coupling term. Consequently, the nonlinear damping of the BAO signal is now described by ${\cal D}_{\rm BAO}$, as shown in Eqs.~(\ref{Eq:P_IR_BAO}) and (\ref{Eq:P_IR_BAO_1loop}), rather than by ${\cal D}$. However, in the context of the $\Lambda$CDM model, the numerical difference between ${\cal D}_{\rm BAO}$ and ${\cal D}$ is known to be minimal~\cite{Blas:2016sfa}.

Figure~\ref{fig:pre_recon} numerically demonstrates the nonlinear effects on the BAO signal in real space, assuming $f=0$. This figure plots the ratio of the wiggle part to the no-wiggle part in the linear IR-resummed model given by Eq.~(\ref{Eq:P_IR_BAO}) at $z=0$ (magenta solid line). For comparison, the linear prediction (black dashed line) and the approximation where ${\cal D}_{\rm BAO}$ is replaced by ${\cal D}$ as provided by~\cite{Sugiyama:2013gza} (blue dotted line) are also plotted. The lower panel of this figure shows the difference between the result from Eq.~(\ref{Eq:P_IR_BAO}) and the approximation where ${\cal D}_{\rm BAO}$ is replaced by ${\cal D}$, defined as ${\rm Diff.}= \left[ {\cal D}^2_{\rm BAO}-{\cal D}^2 \right] P_{\rm w}/P_{\rm nw}$. 

From this figure, we can see that even if we approximate ${\cal D}_{\rm BAO}\approx {\cal D}$ and ignore the wiggles in the mode-coupling term in Eq.~(\ref{Eq:P_MC_lin_BAO}), the numerical effect on the power spectrum is about $0.2\%$. This is the result for the case of $z=0$, where the nonlinear effect is at its maximum, and for actual observations at redshifts $z=0.2-0.2$, this difference will be even smaller.

Therefore, in terms of constructing a model that can realistically explain observations, the approximation ${\cal D}_{\rm BAO}\approx {\cal D}$ is valid. In this case, the mode-coupling term in Eq.~(\ref{Eq:P_MC_lin_BAO}) consists only of the no-wiggle part as
\begin{eqnarray}
    P_{\rm MC}(\VEC{k}) &\approx& \left( 1 - {\cal D}^2(\VEC{k}) \right)[ Z_1(\VEC{k}) ]^2  P_{\rm nw}(k)\;,
\end{eqnarray}
leading to the linear IR-resummed model 
\begin{eqnarray}
    P(\VEC{k}) = \big[ Z_1(\VEC{k}) \big]^2\left[  {\cal D}^2(\VEC{k}) P_{\rm w}(k) + P_{\rm nw}(k)\right]\;.
    \label{Eq:P_IR_BAO_app}
\end{eqnarray}
Similarly, at the 1-loop level, the following model is also approximately satisfied~\cite{Sugiyama:2013gza}:
\begin{eqnarray}
    \hspace{-0.5cm} P(\VEC{k}) &=& 
    {\cal D}^2(\VEC{k}) \nonumber \\
    \hspace{-0.5cm}&\times&\left[ \left[ Z_1(\VEC{k})  \right]^2
        \left( 1 - \ln {\cal D}^2(\VEC{k}) \right)P_{\rm w}(k) + 
    P_{\rm w}^{[13]}(\VEC{k}) \right]  \nonumber \\
    \hspace{-0.5cm}&+& \left[ Z_1(\VEC{k})  \right]^2 P_{\rm nw}(k) 
    + P^{[13]}_{\rm nw}(\VEC{k}) + P^{[22]}(\VEC{k}) \;.
    \label{Eq:P_IR_BAO_1loop_app}
\end{eqnarray}

\section{Post-Reconstruction Case}
\label{Sec:PostRecon}

We apply the method for constructing the IR-resummed power spectrum model from the pre-reconstruction to the post-reconstruction case.

\subsection{Post-reconstruction density fluctuations}

To reconstruct the galaxy distribution, the displacement vector for reconstruction is calculated from the observed galaxy density fluctuations as follows~\cite{Eisenstein:2006nk}:
\begin{eqnarray}
    \hspace{-0.7cm}
    \VEC{s}(\VEC{x}) = 
    \int \frac{d^3p}{(2\pi)^3} e^{i\VEC{p}\cdot\VEC{x}} \left( \frac{i\VEC{p}}{p^2} \right)
    \left(  - \frac{W_{\rm G}(pR_{\rm s})}{b_{1, \rm fid}} \right)
    \widetilde{\delta}_{\rm obs}(\VEC{p}),
    \label{Eq:S}
\end{eqnarray}
where $b_{1, \rm fid}$ is an input fiducial linear bias parameter for reconstruction, $W_{\rm G}(pR_{\rm s}) = \exp\left( -p^2R_{\rm s}^2/2 \right)$ is a Gaussian filter function, and $R_{\rm s}$ is an input smoothing scale. This $\VEC{s}(\VEC{x})$ is derived from the observed galaxy fluctuation $\delta_{\rm obs}(\VEC{x})$. By substituting $\delta_{\rm obs}$ with the nonlinear density fluctuation $\delta$ in Eq.~(\ref{Eq:delta_g}), which includes the bias and RSD effects, we can make a theoretical prediction of $\VEC{s}(\VEC{x})$.

The reconstruction scheme uses $\VEC{s}(\VEC{x})$ to move each of the galaxy data particles and the corresponding random particles. Such an operation is expressed as
\begin{eqnarray}
    1 + \delta_{\rm d}(\VEC{x}) &=& \int d^3x' \left( 1 + \delta(\VEC{x}) \right)
    \delta_{\rm D}\left( \VEC{x} - \VEC{x}' - \VEC{s}(\VEC{x}')  \right) \nonumber \\
    1 + \delta_{\rm s}(\VEC{x}) &=& \int d^3x'  \delta_{\rm D}\left( \VEC{x} - \VEC{x}' - \VEC{s}(\VEC{x}')  \right).
\end{eqnarray}
Here, $\delta_{\rm d}(\VEC{x})$ is the density fluctuation of the reconstructed galaxy data particles, and $\delta_{\rm s}(\VEC{x})$ is the density fluctuation of the reconstructed random particles. The observed post-reconstruction galaxy density fluctuation is then given by~\cite{Sugiyama:2020uil,Shirasaki:2020vkk}
\begin{eqnarray}
    \delta_{\rm rec}(\VEC{x}) &=& \delta_{\rm d}(\VEC{x}) - \delta_{\rm s}(\VEC{x}) \nonumber \\
    &=&\int d^3x' \delta(\VEC{x}') \delta_{\rm D}\left( \VEC{x} - \VEC{x}' - \VEC{s}(\VEC{x}') \right).
    \label{Eq:delta_rec}
\end{eqnarray}

\subsection{IR effects on the post-reconstruction density fluctuation}
\label{Sec:Post_recon_IR}

The displacement vector for reconstruction, $\VEC{s}(\VEC{x})$, is derived from the nonlinear density fluctuation, which includes the shift term that should be considered in the IR limit [e.g., see Eq.~(\ref{Eq:F2})]. This characteristic of $\VEC{s}(\VEC{x})$ is unique, contrasting with the displacement vector $\PP(\VEC{q})$, where the shift term is excluded [e.g., Eq.~(\ref{Eq:L2})]. In this subsection, we focus on examining the non-perturbative behavior of the shift term within $\VEC{s}(\VEC{x})$ and demonstrate its impact on the post-reconstruction density fluctuation.

Following the approach introduced in Section~\ref{Sec:PreRecon}, we decompose the galaxy density fluctuation $\delta(\VEC{x})$ into two components: the IR effect represented by $\YY$, and the short-wavelength density fluctuation $\delta_{\rm (S)}(\VEC{x})$. By substituting Eq.~(\ref{Eq:delta_IR_4}) and Eq.~(\ref{Eq:delta_IR_4_Fourier}) into Eq.~(\ref{Eq:S}) and Eq.~(\ref{Eq:delta_rec}), we obtain
\begin{eqnarray}
    \hspace{-0.3cm}
    \delta_{\rm rec}(\VEC{x}) 
    &=&\int d^3x' \delta_{\rm (S)}(\VEC{x}'-\YY) \delta_{\rm D}\left( \VEC{x} - \VEC{x}' - \VEC{s}_{\rm (S)}(\VEC{x}'-\YY) \right) \nonumber \\
    &=&\int d^3x'' \delta_{\rm (S)}(\VEC{x}'') \delta_{\rm D}\left( \VEC{x}-\YY - \VEC{x}'' - \VEC{s}_{\rm (S)}(\VEC{x}'') \right).
    \label{Eq:delta_rec_IR}
\end{eqnarray}
In the second line, we used the coordinate transformation: $\VEC{x}'' = \VEC{x}'-\YY$. Here, $\VEC{s}_{\rm (S)}(\VEC{x})$ is defined as
\begin{eqnarray}
    \hspace{-0.7cm}
    \VEC{s}_{\rm (S)}(\VEC{x}) = 
    \int \frac{d^3p}{(2\pi)^3} e^{i\VEC{p}\cdot\VEC{x}} \left( \frac{i\VEC{p}}{p^2} \right)
    \left(  - \frac{W_{\rm G}(pR_{\rm s})}{b_{1, \rm fid}} \right)
    \widetilde{\delta}_{\rm (S)}(\VEC{p}).
    \label{Eq:S_S}
\end{eqnarray}
In the IR limit, the short-wavelength displacement vector for reconstruction, $\VEC{s}_{\rm (S)}$, no longer receives contributions from the shift term, leading to its analogy with the displacement vector $\PP$. As demonstrated in Eqs.~(\ref{Eq:PPYY}) and (\ref{Eq:delta_IR_4}), the IR effect within the nonlinear density fluctuation is represented by $\PP$ at the origin. Similar to $\PP$, by decomposing $\VEC{s}_{\rm S}$ into its origin value and other components, we can elucidate the IR effect within the post-reconstruction density fluctuation:
\begin{eqnarray}
    \VEC{s}_{\rm (S)}(\VEC{x}) = \SL + \VEC{s}_{\rm (SS)}(\VEC{x}),
\end{eqnarray}
where $\SL = \VEC{s}_{\rm S}(\VEC{x}=\VEC{0})$. Finally, $\delta_{\rm rec}(\VEC{x})$ can be formally rewritten as
\begin{eqnarray}
    \delta_{\rm rec}(\VEC{x}) = \delta_{\rm (S)rec}(\VEC{x} - \YY_{\rm rec}),
    \label{Eq:delta_rec_main}
\end{eqnarray}
where
\begin{eqnarray}
    \YY_{\rm rec} = \YY + \SL,
\end{eqnarray}
and
\begin{eqnarray}
    \hspace{-0.5cm} 
    \delta_{\rm (S)rec}(\VEC{x}) 
    = \int d^3x' \delta_{\rm (S)}(\VEC{x}') \delta_{\rm D}\left( \VEC{x} - \VEC{x}' - \VEC{s}_{\rm (SS)}(\VEC{x}') \right).
\end{eqnarray}
This result shows that the IR effect manifests in the post-reconstruction density fluctuation as the coordinate transformation via $\YY_{\rm rec}$, like the pre-reconstruction case in Eq.~(\ref{Eq:delta_IR_4}). Consequently, the arguments applicable to the pre-reconstruction IR effect can similarly be applied to the post-reconstruction context.

\subsection{Lagrangian space}

We can also perform the calculation of density fluctuations after reconstruction in Lagrangian space. To facilitate a comparison with the Zel'dovich approximation, set to be discussed in Section~\ref{Sec:Zel}, we provide proof for Eq.~(\ref{Eq:delta_rec_main}) in Lagrangian space.

In Lagrangian space, $\delta_{\rm d}(\VEC{x})$ and $\delta_{\rm s}(\VEC{x})$ are each expressed as
\begin{eqnarray}
    1 + \delta_{\rm d}(\VEC{x}) &=& \int d^3q \left( 1 + \delta_{\rm bias}(\VEC{q}) \right)\nonumber  \\
    &\times& \delta_{\rm D}\left( \VEC{x} - \VEC{q} - \PP(\VEC{q}) - \VEC{s}(\VEC{q}+\PP(\VEC{q}))  \right), \nonumber \\
    1 + \delta_{\rm s}(\VEC{x}) &=& \int d^3q  \delta_{\rm D}\left( \VEC{x} - \VEC{q} - \VEC{s}(\VEC{q})  \right).
    \label{Eq:dd_ds_L}
\end{eqnarray}
Similar to Eq.~(\ref{Eq:delta_rec_IR}), they can be rewritten as
\begin{eqnarray}
    1 + \delta_{\rm d}(\VEC{x}) &=& \int d^3q \left( 1 + \delta_{\rm bias}(\VEC{q}) \right)\nonumber  \\
    &\times& \delta_{\rm D}\left( \VEC{x} - \VEC{q} - \YY - \PP_{\rm (S)}(\VEC{q}) - \VEC{s}_{\rm (S)}(\VEC{q}+\PP_{\rm (S)}(\VEC{q}))  \right), \nonumber \\
    1 + \delta_{\rm s}(\VEC{x}) &=& \int d^3q  \delta_{\rm D}\left( \VEC{x} - \VEC{q} - \VEC{s}_{\rm (S)}(\VEC{q}-\YY)  \right) \nonumber \\
    &=& \int d^3q'  \delta_{\rm D}\left( \VEC{x} - \VEC{q}'-\YY - \VEC{s}_{\rm (S)}(\VEC{q}')  \right),
    \label{Eq:dd_ds_L_S}
\end{eqnarray}
where $\PP_{\rm (S)}(\VEC{q}) = \PP(\VEC{q})-\YY$ in Eq.~(\ref{Eq:PPYY}). For $\delta_{\rm s}(\VEC{x})$, we used the coordinate transformation: $\VEC{q}'=\VEC{q}-\YY$. 

Note that $\VEC{s}(\VEC{q}+\PP(\VEC{q}))$ within $\delta_{\rm d}(\VEC{x})$ does not include the contribution from the shift term in the IR limit. This is because the coordinate transformation, represented by $\VEC{q}+\PP(\VEC{q})$, cancels out the IR effect of $\YY$. As a result, it can be expressed only in terms of short-wavelength modes as $\VEC{s}_{\rm (S)}(\VEC{q}+\PP_{\rm (S)}(\VEC{q}))$.

Furthermore, we decompose $\VEC{s}_{\rm (S)}$ into the value at the origin $\VEC{q}=\VEC{0}$ and other parts:
\begin{eqnarray}
    \VEC{s}_{\rm (S)}(\VEC{q}+\PP_{\rm (S)}(\VEC{q})) 
    &=& \SL + \VEC{s}_{\rm (SS)}(\VEC{q}+\PP_{\rm (S)}(\VEC{q})) \nonumber \\
    \VEC{s}_{\rm (S)}(\VEC{q}) &=&  \SL + \VEC{s}_{\rm (SS)}(\VEC{q}),
\end{eqnarray}
where we used $\PP_{\rm S}(\VEC{q}=\VEC{0})=\VEC{0}$.

Finally, we obtain
\begin{eqnarray}
    \delta_{\rm rec}(\VEC{x}) &=& 
    \delta_{\rm (S)d}(\VEC{x}-\YY_{\rm rec}) -  \delta_{\rm (S)s}(\VEC{x}-\YY_{\rm rec}) \nonumber \\
    &=& 
    \delta_{\rm (S)rec}(\VEC{x}-\YY_{\rm rec}),
    \label{Eq:delta_rec_Lag}
\end{eqnarray}
where
\begin{eqnarray}
    1 + \delta_{\rm (S)d}(\VEC{x}) &=& \int d^3q \left( 1 + \delta_{\rm bias}(\VEC{q}) \right)\nonumber  \\
    &\times& \delta_{\rm D}\left( \VEC{x} - \VEC{q} - \PP_{\rm (S)}(\VEC{q})  - \VEC{s}_{\rm (SS)}(\VEC{q}+\PP_{\rm (S)}(\VEC{q}))  \right), \nonumber \\
    1 + \delta_{\rm (S)s}(\VEC{x}) &=& \int d^3q  \delta_{\rm D}\left( \VEC{x} - \VEC{q} - \VEC{s}_{\rm (SS)}(\VEC{q})  \right).
\end{eqnarray}

\subsection{IR cancellation after reconstruction}
\label{Sec:IRcancel_recon}

We can show that the IR cancellation occurs even after reconstruction:
\begin{eqnarray}
    \left\langle \delta_{\rm rec}(\VEC{x}) \delta_{\rm rec}(\VEC{x}') \right\rangle
    &=& 
    \big\langle \delta_{\rm (S)rec}(\VEC{x}-\YY_{\rm rec}) \delta_{\rm (S)rec}(\VEC{x}'-\YY_{\rm rec}) \big\rangle
    \nonumber \\
    &\to&
    \big\langle \delta_{\rm (S)rec}(\VEC{x}) \delta_{\rm (S)rec}(\VEC{x}') \big\rangle\;.
\end{eqnarray}
Approximating both $\delta_{\rm (S)rec}(\VEC{x})$ and $\YY_{\rm rec}$ as linear, we obtain
\begin{eqnarray}
    \left\langle \delta_{\rm rec}(\VEC{x}) \delta_{\rm rec}(\VEC{x}') \right\rangle
    &\to& 
    \big\langle \delta^{[1]}(\VEC{x}-\YY^{[1]}_{\rm rec}) \delta^{[1]}(\VEC{x}'-\YY^{[1]}_{\rm rec}) \big\rangle
    \nonumber \\
    &=&
    \big\langle \delta^{[1]}(\VEC{x}) \delta^{[1]}(\VEC{x}') \big\rangle \;.
\end{eqnarray}
Under this approximation, the propagator term and the mode-coupling term are, respectively, represented as
\begin{eqnarray}
    G^2(\VEC{k}) P_{\rm m}^{[11]}(k) \to {\cal D}_{\rm rec}^2(\VEC{k}) [Z_1(\VEC{k})]^2P_{\rm m}^{[11]}(k)\;,
    \label{Eq:G_Recon}
\end{eqnarray}
and 
\begin{eqnarray}
    P_{\rm MC}(\VEC{k}) \to ( 1 - {\cal D}_{\rm rec}^2(\VEC{k}) )  [Z_1(\VEC{k})]^2P_{\rm m}^{[11]}(k)\;.
    \label{Eq:MC_Recon}
\end{eqnarray}
These specifically lead to the IR cancellation in Fourier space:
\begin{eqnarray}
    P_{\rm rec}(\VEC{k}) &\to& {\cal D}_{\rm rec}^2(\VEC{k}) [Z_1(\VEC{k})]^2P_{\rm m}^{[11]}(k) \nonumber \\
    &+& ( 1 - {\cal D}_{\rm rec}^2(\VEC{k}) ) [Z_1(\VEC{k})]^2P_{\rm m}^{[11]}(k)\nonumber \\
    &=& [Z_1(\VEC{k})]^2P_{\rm m}^{[11]}(k)\;.
\end{eqnarray}
Here, ${\cal D}_{\rm rec}(\VEC{k})$ is the two-dimensional exponentially decaying function after reconstruction, given by
\begin{eqnarray}
    {\cal D}_{\rm rec}(\VEC{k})
    &=&\exp\left( \frac{1}{2}\left\langle \left( -i\VEC{k}\cdot\YY_{\rm rec}^{[1]} \right)^2\right\rangle_{\rm c} \right), \nonumber \\
    &=&  \exp\left( - \frac{k^2(1-\mu^2)\sigma^2_{\rm rec,\perp}+k^2\mu^2\sigma^2_{\rm rec,\parallel}}{2} \right).
    \label{Eq:Drec}
\end{eqnarray}
The radial and transverse components of the smoothing factors are further decomposed into
\begin{eqnarray}
    \sigma_{\rm rec,\perp}^2 &=& \sigma_{\perp}^2 + \sigma_{\rm ps, \perp}^2 + \sigma_{\rm ss, \perp}^2\;, \nonumber \\
    \sigma_{\rm rec,\parallel}^2 &=& \sigma_{\parallel}^2 + \sigma_{\rm ps, \parallel}^2 + \sigma_{\rm ss, \parallel}^2\;,
    \label{Eq:Sigma2_rec}
\end{eqnarray}
where $\sigma_{\perp}^2$ and $\sigma_{\parallel}^2$ are given in Eq.~(\ref{Eq:sigma_sigma}). The subscript ``ps'' means the correlation between $\YY$ and $\SL$, and ``ss'' denotes the auto-correlation of $\SL$. These smoothing factors are calculated as
\begin{widetext}
    \begin{eqnarray}
        \sigma^2_{\rm ps, \perp}
        &=& \frac{1}{3} \int_{k_{\rm min}}^{k_{\rm max}} \frac{dp}{2\pi^2} \left( -\frac{W_{\rm G}(pR_{\rm s})}{b_{1,\rm fid}} \right)  
        \left[ 2\left( b_1+\frac{f}{5} \right)    \right]P_{\rm m}^{[11]}(p)\;, \nonumber \\
        \sigma^2_{\rm ps, \parallel}
        &=& \frac{1}{3} \int_{k_{\rm min}}^{k_{\rm max}} \frac{dp}{2\pi^2} \left( -\frac{W_{\rm G}(pR_{\rm s})}{b_{1,\rm fid}} \right)
        \Bigg[ 2\left( 1+f \right)\left(  b_1+\frac{3}{5}f\right)
              \Bigg]P_{\rm m}^{[11]}(p)\;, \nonumber \\
        \sigma^2_{\rm ss, \perp}
        &=& \frac{1}{3} \int_{k_{\rm min}}^{k_{\rm max}} \frac{dp}{2\pi^2} \left( -\frac{W_{\rm G}(pR_{\rm s})}{b_{1,\rm fid}} \right)^2
        \left[ \left( b_1^2+\frac{2}{5}b_1f + \frac{3}{35}f^2 \right) P_{\rm m}^{[11]}(p) + \frac{1}{\bar{n}}   \right]\;, \nonumber \\
        \sigma^2_{\rm ss, \parallel}
        &=& \frac{1}{3} \int_{k_{\rm min}}^{k_{\rm max}} \frac{dp}{2\pi^2} \left( -\frac{W_{\rm G}(pR_{\rm s})}{b_{1,\rm fid}} \right)^2 
        \Bigg[ \left( b_1^2 + \frac{42}{35}b_1f +\frac{3}{7}f^2  \right) P_{\rm m}^{[11]}(p) + \frac{1}{\bar{n}}
        \Bigg]\;.
        \label{Eq:ps_ss}
    \end{eqnarray}
\end{widetext}
Here, we summarize the key considerations when performing the above calculations:
\begin{enumerate}
    \item The integral range should depend on the scale at which $\VEC{s}(\VEC{x})$ is measured.
    \item Since $\VEC{s}(\VEC{x})$ is derived from the power spectrum of the measured galaxy density fluctuations, we should consider the shot-noise term $1/\bar{n}$ in $\sigma_{\rm ss, \perp}^2$ and $\sigma_{\rm ss,\parallel}^2$, where $\bar{n}$ is the mean number density~\cite{White:2010qd}.
\end{enumerate}

\subsection{Comparisons with the Zel'dovich approximation}
\label{Sec:Zel}

Previous studies~\cite{Padmanabhan:2008dd,Seo:2015eyw,White:2015eaa,Chen:2019lpf} have often used a simple linear approximation for $\VEC{s}(\VEC{x})$, known as the Zel'dovich approximation for reconstruction. This approximation, however, does not account for nonlinear IR effects in $\VEC{s}(\VEC{x})$. In this subsection, we present a detailed comparison between the Zel'dovich approximation and our findings in Section~\ref{Sec:IRcancel_recon}.

In the Zel'dovich approximation, Eq.~(\ref{Eq:dd_ds_L}) becomes
\begin{eqnarray}
    1 + \delta_{\rm d}(\VEC{x}) &\sim& \int d^3q \left( 1 + \delta_{\rm bias}^{[1]}(\VEC{q}) \right)\nonumber  \\
    &\times& \delta_{\rm D}\left( \VEC{x} - \VEC{q} - \PP^{[1]}(\VEC{q}) - \VEC{s}^{[1]}(\VEC{q})  \right)\;, \nonumber \\
    1 + \delta_{\rm s}(\VEC{x}) &\sim& \int d^3q  \delta_{\rm D}\left( \VEC{x} - \VEC{q} - \VEC{s}^{[1]}(\VEC{q})  \right)\;.
\end{eqnarray}
In the IR limit, they are represented as
\begin{eqnarray}
    \delta_{\rm d}(\VEC{x}) &\to& \delta_{\rm d}^{[1]}\left( \VEC{x} - \YY^{[1]} - \SL^{[1]} \right)\;,\nonumber \\
    \delta_{\rm s}(\VEC{x}) &\to& \delta_{\rm s}^{[1]}\left( \VEC{x} - \SL^{[1]} \right)\;.
\end{eqnarray}
We then derive
\begin{eqnarray}
    \delta_{\rm rec}(\VEC{x}) \to \delta_{\rm d}^{[1]}\left( \VEC{x} - \YY^{[1]} - \SL^{[1]} \right)
    - \delta_{\rm s}^{[1]}\left( \VEC{x} - \SL^{[1]} \right)\;.
\label{Eq:delta_rec_Zel}
\end{eqnarray}
The key difference of this equation from Eq.~(\ref{Eq:delta_rec_Lag}) is how $\VEC{s}(\VEC{q})$ is treated within $\delta_{\rm s}(\VEC{x})$. In Eq.~(\ref{Eq:delta_rec_Lag}), non-perturbative IR effects are incorporated into $\VEC{s}(\VEC{q})$, revealing both $\YY$ and $\SL$ effects in $\delta_{\rm s}(\VEC{x})$. In contrast, the Zel'dovich approximation assumes a linear $\VEC{s}(\VEC{q})$, resulting in only the $\SL$ effect.

In Fourier space, Eq.~(\ref{Eq:delta_rec_Zel}) becomes
\begin{eqnarray}
    \hspace{-0.7cm} 
    && 
    \widetilde{\delta}_{\rm rec}(\VEC{k}) \nonumber \\ 
    \hspace{-0.7cm} 
   &\to&
   \Big[  e^{-i\VEC{k}\cdot\left( \YY^{[1]} + \SL^{[1]} \right)}Z_{{\rm d},1}(\VEC{k}) 
   - e^{-i\VEC{k}\cdot\SL^{[1]} }Z_{{\rm s},1}(\VEC{k})\Big] \widetilde{\delta}_{\rm m}^{\, [1]}(\VEC{k})\;,
    \label{Eq:Zel}
\end{eqnarray}
where
\begin{eqnarray}
    Z_{{\rm d},1}(\VEC{k}) &=& \left[ 1 + \left(- \frac{W_{\rm G}(k R_{\rm s})}{b_{1, \rm fid}}  \right)\right]
    Z_1(\VEC{k})\;, \nonumber \\
    Z_{{\rm s},1}(\VEC{k}) &=& \left(  - \frac{W_{\rm G}(k R_{\rm s})}{b_{1, \rm fid}} \right) \, Z_1(\VEC{k})\;.
    \label{Eq:Zdr}
\end{eqnarray}
The propagator is then given by
\begin{eqnarray}
    G(\VEC{k}) =
    {\cal D}_{\rm rec}(\VEC{k}) Z_{{\rm d},1}(\VEC{k})
    - {\cal D}_{\rm ss}(\VEC{k}) Z_{{\rm s},1}(\VEC{k})\;,
    \label{Eq:G_Zel}
\end{eqnarray}
with
\begin{eqnarray}
    {\cal D}_{\rm ss}(\VEC{k})
    =\exp\left( \frac{1}{2}\left\langle \left( -i\VEC{k}\cdot\SL^{[1]} \right)^2\right\rangle_{\rm c} \right)\;.
\end{eqnarray}
Equation~(\ref{Eq:G_Zel}) shows that the propagator in the Zel'dovich approximation comprises two Gaussian decay functions: ${\cal D}_{\rm rec}(\VEC{k})$ and ${\cal D}_{\rm ss}(\VEC{k})$. This contrasts with Eq.~(\ref{Eq:G_Recon}), where only ${\cal D}_{\rm rec}(\VEC{k})$ appears. The specific expression for ${\cal D}_{\rm ss}(\VEC{k})$ is 
\begin{eqnarray}
    {\cal D}_{\rm ss}(\VEC{k}) = \exp\left( - \frac{k^2(1-\mu^2)\sigma^2_{\rm ss,\perp}+k^2\mu^2\sigma^2_{\rm ss,\parallel}}{2} \right),
    \label{Eq:Dss}
\end{eqnarray}
where the radial and transverse components of the smoothing factors are given in Eq.~(\ref{Eq:ps_ss}).

We can also demonstrate that Eq.~(\ref{Eq:delta_rec_Zel}) breaks the IR cancellation:
\begin{eqnarray}
    \hspace{-0.3cm}
    \langle \delta_{\rm rec}(\VEC{x}) \delta_{\rm rec}(\VEC{x}') \rangle
    &\to& \langle \delta_{\rm d}^{[1]}(\VEC{x})\delta_{\rm d}^{[1]}(\VEC{x}') \rangle \nonumber \\
    &-&   \langle \delta_{\rm d}^{[1]}(\VEC{x}- \YY^{[1]})\delta^{[1]}_{\rm s}(\VEC{x}') \rangle\nonumber \\
    &-&   \langle \delta_{\rm s}^{[1]}(\VEC{x})\delta^{[1]}_{\rm d}(\VEC{x}'- \YY^{[1]}) \rangle\nonumber \\
    &+&   \langle \delta_{\rm s}^{[1]}(\VEC{x})\delta^{[1]}_{\rm s}(\VEC{x}') \rangle \;.
\end{eqnarray}
In the cross-correlation functions between $\delta_{\rm d}(\VEC{x})$ and $\delta_{\rm s}(\VEC{x}')$, the IR cancellation does not occur, and the contribution of $\YY^{[1]}$ remains. In Fourier space, the cross-power spectrum between $\widetilde{\delta}_{\rm d}(\VEC{k})$ and $\widetilde{\delta}_{\rm s}(\VEC{k}')$ is represented as
\begin{eqnarray}
    P_{\rm ds}(\VEC{k}) \to {\cal D}(\VEC{k}) Z_{{\rm d},1}(\VEC{k})Z_{{\rm s},1}(\VEC{k}) P_{\rm m}^{[11]}(k)\;.
\end{eqnarray}
Hence, breaking the IR cancellation leads to an exponential decay of the power spectrum. This finding deviates from our result where the IR cancellation is satisfied even after reconstruction.

\subsection{Comparison with the 1-loop solution in SPT}
\label{Sec:SPT_1loop_rec}

In this subsection, we investigate the IR limit properties of the 1-loop solution in SPT for the post-reconstruction power spectrum.

In Fourier space, the $n$th-order term of the redshift-space density fluctuations after reconstruction is given by
\begin{eqnarray}
    \widetilde{\delta}_{\rm rec}^{\,[n]}(\VEC{k}) &=& \int \frac{d^3p_1}{(2\pi)^3}\cdots\frac{d^3p_n}{(2\pi)^3}
    (2\pi)^3\delta_{\rm D}(\VEC{k}-\VEC{p}_{[1,n]}) \nonumber\\ 
    &\times& Z_{{\rm rec},n}(\VEC{p}_1,\dots,\VEC{p}_n) 
    \widetilde{\delta}_{\rm m}^{\,[1]}(\VEC{p}_1)\cdots\widetilde{\delta}_{\rm m}^{\,[1]}(\VEC{p}_n).
    \label{Eq:delta_n_red_rec}
\end{eqnarray}
The function $Z_{{\rm rec},n}(\VEC{p}_1,\cdots,\VEC{p}_n)$ denotes nonlinear kernel functions accounting for the bias effect, the RSD effect, and the reconstruction effect. Consequently, the post-reconstruction terms corresponding to $P^{[22]}$ and $P^{[13]}$ are expressed as
\begin{eqnarray}
    P_{\rm rec}^{[22]}(\VEC{k}) &=& 2 \int \frac{d^3p_1}{(2\pi)^3}\int \frac{d^3p_2}{(2\pi)^3} 
(2\pi)^3\delta_{\rm D}(\VEC{k}-\VEC{p}_{[1,2]})
\nonumber \\
&\times&
[Z_{{\rm rec},2}(\VEC{p}_1,\VEC{p}_2)]^2 P_{\rm m}^{[11]}(p_1)P_{\rm m}^{[11]}(p_2), \nonumber \\
P_{\rm rec}^{[13]}(\VEC{k}) &=& 6 Z_1(\VEC{k})\, P_{\rm m}^{[11]}(k)  \nonumber \\
&\times& \int \frac{d^3p}{(2\pi)^3}Z_{{\rm rec},3}(\VEC{k},\VEC{p},-\VEC{p}) P_{\rm m}^{[11]}(p).
    \label{Eq:P22_P13_rec}
\end{eqnarray}
In the IR limit, both $Z_{{\rm rec},2}(\VEC{k},\VEC{p})$ and $Z_{{\rm rec},3}(\VEC{k},\VEC{p},-\VEC{p})$ are approximated as (see Appendix~\ref{Sec:SPT_REC})
\begin{eqnarray}
    Z_{{\rm rec},2}(\VEC{k},\VEC{p}) &\xrightarrow[p\to0]{}&
    \frac{1}{2} Z_1(\VEC{k})    \Bigg\{ \left( \frac{\VEC{k}\cdot\MAT{R}_1\cdot\VEC{p}}{p^2} \right) \nonumber \\
    &+& \left( \frac{\VEC{k}\cdot\VEC{p}}{p^2} \right) \left( -\frac{W(pR_{\rm s})}{b_{\rm 1,fid}} \right) Z_1(\VEC{p}) \Bigg\}
    \nonumber \\
    \hspace{-0.4cm}   Z_{ {\rm rec},3}(\VEC{k},\VEC{p},-\VEC{p})&\xrightarrow[p\to0]{}&
    - \frac{1}{3!}Z_1(\VEC{k}) \Bigg\{ \left( \frac{\VEC{k}\cdot\MAT{R}_1\cdot\VEC{p}}{p^2} \right) 
    \nonumber \\
    &+& \hspace{-0.2cm}\left( \frac{\VEC{k}\cdot\VEC{p}}{p^2} \right) 
    \left( -\frac{W(pR_{\rm s})}{b_{\rm 1,fid}} \right) Z_1(\VEC{p}) \Bigg\}^2 \;.
    \label{Eq:Z2_Z3_IR}
\end{eqnarray}
Using Eq.~(\ref{Eq:Z2_Z3_IR}), $P_{\rm rec}^{[22]}$ and $P_{\rm rec}^{[13]}$ in the IR limit are described as
\begin{eqnarray}
    P_{\rm rec, IR}^{[22]}(\VEC{k}) &=&
    \left( k^2(1-\mu^2)\sigma_{\rm rec,\perp}^2 + k^2\mu^2 \sigma^2_{\rm rec,\parallel} \right) 
    \nonumber \\
    &\times&
    [Z_1(\VEC{k})]^2P_{\rm m}^{[11]}(k),\nonumber \\
    P_{\rm rec, IR}^{[13]}(\VEC{k}) &=&
    -\left( k^2(1-\mu^2)\sigma_{\rm rec,\perp}^2 + k^2\mu^2 \sigma^2_{\rm rec,\parallel} \right) 
    \nonumber \\
    &\times&
    [Z_1(\VEC{k})]^2P_{\rm m}^{[11]}(k).
    \label{Eq:P22_P13_IR_rec}
\end{eqnarray}
These terms cancel each other out, leading to the post-reconstruction 1-loop power spectrum approaching zero in the IR limit: $P_{\rm rec}^{\rm 1\mathchar`-loop}(\VEC{k})\to0$. This result is consistent with those obtained by expanding both the propagator term and the mode-coupling term up to the 1-loop order, as presented in Eqs.~(\ref{Eq:G_Recon}) and (\ref{Eq:MC_Recon}). Given that SPT ensures accurate calculations for each respective order, this finding further reinforces the validity of our main result: i.e., the IR cancellation occurs even after reconstruction.

\subsection{IR-resummed model for \\ the post-reconstruction power spectrum}
\label{Sec:Main}

As discussed in Section~\ref{Sec:IRresummed_pre}, the function ${\cal E}(\VEC{k},-\VEC{k})={\cal D}^{-2}(\VEC{k})$ in Eq.~(\ref{Eq:cal_E}), resulting from the mode-coupling integrals, plays an essential role in describing the nonlinearity of the BAO signal. In particular, focusing on the wiggle part, the IR-resummed model can be constructed by replacing ${\cal D}^{-2}$ with ${\cal A}^{2}$ in Eq.~(\ref{Eq:A}).

After reconstruction, the same calculations used to derive the IR-resummed model before reconstruction can be repeated. The post-reconstruction ${\cal A}(\VEC{k})$ is denoted as ${\cal A}_{\rm rec}(\VEC{k})$. This ${\cal A}_{\rm rec}(\VEC{k})$ can be calculated by replacing $P_{\rm m}^{[11]}(p)$ in the smoothing parameters $\sigma_{\rm rec,\,\perp}^2$ and $\sigma_{\rm rec,\,\parallel}^2$, which characterize ${\cal D}^{-1}_{\rm rec}(\VEC{k})$ in Eq.~(\ref{Eq:Drec}), with $j_0(pr_{\rm BAO})\,P_{\rm m}^{[11]}(p)$. The function describing the nonlinear damping of BAO after reconstruction then becomes
\begin{eqnarray}
    {\cal D}_{\rm BAO,\,rec}(\VEC{k}) = {\cal D}_{\rm rec}(\VEC{k})\,{\cal A}_{\rm rec}(\VEC{k})\;,
\end{eqnarray}
and the linear IR-resummed model of the post-reconstruction power spectrum is obtained as
\begin{eqnarray}
    \hspace{-0.7cm} P_{\rm rec}(\VEC{k}) = \big[ Z_1(\VEC{k}) \big]^2 \left[  {\cal D}_{\rm BAO,\,rec}^2(\VEC{k}) P_{\rm w}(k) + P_{\rm nw}(k)\right]\;.
    \label{Eq:P_IR_BAO_recon}
\end{eqnarray}
This form is identical to the one before reconstruction, with the exception of the values of the smoothing parameters that characterize the exponential decay of the BAO signal. Template models using a single Gaussian function have been widely used in post-reconstruction BAO analysis (e.g., see~\cite{BOSS:2016wmc,Hinton:2016atz,eBOSS:2020yzd,DESI:2023bgx}). The findings in this paper confirm the validity of these previous post-reconstruction BAO analyses.

At the 1-loop level, we obtain
\begin{eqnarray}
    P_{\rm rec}(\VEC{k}) &=& 
    {\cal D}_{\rm BAO,\, rec}^2(\VEC{k}) \nonumber \\
    \hspace{-0.5cm}&\times&\left[ \left[ Z_1(\VEC{k})  \right]^2
        \left( 1 - \ln {\cal D}^2_{\rm BAO,\, rec}(\VEC{k}) \right)P_{\rm w}(k) + 
    P_{\rm rec,\,w}^{[13]}(\VEC{k}) \right]  \nonumber \\
    \hspace{-0.5cm}&+& \left[ Z_1(\VEC{k})  \right]^2 P_{\rm nw}(k) 
    + P^{[13]}_{\rm rec,\,nw}(\VEC{k}) + P_{\rm rec}^{[22]}(\VEC{k}) \;,
    \label{Eq:P_IR_BAO_recon_1loop}
\end{eqnarray}
where
\begin{eqnarray}
    P^{[13]}_{\rm rec,\,w}(\VEC{k}) &=& \left( P_{\rm w}(k) / P_{\rm m}^{[11]}(k) \right) P_{\rm rec}^{[13]}(\VEC{k})  \;, \nonumber \\
    P^{[13]}_{\rm rec,\,nw}(\VEC{k}) &=& \left( P_{\rm nw}(k) / P_{\rm m}^{[11]}(k) \right) P_{\rm rec}^{[13]}(\VEC{k})  \;.
\end{eqnarray}

\subsection{BAO signal in the mode-coupling term}

\begin{figure}[t]
    \centering
    \includegraphics[width=\columnwidth]{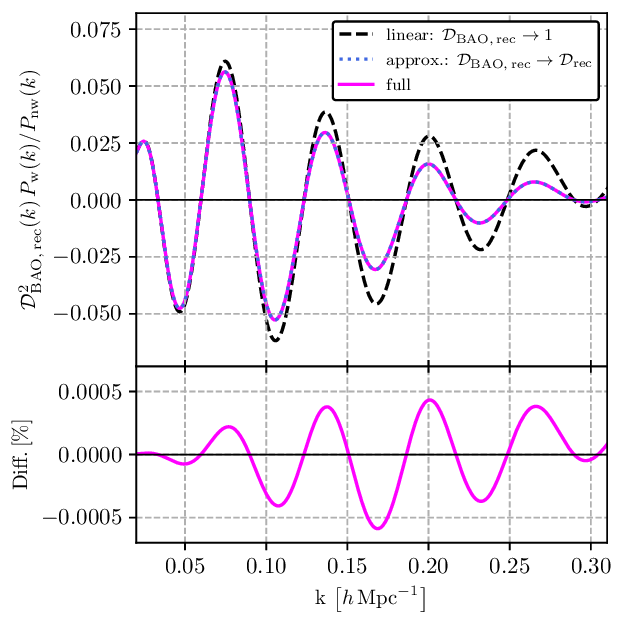}
    \caption{Same as Figure~\ref{fig:pre_recon}, except that the functions ${\cal D}_{\rm BAO}$ and ${\cal D}$ have been replaced by ${\cal D}_{\rm BAO,\,rec}$ and ${\cal D}_{\rm rec}$, respectively.}
    \label{fig:post_recon}
\end{figure}

To investigate the behavior of the BAO signal in the mode-coupling term after reconstruction, we plot Figure~\ref{fig:post_recon} similar to Figure~\ref{fig:pre_recon} before reconstruction, assuming $f=0$, $b_1=b_{1, {\rm fid}}=2$, $\bar{n}=3.0\times10^{-4}(\hMpc)^{-3}$, and $R_{\rm s}=15\hMpc$.

This figure shows that the difference between the two results obtained with ${\cal D}_{\rm BAO,\,rec}$ and with the ${\cal D}_{\rm BAO,\,rec}\approx{\cal D}_{\rm rec}$ approximation on the power spectrum is extremely small, about ${\rm Diff.}=0.0005\%$. This is due to the fact that the post-reconstruction smoothing parameters $\sigma^2_{\rm rec, \perp}$ and $\sigma^2_{\rm rec, \parallel}$ in Eq.~(\ref{Eq:Sigma2_rec}) are constructed so that they become small in the limit of $p\to0$ in Eq.~(\ref{Eq:ps_ss}), i.e., in the large-scale limit, and the contribution from large scales of $r_{\rm BAO}\sim110\hMpc$ is strongly suppressed compared to before reconstruction.

Therefore, after reconstruction, the approximation ${\cal D}_{\rm BAO,\,rec}\approx {\cal D}_{\rm rec}$ holds very well, and the model can be constructed while effectively ignoring the wiggle part in the mode-coupling term. Under this approximation, the following models are provided at the linear and 1-loop levels, respectively:
\begin{eqnarray}
    P_{\rm rec}(\VEC{k}) = \big[ Z_1(\VEC{k}) \big]^2\left[  {\cal D}_{\rm rec}^2(\VEC{k}) P_{\rm w}(k) + P_{\rm nw}(k)\right]\;,
    \label{Eq:P_IR_BAO_recon_app}
\end{eqnarray}
and
\begin{eqnarray}
    \hspace{-0.5cm} P_{\rm rec}(\VEC{k}) &=& 
    {\cal D}_{\rm rec}^2(\VEC{k}) \nonumber \\
    \hspace{-0.5cm}&\times&\left[ \left[ Z_1(\VEC{k})  \right]^2
        \left( 1 - \ln {\cal D}_{\rm rec}^2(\VEC{k}) \right)P_{\rm w}(k) + 
    P_{\rm rec,\,w}^{[13]}(\VEC{k}) \right]  \nonumber \\
    \hspace{-0.5cm}&+& \left[ Z_1(\VEC{k})  \right]^2 P_{\rm nw}(k) 
    + P^{[13]}_{\rm rec,\,nw}(\VEC{k}) + P_{\rm rec}^{[22]}(\VEC{k}) \;.
    \label{Eq:P_IR_BAO_1loop_recon_app}
\end{eqnarray}

\section{Cross-Power Spectrum between Pre- and Post-Reconstruction Density Fluctuations}
\label{Sec:Cross}

We now turn to the cross-power spectrum between the pre- and post reconstruction density fluctuations. In this section, we demonstrate that the pre- and post-reconstruction cross-power spectrum does not occur with the IR cancellation, resulting in an overall exponential decay.

In the IR limit, the cross 2PCF of $\delta(\VEC{x})$ and $\delta_{\rm rec}(\VEC{x})$ can be expressed as
\begin{eqnarray}
    \langle \delta(\VEC{x})\delta_{\rm rec}(\VEC{x}') \rangle
    &\to& \big\langle \delta^{[1]}(\VEC{x}-\YY^{[1]})\delta_{\rm rec}^{[1]}(\VEC{x}'-\YY_{\rm rec}^{[1]})\big\rangle\;.
    \label{Eq:cross_2PCF}
\end{eqnarray}
Because of the statistical translation symmetry, this expression transforms to
\begin{eqnarray}
    \langle \delta(\VEC{x})\delta_{\rm rec}(\VEC{x}') \rangle
    &\to& \big\langle \delta^{[1]}(\VEC{x}+\SL^{[1]})\delta_{\rm rec}^{[1]}(\VEC{x}')\big\rangle\;.
    \label{Eq:cross_2PCF_ver2}
\end{eqnarray}
As a result, the contributions from the IR effect related to $\SL^{[1]}$ are retained in the cross 2PCF.

In Fourier space, the cross-power spectrum of $\widetilde{\delta}(\VEC{k})$ and $\widetilde{\delta}_{\rm rec}(\VEC{k})$ in the IR limit is given by
\begin{eqnarray}
    &&P_{\rm cross}(\VEC{k}) \nonumber \\
    &\to& 
    {\cal D}(\VEC{k}) {\cal D}_{\rm rec}(\VEC{k}) {\cal E}_{\rm pd}(\VEC{k},-\VEC{k}) \big[Z_1(\VEC{k})\big]^2 P_{\rm m}^{[11]}(k),
    \label{Eq:P_cross}
\end{eqnarray}
where the exponential function ${\cal E}_{\rm pd}(\VEC{k},\VEC{k}')$ is defined as
\begin{eqnarray}
    {\cal E}_{\rm pd}(\VEC{k},\VEC{k}') &=&
    \exp\left( \left\langle \left( -i\VEC{k}\cdot \YY^{[1]} \right)
    \left( -i\VEC{k}'\cdot \YY_{\rm rec}^{[1]}\right)\right\rangle_{\rm c}  \right),
\end{eqnarray}
This function satisfies the following relation:
\begin{eqnarray}
    {\cal E}_{\rm pd}(\VEC{k},-\VEC{k})= {\cal D}_{\rm ss}(\VEC{k}) {\cal D}^{-1}_{\rm rec}(\VEC{k}){\cal D}^{-1}(\VEC{k}).
    \label{Eq:EX}
\end{eqnarray}
Substituting Eq.~(\ref{Eq:EX}) into Eq.~(\ref{Eq:P_cross}) leads to
\begin{eqnarray}
    P_{\rm cross}(\VEC{k})
    &\to& {\cal D}_{\rm ss}(\VEC{k})\big[Z_1(\VEC{k})\big]^2 P_{\rm m}^{[11]}(k).
\end{eqnarray}
Thus, the contribution from the remaining IR effect in the cross 2PCF, shown in Eq.~(\ref{Eq:cross_2PCF_ver2}), appears as the exponential decay function in the cross-power spectrum. 

To account for the nonlinear damping effect of BAO, as was done in Eqs.~(\ref{Eq:P_IR_BAO}) and (\ref{Eq:P_IR_BAO_recon}), we calculate ${\cal A}_{\rm ss}$ by replacing $P_{\rm m}^{[11]}(p)$ in the smoothing parameters $\sigma^2_{\rm ss,\,\perp}$ and $\sigma^2_{\rm ss,\,\parallel}$ with $j_0(pr_{\rm BAO})P_{\rm m}^{[11]}(p)$ and substituting the new smoothing parameters obtained there into ${\cal D}^{-1}_{\rm ss}$. Then, when focusing on the wiggle part, we perform the replacement
\begin{eqnarray}
    {\cal E}_{\rm pd}(\VEC{k},-\VEC{k}) \xrightarrow[\rm replacement]{} {\cal A}^{-1}_{\rm ss}(\VEC{k}) {\cal A}_{\rm rec}(\VEC{k}){\cal A}(\VEC{k}),
\end{eqnarray}
and when considering the no-wiggle part, we leave it as it is. Through these operations, we can finally obtain
\begin{eqnarray}
    P_{\rm cross}(\VEC{k}) 
     &\to& 
    \big[Z_1(\VEC{k})\big]^2
    \big[ {\cal D}_{\rm BAO}(\VEC{k}){\cal D}_{\rm BAO,\,rec}(\VEC{k}){\cal A}^{-1}_{\rm ss}(\VEC{k}) P_{\rm w}(k) \nonumber \\
    && \hspace{1.5cm}+ {\cal D}_{\rm ss}(\VEC{k}) P_{\rm nw}(\VEC{k}) \big]\;.
    \label{Eq:P_cross_IR}
\end{eqnarray}

\begin{figure}[t]
    \centering
    \includegraphics[width=\columnwidth]{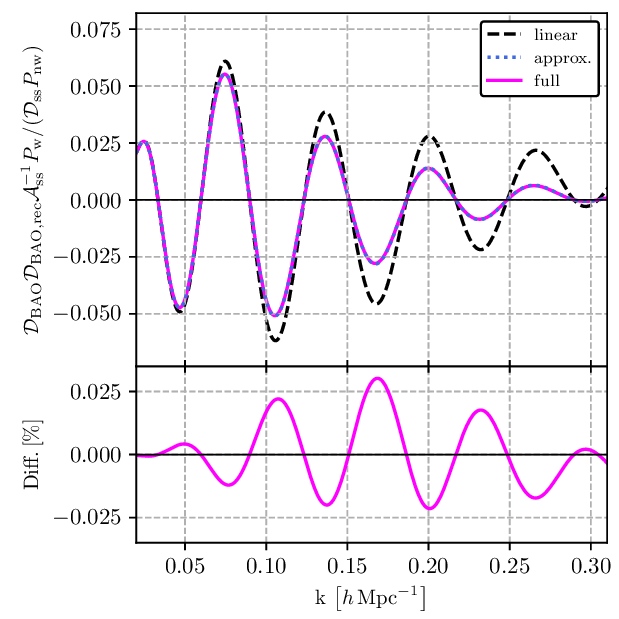}
    \caption{Ratio of the wiggle part to the no-wiggle part in the cross-power spectrum model between before and after reconstruction at $z=0$ in real space. The upper panel shows the result in Eq.~(\ref{Eq:P_cross_IR}) (magenta solid line), the linear theory (black dashed line), and the approximation in Eq.~(\ref{Eq:P_cross_IR_app}) (blue dotted line). The lower panel plots the difference between Eq.~(\ref{Eq:P_cross_IR}) and Eq.~(\ref{Eq:P_cross_IR_app}).}
    \label{fig:cross}
\end{figure}

In the same way as Eq.~(\ref{Eq:P_IR_BAO_app}) and Eq.~(\ref{Eq:P_IR_BAO_recon_app}), when we ignore the BAO signal in the mode-coupling term, set ${\cal D}_{\rm BAO}={\cal D}$, ${\cal D}_{\rm BAO,\,rec}={\cal D}_{\rm rec}$, and ${\cal A}_{\rm ss}^{-1}=1$, we obtain 
\begin{eqnarray}
    \hspace{-0.8cm}
    && P_{\rm cross}(\VEC{k})  \nonumber \\
    \hspace{-0.8cm}
    &\approx& 
    \big[Z_1(\VEC{k})\big]^2
    \big[ {\cal D}(\VEC{k}){\cal D}_{\rm rec}(\VEC{k}) P_{\rm w}(k) + {\cal D}_{\rm ss}(\VEC{k}) P_{\rm nw}(\VEC{k}) \big]\;.
    \label{Eq:P_cross_IR_app}
\end{eqnarray}
Figure~\ref{fig:cross} shows the difference between the models of Eq.~(\ref{Eq:P_cross_IR}) and Eq.~(\ref{Eq:P_cross_IR_app}), and the difference is very small, at $0.025\%$. Therefore, Eq.~(\ref{Eq:P_cross_IR_app}) can be considered a good approximation.

\subsection{Shot noise in the cross-power spectrum}

The shot-noise term in the power spectrum arises when the same particle is counted multiple times during power spectrum calculations. Usually, the shot-noise term is absent in the cross-power spectrum of different density fluctuations. However, in the case of the cross-power spectrum between the pre- and post-reconstruction density fluctuations, the same galaxies can be identified, making the shot-noise term significant.

In the discrete picture, the number density in Fourier space is given by
\begin{eqnarray}
    n(\VEC{k}) = \sum_i^{N_{\rm g}} e^{-i\VEC{k}\cdot\VEC{x}_i},
\end{eqnarray}
where $\VEC{x}_i$ denotes the position of the $i$th galaxy and $N_{\rm g}$ represents the total number of galaxies. The auto power spectrum in this discrete picture is expressed as
\begin{eqnarray}
    \frac{V}{N^2_{\rm g}}|n(\VEC{k})|^2
    &=&\frac{V}{N^2_{\rm g}}\sum_{i\neq j} e^{-i\VEC{k}\cdot(\VEC{x}_i-\VEC{x}_j)}
    + \frac{V}{N^2_{\rm g}}\sum_{i=j}^{N_{\rm g}} \nonumber \\
    &=& P(\VEC{k}) + \frac{1}{\bar{n}},
\end{eqnarray}
where $V$ denotes the survey volume. The first and second terms on the right-hand side correspond to the power spectrum and the shot-noise term, respectively,
\begin{eqnarray}
    P(\VEC{k}) &=&\frac{V}{N^2_{\rm g}}\sum_{i\neq j} e^{-i\VEC{k}\cdot(\VEC{x}_i-\VEC{x}_j)}, \nonumber \\
    \frac{1}{\bar{n}} &=& \frac{V}{N^2_{\rm g}}\sum_{i=j}^{N_{\rm g}},
\end{eqnarray}
where $\bar{n}=N_{\rm g}/V$ is the mean number density.

Let $n_{\rm rec}(\VEC{k})$ be the post-reconstruction density field in Fourier space. It can be expressed in the discrete picture as
\begin{eqnarray}
    n_{\rm rec}(\VEC{k}) = \sum_i^{N_{\rm g}} e^{-i\VEC{k}\cdot\VEC{x}_{{\rm rec},i}},
\end{eqnarray}
where 
\begin{eqnarray}
    \VEC{x}_{{\rm rec},i} = \VEC{x}_i + \VEC{s}(\VEC{x}_i).
\end{eqnarray}
Therefore, the shot-noise term in the cross-power spectrum is given by
\begin{eqnarray}
    \frac{V}{N^2_{\rm g}}\sum_{i=j}^{N_{\rm g}} 
    e^{-i\VEC{k}\cdot\left( \VEC{x}_i -\VEC{x}_{ {\rm rec},i} \right)}
    &=&    \frac{V}{N^2_{\rm g}}\sum_{i=j}^{N_{\rm g}} 
    e^{i\VEC{k}\cdot\VEC{s}(\VEC{x}_i)}
\end{eqnarray}
Calculating the ensemble average of the above equation and taking the continuous limit, we obtain
\begin{eqnarray}
    \left\langle \frac{V}{N^2_{\rm g}}\sum_{i=j}^{N_{\rm g}} 
    e^{i\VEC{k}\cdot\VEC{s}(\VEC{x}_i)} \right\rangle
    &=&    \frac{1}{N_{\rm g}}\int d^3x
     \left\langle e^{i\VEC{k}\cdot\VEC{s}(\VEC{x})} \right\rangle \nonumber \\
     &=&    \frac{1}{\bar{n}}
     \left\langle e^{i\VEC{k}\cdot\VEC{s}(\VEC{x}=\VEC{0})} \right\rangle \nonumber \\
     &\simeq&
     \frac{1}{\bar{n}}
     \left\langle e^{i\VEC{k}\cdot\SL^{[1]}} \right\rangle.
\end{eqnarray}
In the final line, we made an approximation of the linear $\VEC{s}(\VEC{x})$. Finally, we derive
\begin{eqnarray}
    \left\langle \frac{V}{N_{\rm g}^2}n(\VEC{k}) n_{\rm rec}^*(\VEC{k}) \right\rangle
    = P_{\rm cross}(\VEC{k}) + {\cal D}_{\rm ss}(\VEC{k}) \left(\frac{1}{\bar{n}}  \right).
\end{eqnarray}
This result indicates that the shot-noise term is proportional to ${\cal D}_{\rm ss}(\VEC{k})$ and exponentially decays, as is $P_{\rm cross}(\VEC{k})$.

\section{Conclusions}
\label{Sec:Conclusions}

In this paper, we present a new theoretical model addressing the resummation of infrared (IR) effects in the power spectrum of galaxy density fluctuations after reconstruction. Our model accurately describes the nonlinear damping of the post-reconstruction BAO signal, while including the 1-loop correction term in SPT. The first main results are summarized in Eqs.~(\ref{Eq:P_IR_BAO_recon}) and (\ref{Eq:P_IR_BAO_recon_1loop}).

We point out in Section~\ref{Sec:Post_recon_IR} that it is crucial to consider the nonlinear IR effects contained within the displacement vector for reconstruction $\VEC{s}$ (\ref{Eq:S}). As a result, the IR effects on the post-reconstruction density fluctuation can be described as a coordinate transformation of the density fluctuations (\ref{Eq:delta_rec_main}), like the pre-reconstruction case (\ref{Eq:delta_IR_4}). This result leads to the cancellation of the IR effects on the power spectrum in the IR limit even after reconstruction. Furthermore, the nonlinear behavior of BAO is characterized by the two-dimensional Gaussian decaying function of BAO, supporting the theoretical foundation of the commonly used single two-dimensional Gaussian decay function in post-reconstruction power spectrum and 2PCF analyses~\cite[e.g.,][]{BOSS:2016wmc,Hinton:2016atz,eBOSS:2020yzd,DESI:2023bgx}. 

We compare our findings regarding the post-reconstruction infrared (IR) effects with those obtained using the Zel'dovich approximation and SPT in Section~\ref{Sec:Zel} and Section\ref{Sec:SPT_1loop_rec}. Notably, in the Zel'dovich approximation, which linearly approximates $\VEC{s}$, IR cancellation does not occur, and two Gaussian damping functions are required to describe the nonlinear BAO effects. Conversely, in the post-reconstruction SPT, IR cancellation is observed in the IR limit, corroborating our results. 

We further clarify the behavior of the cross-power spectrum between pre- and post-reconstruction density fluctuations, noting its overall exponential decay due to the absence of IR cancellation. This observation also applies to the corresponding shot-noise term. While these phenomena have been observed in previous simulation studies~\cite{Wang:2022nlx}, our work thus provides a mathematical foundation for these observations. Additionally, as our second main result, we present a model, as shown in Eq.~(\ref{Eq:P_cross_IR}), to elucidate the nonlinear effects of BAO on the cross-power spectrum.

In conclusion, our work presents a novel direction in developing theoretical models for the post-reconstruction power spectrum beyond standard perturbation theory. The IR-resummed models presented in this paper, which consider the galaxy bias and RSD effects, are directly applicable to the analysis of real galaxy data.

\qquad

\textit{Note added.}---Recently, we learned that the DESI Collaboration~\cite{Chen:2024tfp} was planning to submit a study with conclusions similar to ours, suggesting that the BAO signal in the post-reconstruction power spectrum should be described by a single Gaussian damping function. To ensure a coordinated approach, we arranged to submit our paper close to the DESI Collaborators' paper submission. Their study investigates the theoretical and numerical systematics of the post-reconstruction BAO signal in detail. In contrast, our work focuses on the IR effects before and after reconstruction, providing a more detailed theoretical analysis. Therefore, these studies complement each other well, each enriching the understanding of the topic from a different perspective. We refer readers interested in the numerical aspects to the paper by the DESI Collaboration.

\begin{acknowledgments}
    N.S. acknowledges financial support from JSPS KAKENHI Grant No. 19K14703. 
    N.S. thanks Shun Saito, Hee-Jong Seo, Florian Beutler, and Shi-Fan Stephen Chen for useful comments and discussion.
    N.S. thanks the referee for bringing up valuable comments and heavily improving the paper's quality.
The numerical calculations in this paper used \textsc{CLASS}~\cite{Blas:2011rf} and \textsc{FFTLog}~\cite{Hamilton:2000MNRAS.312..257H,Fang:2019xat}.

\end{acknowledgments}

\appendix

\section{Degenerate Higher-Order Scalar-Tensor (DHOST) Theories}
\label{Sec:DHOST}

In DHOST theories, the second-order kernel density fluctuation of dark matter is given by~\cite{Hirano:2018uar}
\begin{eqnarray}
    \delta_{\rm m}^{[2]}(\VEC{x})
    &=& \left( \kappa - \frac{4}{21}\lambda \right)
    \big[ \delta_{\rm m}^{[1]}(\VEC{x}) \big]^2
    - \kappa\,\PP^{[1]}(\VEC{x}) \cdot \nabla \delta_{\rm m}^{[1]}(\VEC{x}) \nonumber \\
    &+&
    \frac{2}{7}\,\lambda\, \left[\left( \frac{\partial_i\partial_j}{\partial^2}-\frac{1}{3}\delta_{ij} \right) \delta_{\rm m}^{[1]}(\VEC{x}) \right] ^2.
    \label{Eq:delta_2_DHOST}
\end{eqnarray}
Here, in $\Lambda$CDM assuming $f^2=\Omega_{\rm m}$, $\kappa=\lambda=1$.
From Eqs.~(\ref{Eq:delta_2_Sp}) and (\ref{Eq:delta_2_S}), the short-wavelength density fluctuation then becomes
\begin{eqnarray}
    \delta_{\rm (S)m}^{[2]}(\VEC{x})
    &=& \left( \kappa - \frac{4}{21}\lambda \right)
    \big[ \delta_{\rm m}^{[1]}(\VEC{x}) \big]^2
    - \PP_{\rm (S)}^{[1]}(\VEC{x}) \cdot \nabla \delta_{\rm m}^{[1]}(\VEC{x}) 
    \nonumber \\
    &-& \Delta \kappa\,\PP^{[1]}(\VEC{x}) \cdot \nabla \delta_{\rm m}^{[1]}(\VEC{x}) 
\nonumber \\
    &+&
    \frac{2}{7}\,\lambda\, \left[\left( \frac{\partial_i\partial_j}{\partial^2}-\frac{1}{3}\delta_{ij} \right) \delta_{\rm m}^{[1]}(\VEC{x}) \right] ^2.
    \label{Eq:delta_2_S_DHOST}
\end{eqnarray}
where $\kappa = 1 + \Delta \kappa$. There remains a shift term, $\Delta \kappa\,\PP^{[1]}(\VEC{x}) \cdot \nabla \delta_{\rm m}^{[1]}(\VEC{x}) $, in $\delta^{[2]}_{\rm (S)m}(\VEC{x})$ that does not go to zero in the IR limit. Therefore, it is not possible to simply separate IR effects from the rest based on Eq.~(\ref{Eq:delta_IR_4}). This unique feature of DHOST theories leads to a violation of the IR cancellation~\cite{Hirano:2020dom}. This deviation in the coefficient of the shift term from $1$ in DHOST theories is also related to the consistency relation for the large-scale structure~\cite{Crisostomi:2019vhj,Lewandowski:2019txi,Sugiyama:2023zvd}.

\section{Standard Perturbation Theory}
\label{Sec:SPT}

In this appendix, we show the specific form of the nonlinear kernel functions $Z_n$ given in Eq.~(\ref{Eq:delta_n_red}) up to the third order in SPT and investigate their behaviors in the IR limit.

In Fourier space, Eq.~(\ref{Eq:delta_g}) becomes
\begin{eqnarray}
    \widetilde{\delta}_{\rm g}(\VEC{k}) 
    =\hspace{-0.15cm}
    \int \hspace{-0.1cm}d^3q e^{-i\VEC{k}\cdot\VEC{q}}
    \Big\{\left[ 1 + \delta_{\rm bias}(\VEC{q}) \right] e^{-i\VEC{k}\cdot\PP_{\rm red}(\VEC{q})} - 1\Big\}\;.
\end{eqnarray}
We assume that the $n$th-order of the biased density fluctuation can be represented as
\begin{eqnarray}
    \widetilde{\delta}_{\rm bias}^{\,[n]}(\VEC{k}) &=& \int \frac{d^3p_1}{(2\pi)^3}\cdots\frac{d^3p_n}{(2\pi)^3}
    (2\pi)^3\delta_{\rm D}(\VEC{k}-\VEC{p}_{[1,n]}) \nonumber\\ 
    &\times& B_n(\VEC{p}_1,\dots,\VEC{p}_n) 
    \widetilde{\delta}_{\rm m}^{\,[1]}(\VEC{p}_1)\cdots\widetilde{\delta}_{\rm m}^{\,[1]}(\VEC{p}_n)\;,
\end{eqnarray}
where $B_1(\VEC{p}_1) = (b_1-1)$ with $b_1$ being the Eulerian linear bias. We follow the standard bias theory (e..g,~\cite{Desjacques:2016bnm}) and assume that $B_{n\geq2}(\VEC{p}_1,\dots,\VEC{p}_n)$ do not include the shift terms. Then, the nonlinear kernel functions up to the third order are given by~\cite{Scoccimarro:1999ed,Matsubara:2008wx}
\begin{eqnarray}
    Z_1(\VEC{k}) = B_1(\VEC{k}) + \VEC{k}\cdot\VEC{L}_{ {\rm red},1}(\VEC{k}) \;,
\end{eqnarray}
\begin{eqnarray}
    Z_2(\VEC{p}_1,\VEC{p}_2)
    &=& B_2(\VEC{p}_1,\VEC{p}_2) + \big[\VEC{k}\cdot\VEC{L}_{ {\rm red},2}(\VEC{p}_1,\VEC{p}_2)\big] \nonumber \\
    &+& \frac{1}{2} \big\{ B_1(\VEC{p}_1) \big[\VEC{k}\cdot\VEC{L}_{ {\rm red},1}(\VEC{p}_2)\big] + \mbox{1 perm.}\big\} \nonumber \\
    &+& \frac{1}{2} \big[\VEC{k}\cdot\VEC{L}_{ {\rm red},1}(\VEC{p}_1)\big] 
    \big[ \VEC{k}\cdot\VEC{L}_{ {\rm red},1}(\VEC{p}_2)\big]\;,
\end{eqnarray}
and
\begin{eqnarray}
    &&Z_3(\VEC{p}_1,\VEC{p}_2,\VEC{p}_3) \nonumber \\
    &=& B_3(\VEC{p}_1,\VEC{p}_2,\VEC{p}_3) 
    + \big[\VEC{k}\cdot\VEC{L}_{ {\rm red},3}(\VEC{p}_1,\VEC{p}_2,\VEC{p}_3) \big]\nonumber \\
    &+& \frac{1}{3} \big\{ B_2(\VEC{p}_1,\VEC{p}_2) 
    \big[\VEC{k}\cdot\VEC{L}_{ {\rm red},1}(\VEC{p}_3)\big] + \mbox{2 perms.}\big\} \nonumber \\
    &+& \frac{1}{3} \big\{ B_1(\VEC{p}_1) 
    \big[\VEC{k}\cdot\VEC{L}_{ {\rm red},2}(\VEC{p}_2,\VEC{p}_3)\big] + \mbox{2 perms.}\big\} \nonumber \\
    &+& \frac{1}{6} \big\{ B_1(\VEC{p}_1)
    \big[\VEC{k}\cdot\VEC{L}_{ {\rm red},1}(\VEC{p}_2)\big]
    \big[\VEC{k}\cdot\VEC{L}_{ {\rm red},1}(\VEC{p}_3)\big]
    + \mbox{2 perms.}\big\} \nonumber \\
    &+& \frac{1}{6}\big[ \VEC{k}\cdot\VEC{L}_{ {\rm red},1}(\VEC{p}_1)\big]
    \big[ \VEC{k}\cdot\VEC{L}_{ {\rm red},1}(\VEC{p}_2)\big]
    \big[\VEC{k}\cdot\VEC{L}_{ {\rm red},1}(\VEC{p}_3)\big]\;,
\end{eqnarray}
where $\VEC{k}=\VEC{p}_1+\VEC{p}_2$ for $Z_2$, $\VEC{k}=\VEC{p}_1+\VEC{p}_2+\VEC{p}_3$ for $Z_3$, and $L_{ {\rm red}, n}(\VEC{p}_1,\cdots,\VEC{p}_n)$ are calculated through the transformation matrix given in Eq.~(\ref{Eq:R}):
\begin{eqnarray}
    \VEC{L}_{ {\rm red},n}(\VEC{p}_1,\cdots,\VEC{p}_n)= \MAT{R}_n\cdot\VEC{L}_n(\VEC{p}_1,\cdots,\VEC{p}_n).
\end{eqnarray}

Since $\VEC{L}_{n\geq2}(\VEC{p}_1,\cdots,\VEC{p}_n)$ do not include the shift term, the functions $Z_2(\VEC{k},\VEC{p})$ and $Z_3(\VEC{k},\VEC{p},-\VEC{p})$, which are needed to compute the 1-loop power spectrum in the IR limit, are approximated as
\begin{eqnarray}
    Z_2(\VEC{k},\VEC{p}) &\xrightarrow[p\to0]{}&
    \frac{1}{2} \left( \frac{\VEC{k}\cdot \MAT{R}_1 \cdot \VEC{p}}{p^2} \right) Z_1(\VEC{k})\; , \nonumber \\
    \hspace{-0.3cm}
    Z_3(\VEC{k},\VEC{p},-\VEC{p}) &\xrightarrow[p\to0]{}&
    - \frac{1}{3!} \left( \frac{\VEC{k}\cdot \MAT{R}_1 \cdot \VEC{p}}{p^2} \right)^2 Z_1(\VEC{k})\;.
\end{eqnarray}

\section{Standard Perturbation Theory \\ after Reconstruction}
\label{Sec:SPT_REC}

We investigate the IR limit properties of the post-reconstruction density fluctuations up to the third order.

In Fourier space, Eq.~(\ref{Eq:delta_rec}) becomes
\begin{eqnarray}
    \widetilde{\delta}_{\rm rec}(\VEC{k}) 
    = \int d^3x e^{-i\VEC{k}\cdot\VEC{x}} e^{-i\VEC{k}\cdot\VEC{s}(\VEC{x})}\delta(\VEC{x}).
\end{eqnarray}
The nonlinear kernel function up to the third order are then given by~\cite{Hikage:2019ihj}
\begin{widetext}
\begin{eqnarray}
    Z_{ {\rm rec},1}(\VEC{p}_1) 
    &=& Z_1(\VEC{p}_1) \;, \nonumber \\
    Z_{ {\rm rec},2}(\VEC{p}_1,\VEC{p}_2) 
    &=& Z_2(\VEC{p}_1,\VEC{p}_2) 
    + \frac{1}{2} \left\{ \left( \frac{\VEC{k}\cdot\VEC{p}_1}{p_1^2} \right)\left( -\frac{W_{\rm G}(p_1R_{\rm s})}{b_{\rm 1,fid}} \right) 
    + \left( \frac{\VEC{k}\cdot\VEC{p}_2}{p_2^2} \right)\left( -\frac{W_{\rm G}(p_2R_{\rm s})}{b_{\rm 1,fid}} \right)  \right\}
    Z_1(\VEC{p}_1) Z_1(\VEC{p}_2) \;, \nonumber \\
    Z_{ {\rm rec},3}(\VEC{p}_1,\VEC{p}_2,\VEC{p}_3) 
    &=& Z_3(\VEC{p}_1,\VEC{p}_2,\VEC{p}_3) \nonumber \\
    &+& \frac{1}{3} \left\{ \left( \frac{\VEC{k}\cdot\VEC{p}_{[1,2]}}{p_{[1,2]}^2} \right)
    \left( -\frac{W_{\rm G}(p_{[1,2]}R_{\rm s})}{b_{\rm 1,fid}} \right) Z_2(\VEC{p}_1,\VEC{p}_2)Z_1(\VEC{p}_3) + \mbox{2 perms.} \right\}
    \nonumber \\
    &+& \frac{1}{3} \left\{ \left( \frac{\VEC{k}\cdot\VEC{p}_{1}}{p_{1}^2} \right)
    \left( -\frac{W_{\rm G}(p_{1}R_{\rm s})}{b_{\rm 1,fid}} \right) Z_1(\VEC{p}_1)Z_2(\VEC{p}_2,\VEC{p}_3) + \mbox{2 perms.} \right\}
    \nonumber \\
    &+& \frac{1}{6} \left\{ \left( \frac{\VEC{k}\cdot\VEC{p}_{1}}{p_{1}^2} \right)
    \left( -\frac{W_{\rm G}(p_{1}R_{\rm s})}{b_{\rm 1,fid}} \right) 
     \left( \frac{\VEC{k}\cdot\VEC{p}_{2}}{p_{2}^2} \right)
    \left( -\frac{W_{\rm G}(p_{2}R_{\rm s})}{b_{\rm 1,fid}} \right) 
Z_1(\VEC{p}_1)Z_1(\VEC{p}_2)Z_1(\VEC{p}_3) + \mbox{2 perms.}\right\}\;,
\end{eqnarray}
where $\VEC{k}=\VEC{p}_1+\VEC{p}_2$ for $Z_{ {\rm rec},2}$ and $\VEC{k}=\VEC{p}_1+\VEC{p}_2+\VEC{p}_3$ for $Z_{ {\rm rec},3}$. The above equations include the bias effects as a slight extension of those given by \citet{Hikage:2019ihj}. Approximations in the IR limit of $Z_{ {\rm rec},2}$ and $Z_{ {\rm rec},3}$ in the form needed to compute the 1-loop power spectrum are given by
\end{widetext}
\begin{widetext}
\begin{eqnarray}
    Z_{ {\rm rec},2}(\VEC{k},\VEC{p}) 
    &\xrightarrow[p\to0]{}& 
    \frac{1}{2}     \left\{ \left( \frac{\VEC{k}\cdot\MAT{R}_1\cdot\VEC{p}}{p^2} \right) 
    + \left( \frac{\VEC{k}\cdot\VEC{p}}{p^2} \right) \left( -\frac{W_{\rm G}(pR_{\rm s})}{b_{\rm 1,fid}} \right) 
    Z_1(\VEC{p}) \right\}Z_1(\VEC{k})\;,
    \nonumber \\
    Z_{ {\rm rec},3}(\VEC{k},\VEC{p},-\VEC{p}) 
    &\xrightarrow[p\to0]{}& 
    -\frac{1}{3!}
    \left\{ \left( \frac{\VEC{k}\cdot\MAT{R}_1\cdot\VEC{p}}{p^2} \right) 
    + \left( \frac{\VEC{k}\cdot\VEC{p}}{p^2} \right) \left( -\frac{W_{\rm G}(pR_{\rm s})}{b_{\rm 1,fid}} \right)
    Z_1(\VEC{p}) \right\}^2 Z_1(\VEC{k})\;.
\end{eqnarray}
\end{widetext}

%\nocite{*}
\bibliography{ms}% Produces the bibliography via BibTeX.

\begin{thebibliography}{83}%
\makeatletter
\providecommand \@ifxundefined [1]{%
 \@ifx{#1\undefined}
}%
\providecommand \@ifnum [1]{%
 \ifnum #1\expandafter \@firstoftwo
 \else \expandafter \@secondoftwo
 \fi
}%
\providecommand \@ifx [1]{%
 \ifx #1\expandafter \@firstoftwo
 \else \expandafter \@secondoftwo
 \fi
}%
\providecommand \natexlab [1]{#1}%
\providecommand \enquote  [1]{``#1''}%
\providecommand \bibnamefont  [1]{#1}%
\providecommand \bibfnamefont [1]{#1}%
\providecommand \citenamefont [1]{#1}%
\providecommand \href@noop [0]{\@secondoftwo}%
\providecommand \href [0]{\begingroup \@sanitize@url \@href}%
\providecommand \@href[1]{\@@startlink{#1}\@@href}%
\providecommand \@@href[1]{\endgroup#1\@@endlink}%
\providecommand \@sanitize@url [0]{\catcode `\\12\catcode `\$12\catcode
  `\&12\catcode `\#12\catcode `\^12\catcode `\_12\catcode `\%12\relax}%
\providecommand \@@startlink[1]{}%
\providecommand \@@endlink[0]{}%
\providecommand \url  [0]{\begingroup\@sanitize@url \@url }%
\providecommand \@url [1]{\endgroup\@href {#1}{\urlprefix }}%
\providecommand \urlprefix  [0]{URL }%
\providecommand \Eprint [0]{\href }%
\providecommand \doibase [0]{http://dx.doi.org/}%
\providecommand \selectlanguage [0]{\@gobble}%
\providecommand \bibinfo  [0]{\@secondoftwo}%
\providecommand \bibfield  [0]{\@secondoftwo}%
\providecommand \translation [1]{[#1]}%
\providecommand \BibitemOpen [0]{}%
\providecommand \bibitemStop [0]{}%
\providecommand \bibitemNoStop [0]{.\EOS\space}%
\providecommand \EOS [0]{\spacefactor3000\relax}%
\providecommand \BibitemShut  [1]{\csname bibitem#1\endcsname}%
\let\auto@bib@innerbib\@empty
%</preamble>
\bibitem [{\citenamefont {Eisenstein}\ \emph
  {et~al.}(2007{\natexlab{a}})\citenamefont {Eisenstein}, \citenamefont {Seo},
  \citenamefont {Sirko},\ and\ \citenamefont {Spergel}}]{Eisenstein:2006nk}%
  \BibitemOpen
  \bibfield  {author} {\bibinfo {author} {\bibfnamefont {D.~J.}\ \bibnamefont
  {Eisenstein}}, \bibinfo {author} {\bibfnamefont {H.-j.}\ \bibnamefont {Seo}},
  \bibinfo {author} {\bibfnamefont {E.}~\bibnamefont {Sirko}}, \ and\ \bibinfo
  {author} {\bibfnamefont {D.}~\bibnamefont {Spergel}},\ }\href {\doibase
  10.1086/518712} {\bibfield  {journal} {\bibinfo  {journal} {Astrophys. J.}\
  }\textbf {\bibinfo {volume} {664}},\ \bibinfo {pages} {675} (\bibinfo {year}
  {2007}{\natexlab{a}})},\ \Eprint {http://arxiv.org/abs/astro-ph/0604362}
  {arXiv:astro-ph/0604362} \BibitemShut {NoStop}%
\bibitem [{\citenamefont {Sunyaev}\ and\ \citenamefont
  {Zeldovich}(1970)}]{Sunyaev:1970eu}%
  \BibitemOpen
  \bibfield  {author} {\bibinfo {author} {\bibfnamefont {R.~A.}\ \bibnamefont
  {Sunyaev}}\ and\ \bibinfo {author} {\bibfnamefont {Y.~B.}\ \bibnamefont
  {Zeldovich}},\ }\href@noop {} {\bibfield  {journal} {\bibinfo  {journal}
  {Astrophys. Space Sci.}\ }\textbf {\bibinfo {volume} {7}},\ \bibinfo {pages}
  {3} (\bibinfo {year} {1970})}\BibitemShut {NoStop}%
\bibitem [{\citenamefont {Peebles}\ and\ \citenamefont
  {Yu}(1970)}]{Peebles:1970ag}%
  \BibitemOpen
  \bibfield  {author} {\bibinfo {author} {\bibfnamefont {P.~J.~E.}\
  \bibnamefont {Peebles}}\ and\ \bibinfo {author} {\bibfnamefont {J.~T.}\
  \bibnamefont {Yu}},\ }\href {\doibase 10.1086/150713} {\bibfield  {journal}
  {\bibinfo  {journal} {Astrophys. J.}\ }\textbf {\bibinfo {volume} {162}},\
  \bibinfo {pages} {815} (\bibinfo {year} {1970})}\BibitemShut {NoStop}%
\bibitem [{\citenamefont {Hikage}\ \emph
  {et~al.}(2020{\natexlab{a}})\citenamefont {Hikage}, \citenamefont
  {Takahashi},\ and\ \citenamefont {Koyama}}]{Hikage:2020fte}%
  \BibitemOpen
  \bibfield  {author} {\bibinfo {author} {\bibfnamefont {C.}~\bibnamefont
  {Hikage}}, \bibinfo {author} {\bibfnamefont {R.}~\bibnamefont {Takahashi}}, \
  and\ \bibinfo {author} {\bibfnamefont {K.}~\bibnamefont {Koyama}},\ }\href
  {\doibase 10.1103/PhysRevD.102.083514} {\bibfield  {journal} {\bibinfo
  {journal} {Phys. Rev. D}\ }\textbf {\bibinfo {volume} {102}},\ \bibinfo
  {pages} {083514} (\bibinfo {year} {2020}{\natexlab{a}})},\ \Eprint
  {http://arxiv.org/abs/2007.13998} {arXiv:2007.13998 [astro-ph.CO]}
  \BibitemShut {NoStop}%
\bibitem [{\citenamefont {{Wang}}\ \emph {et~al.}(2022)\citenamefont {{Wang}},
  \citenamefont {{Zhao}}, \citenamefont {{Koyama}}, \citenamefont {{Percival}},
  \citenamefont {{Takahashi}}, \citenamefont {{Hikage}}, \citenamefont
  {{Gil-Mar{\'\i}n}}, \citenamefont {{Hahn}}, \citenamefont {{Zhao}},
  \citenamefont {{Zhang}}, \citenamefont {{Mu}}, \citenamefont {{Yu}},
  \citenamefont {{Zhu}},\ and\ \citenamefont {{Ge}}}]{Wang:2022nlx}%
  \BibitemOpen
  \bibfield  {author} {\bibinfo {author} {\bibfnamefont {Y.}~\bibnamefont
  {{Wang}}}, \bibinfo {author} {\bibfnamefont {G.-B.}\ \bibnamefont {{Zhao}}},
  \bibinfo {author} {\bibfnamefont {K.}~\bibnamefont {{Koyama}}}, \bibinfo
  {author} {\bibfnamefont {W.~J.}\ \bibnamefont {{Percival}}}, \bibinfo
  {author} {\bibfnamefont {R.}~\bibnamefont {{Takahashi}}}, \bibinfo {author}
  {\bibfnamefont {C.}~\bibnamefont {{Hikage}}}, \bibinfo {author}
  {\bibfnamefont {H.}~\bibnamefont {{Gil-Mar{\'\i}n}}}, \bibinfo {author}
  {\bibfnamefont {C.}~\bibnamefont {{Hahn}}}, \bibinfo {author} {\bibfnamefont
  {R.}~\bibnamefont {{Zhao}}}, \bibinfo {author} {\bibfnamefont
  {W.}~\bibnamefont {{Zhang}}}, \bibinfo {author} {\bibfnamefont
  {X.}~\bibnamefont {{Mu}}}, \bibinfo {author} {\bibfnamefont {Y.}~\bibnamefont
  {{Yu}}}, \bibinfo {author} {\bibfnamefont {H.-M.}\ \bibnamefont {{Zhu}}}, \
  and\ \bibinfo {author} {\bibfnamefont {F.}~\bibnamefont {{Ge}}},\ }\href
  {\doibase 10.48550/arXiv.2202.05248} {\bibfield  {journal} {\bibinfo
  {journal} {arXiv e-prints}\ ,\ \bibinfo {eid} {arXiv:2202.05248}} (\bibinfo
  {year} {2022})},\ \Eprint {http://arxiv.org/abs/2202.05248} {arXiv:2202.05248
  [astro-ph.CO]} \BibitemShut {NoStop}%
\bibitem [{\citenamefont {Shirasaki}\ \emph {et~al.}(2021)\citenamefont
  {Shirasaki}, \citenamefont {Sugiyama}, \citenamefont {Takahashi},\ and\
  \citenamefont {Kitaura}}]{Shirasaki:2020vkk}%
  \BibitemOpen
  \bibfield  {author} {\bibinfo {author} {\bibfnamefont {M.}~\bibnamefont
  {Shirasaki}}, \bibinfo {author} {\bibfnamefont {N.~S.}\ \bibnamefont
  {Sugiyama}}, \bibinfo {author} {\bibfnamefont {R.}~\bibnamefont {Takahashi}},
  \ and\ \bibinfo {author} {\bibfnamefont {F.-S.}\ \bibnamefont {Kitaura}},\
  }\href {\doibase 10.1103/PhysRevD.103.023506} {\bibfield  {journal} {\bibinfo
   {journal} {Phys. Rev. D}\ }\textbf {\bibinfo {volume} {103}},\ \bibinfo
  {pages} {023506} (\bibinfo {year} {2021})},\ \Eprint
  {http://arxiv.org/abs/2010.04567} {arXiv:2010.04567 [astro-ph.CO]}
  \BibitemShut {NoStop}%
\bibitem [{\citenamefont {Bernardeau}\ \emph {et~al.}(2002)\citenamefont
  {Bernardeau}, \citenamefont {Colombi}, \citenamefont {Gaztanaga},\ and\
  \citenamefont {Scoccimarro}}]{Bernardeau:2001qr}%
  \BibitemOpen
  \bibfield  {author} {\bibinfo {author} {\bibfnamefont {F.}~\bibnamefont
  {Bernardeau}}, \bibinfo {author} {\bibfnamefont {S.}~\bibnamefont {Colombi}},
  \bibinfo {author} {\bibfnamefont {E.}~\bibnamefont {Gaztanaga}}, \ and\
  \bibinfo {author} {\bibfnamefont {R.}~\bibnamefont {Scoccimarro}},\ }\href
  {\doibase 10.1016/S0370-1573(02)00135-7} {\bibfield  {journal} {\bibinfo
  {journal} {Phys. Rept.}\ }\textbf {\bibinfo {volume} {367}},\ \bibinfo
  {pages} {1} (\bibinfo {year} {2002})},\ \Eprint
  {http://arxiv.org/abs/astro-ph/0112551} {arXiv:astro-ph/0112551} \BibitemShut
  {NoStop}%
\bibitem [{\citenamefont {Carlson}\ \emph {et~al.}(2013)\citenamefont
  {Carlson}, \citenamefont {Reid},\ and\ \citenamefont
  {White}}]{Carlson:2012bu}%
  \BibitemOpen
  \bibfield  {author} {\bibinfo {author} {\bibfnamefont {J.}~\bibnamefont
  {Carlson}}, \bibinfo {author} {\bibfnamefont {B.}~\bibnamefont {Reid}}, \
  and\ \bibinfo {author} {\bibfnamefont {M.}~\bibnamefont {White}},\ }\href
  {\doibase 10.1093/mnras/sts457} {\bibfield  {journal} {\bibinfo  {journal}
  {Mon. Not. Roy. Astron. Soc.}\ }\textbf {\bibinfo {volume} {429}},\ \bibinfo
  {pages} {1674} (\bibinfo {year} {2013})},\ \Eprint
  {http://arxiv.org/abs/1209.0780} {arXiv:1209.0780 [astro-ph.CO]} \BibitemShut
  {NoStop}%
\bibitem [{\citenamefont {Wang}\ \emph {et~al.}(2014)\citenamefont {Wang},
  \citenamefont {Reid},\ and\ \citenamefont {White}}]{Wang:2013hwa}%
  \BibitemOpen
  \bibfield  {author} {\bibinfo {author} {\bibfnamefont {L.}~\bibnamefont
  {Wang}}, \bibinfo {author} {\bibfnamefont {B.}~\bibnamefont {Reid}}, \ and\
  \bibinfo {author} {\bibfnamefont {M.}~\bibnamefont {White}},\ }\href
  {\doibase 10.1093/mnras/stt1916} {\bibfield  {journal} {\bibinfo  {journal}
  {Mon. Not. Roy. Astron. Soc.}\ }\textbf {\bibinfo {volume} {437}},\ \bibinfo
  {pages} {588} (\bibinfo {year} {2014})},\ \Eprint
  {http://arxiv.org/abs/1306.1804} {arXiv:1306.1804 [astro-ph.CO]} \BibitemShut
  {NoStop}%
\bibitem [{\citenamefont {Crocce}\ and\ \citenamefont
  {Scoccimarro}(2006)}]{Crocce:2005xy}%
  \BibitemOpen
  \bibfield  {author} {\bibinfo {author} {\bibfnamefont {M.}~\bibnamefont
  {Crocce}}\ and\ \bibinfo {author} {\bibfnamefont {R.}~\bibnamefont
  {Scoccimarro}},\ }\href {\doibase 10.1103/PhysRevD.73.063519} {\bibfield
  {journal} {\bibinfo  {journal} {Phys. Rev.}\ }\textbf {\bibinfo {volume}
  {D73}},\ \bibinfo {pages} {063519} (\bibinfo {year} {2006})},\ \Eprint
  {http://arxiv.org/abs/astro-ph/0509418} {arXiv:astro-ph/0509418 [astro-ph]}
  \BibitemShut {NoStop}%
%%CITATION = ASTRO-PH/0509418;%%
\bibitem [{\citenamefont {Taruya}\ \emph {et~al.}(2010)\citenamefont {Taruya},
  \citenamefont {Nishimichi},\ and\ \citenamefont {Saito}}]{Taruya:2010mx}%
  \BibitemOpen
  \bibfield  {author} {\bibinfo {author} {\bibfnamefont {A.}~\bibnamefont
  {Taruya}}, \bibinfo {author} {\bibfnamefont {T.}~\bibnamefont {Nishimichi}},
  \ and\ \bibinfo {author} {\bibfnamefont {S.}~\bibnamefont {Saito}},\ }\href
  {\doibase 10.1103/PhysRevD.82.063522} {\bibfield  {journal} {\bibinfo
  {journal} {Phys. Rev. D}\ }\textbf {\bibinfo {volume} {82}},\ \bibinfo
  {pages} {063522} (\bibinfo {year} {2010})},\ \Eprint
  {http://arxiv.org/abs/1006.0699} {arXiv:1006.0699 [astro-ph.CO]} \BibitemShut
  {NoStop}%
\bibitem [{\citenamefont {Baumann}\ \emph {et~al.}(2012)\citenamefont
  {Baumann}, \citenamefont {Nicolis}, \citenamefont {Senatore},\ and\
  \citenamefont {Zaldarriaga}}]{Baumann:2010tm}%
  \BibitemOpen
  \bibfield  {author} {\bibinfo {author} {\bibfnamefont {D.}~\bibnamefont
  {Baumann}}, \bibinfo {author} {\bibfnamefont {A.}~\bibnamefont {Nicolis}},
  \bibinfo {author} {\bibfnamefont {L.}~\bibnamefont {Senatore}}, \ and\
  \bibinfo {author} {\bibfnamefont {M.}~\bibnamefont {Zaldarriaga}},\ }\href
  {\doibase 10.1088/1475-7516/2012/07/051} {\bibfield  {journal} {\bibinfo
  {journal} {JCAP}\ }\textbf {\bibinfo {volume} {07}},\ \bibinfo {pages} {051}
  (\bibinfo {year} {2012})},\ \Eprint {http://arxiv.org/abs/1004.2488}
  {arXiv:1004.2488 [astro-ph.CO]} \BibitemShut {NoStop}%
\bibitem [{\citenamefont {Carrasco}\ \emph {et~al.}(2012)\citenamefont
  {Carrasco}, \citenamefont {Hertzberg},\ and\ \citenamefont
  {Senatore}}]{Carrasco:2012cv}%
  \BibitemOpen
  \bibfield  {author} {\bibinfo {author} {\bibfnamefont {J.~J.~M.}\
  \bibnamefont {Carrasco}}, \bibinfo {author} {\bibfnamefont {M.~P.}\
  \bibnamefont {Hertzberg}}, \ and\ \bibinfo {author} {\bibfnamefont
  {L.}~\bibnamefont {Senatore}},\ }\href {\doibase 10.1007/JHEP09(2012)082}
  {\bibfield  {journal} {\bibinfo  {journal} {JHEP}\ }\textbf {\bibinfo
  {volume} {09}},\ \bibinfo {pages} {082} (\bibinfo {year} {2012})},\ \Eprint
  {http://arxiv.org/abs/1206.2926} {arXiv:1206.2926 [astro-ph.CO]} \BibitemShut
  {NoStop}%
\bibitem [{\citenamefont {Eisenstein}\ \emph {et~al.}(2011)\citenamefont
  {Eisenstein} \emph {et~al.}}]{Eisenstein:2011sa}%
  \BibitemOpen
  \bibfield  {author} {\bibinfo {author} {\bibfnamefont {D.~J.}\ \bibnamefont
  {Eisenstein}} \emph {et~al.} (\bibinfo {collaboration} {SDSS}),\ }\href
  {\doibase 10.1088/0004-6256/142/3/72} {\bibfield  {journal} {\bibinfo
  {journal} {Astron. J.}\ }\textbf {\bibinfo {volume} {142}},\ \bibinfo {pages}
  {72} (\bibinfo {year} {2011})},\ \Eprint {http://arxiv.org/abs/1101.1529}
  {arXiv:1101.1529 [astro-ph.IM]} \BibitemShut {NoStop}%
%%CITATION = ARXIV:1101.1529;%%
\bibitem [{\citenamefont {Bolton}\ \emph {et~al.}(2012)\citenamefont {Bolton}
  \emph {et~al.}}]{Bolton:2012hz}%
  \BibitemOpen
  \bibfield  {author} {\bibinfo {author} {\bibfnamefont {A.~S.}\ \bibnamefont
  {Bolton}} \emph {et~al.} (\bibinfo {collaboration} {Cutler Group, LP}),\
  }\href {\doibase 10.1088/0004-6256/144/5/144} {\bibfield  {journal} {\bibinfo
   {journal} {Astron. J.}\ }\textbf {\bibinfo {volume} {144}},\ \bibinfo
  {pages} {144} (\bibinfo {year} {2012})},\ \Eprint
  {http://arxiv.org/abs/1207.7326} {arXiv:1207.7326 [astro-ph.CO]} \BibitemShut
  {NoStop}%
%%CITATION = ARXIV:1207.7326;%%
\bibitem [{\citenamefont {Dawson}\ \emph {et~al.}(2013)\citenamefont {Dawson}
  \emph {et~al.}}]{Dawson:2012va}%
  \BibitemOpen
  \bibfield  {author} {\bibinfo {author} {\bibfnamefont {K.~S.}\ \bibnamefont
  {Dawson}} \emph {et~al.} (\bibinfo {collaboration} {BOSS}),\ }\href {\doibase
  10.1088/0004-6256/145/1/10} {\bibfield  {journal} {\bibinfo  {journal}
  {Astron. J.}\ }\textbf {\bibinfo {volume} {145}},\ \bibinfo {pages} {10}
  (\bibinfo {year} {2013})},\ \Eprint {http://arxiv.org/abs/1208.0022}
  {arXiv:1208.0022 [astro-ph.CO]} \BibitemShut {NoStop}%
%%CITATION = ARXIV:1208.0022;%%
\bibitem [{\citenamefont {Alam}\ \emph {et~al.}(2015)\citenamefont {Alam} \emph
  {et~al.}}]{Alam:2015mbd}%
  \BibitemOpen
  \bibfield  {author} {\bibinfo {author} {\bibfnamefont {S.}~\bibnamefont
  {Alam}} \emph {et~al.} (\bibinfo {collaboration} {SDSS-III}),\ }\href
  {\doibase 10.1088/0067-0049/219/1/12} {\bibfield  {journal} {\bibinfo
  {journal} {Astrophys. J. Suppl.}\ }\textbf {\bibinfo {volume} {219}},\
  \bibinfo {pages} {12} (\bibinfo {year} {2015})},\ \Eprint
  {http://arxiv.org/abs/1501.00963} {arXiv:1501.00963 [astro-ph.IM]}
  \BibitemShut {NoStop}%
%%CITATION = ARXIV:1501.00963;%%
\bibitem [{\citenamefont {Schmittfull}\ \emph
  {et~al.}(2015{\natexlab{a}})\citenamefont {Schmittfull}, \citenamefont
  {Feng}, \citenamefont {Beutler}, \citenamefont {Sherwin},\ and\ \citenamefont
  {Chu}}]{Schmittfull:2015mja}%
  \BibitemOpen
  \bibfield  {author} {\bibinfo {author} {\bibfnamefont {M.}~\bibnamefont
  {Schmittfull}}, \bibinfo {author} {\bibfnamefont {Y.}~\bibnamefont {Feng}},
  \bibinfo {author} {\bibfnamefont {F.}~\bibnamefont {Beutler}}, \bibinfo
  {author} {\bibfnamefont {B.}~\bibnamefont {Sherwin}}, \ and\ \bibinfo
  {author} {\bibfnamefont {M.~Y.}\ \bibnamefont {Chu}},\ }\href {\doibase
  10.1103/PhysRevD.92.123522} {\bibfield  {journal} {\bibinfo  {journal} {Phys.
  Rev. D}\ }\textbf {\bibinfo {volume} {92}},\ \bibinfo {pages} {123522}
  (\bibinfo {year} {2015}{\natexlab{a}})},\ \Eprint
  {http://arxiv.org/abs/1508.06972} {arXiv:1508.06972 [astro-ph.CO]}
  \BibitemShut {NoStop}%
\bibitem [{\citenamefont {Hikage}\ \emph {et~al.}(2017)\citenamefont {Hikage},
  \citenamefont {Koyama},\ and\ \citenamefont {Heavens}}]{Hikage:2017tmm}%
  \BibitemOpen
  \bibfield  {author} {\bibinfo {author} {\bibfnamefont {C.}~\bibnamefont
  {Hikage}}, \bibinfo {author} {\bibfnamefont {K.}~\bibnamefont {Koyama}}, \
  and\ \bibinfo {author} {\bibfnamefont {A.}~\bibnamefont {Heavens}},\ }\href
  {\doibase 10.1103/PhysRevD.96.043513} {\bibfield  {journal} {\bibinfo
  {journal} {Phys. Rev. D}\ }\textbf {\bibinfo {volume} {96}},\ \bibinfo
  {pages} {043513} (\bibinfo {year} {2017})},\ \Eprint
  {http://arxiv.org/abs/1703.07878} {arXiv:1703.07878 [astro-ph.CO]}
  \BibitemShut {NoStop}%
\bibitem [{\citenamefont {Hikage}\ \emph
  {et~al.}(2020{\natexlab{b}})\citenamefont {Hikage}, \citenamefont {Koyama},\
  and\ \citenamefont {Takahashi}}]{Hikage:2019ihj}%
  \BibitemOpen
  \bibfield  {author} {\bibinfo {author} {\bibfnamefont {C.}~\bibnamefont
  {Hikage}}, \bibinfo {author} {\bibfnamefont {K.}~\bibnamefont {Koyama}}, \
  and\ \bibinfo {author} {\bibfnamefont {R.}~\bibnamefont {Takahashi}},\ }\href
  {\doibase 10.1103/PhysRevD.101.043510} {\bibfield  {journal} {\bibinfo
  {journal} {Phys. Rev. D}\ }\textbf {\bibinfo {volume} {101}},\ \bibinfo
  {pages} {043510} (\bibinfo {year} {2020}{\natexlab{b}})},\ \Eprint
  {http://arxiv.org/abs/1911.06461} {arXiv:1911.06461 [astro-ph.CO]}
  \BibitemShut {NoStop}%
\bibitem [{\citenamefont {Padmanabhan}\ \emph {et~al.}(2009)\citenamefont
  {Padmanabhan}, \citenamefont {White},\ and\ \citenamefont
  {Cohn}}]{Padmanabhan:2008dd}%
  \BibitemOpen
  \bibfield  {author} {\bibinfo {author} {\bibfnamefont {N.}~\bibnamefont
  {Padmanabhan}}, \bibinfo {author} {\bibfnamefont {M.}~\bibnamefont {White}},
  \ and\ \bibinfo {author} {\bibfnamefont {J.~D.}\ \bibnamefont {Cohn}},\
  }\href {\doibase 10.1103/PhysRevD.79.063523} {\bibfield  {journal} {\bibinfo
  {journal} {Phys. Rev. D}\ }\textbf {\bibinfo {volume} {79}},\ \bibinfo
  {pages} {063523} (\bibinfo {year} {2009})},\ \Eprint
  {http://arxiv.org/abs/0812.2905} {arXiv:0812.2905 [astro-ph]} \BibitemShut
  {NoStop}%
\bibitem [{\citenamefont {Seo}\ \emph {et~al.}(2016)\citenamefont {Seo},
  \citenamefont {Beutler}, \citenamefont {Ross},\ and\ \citenamefont
  {Saito}}]{Seo:2015eyw}%
  \BibitemOpen
  \bibfield  {author} {\bibinfo {author} {\bibfnamefont {H.-J.}\ \bibnamefont
  {Seo}}, \bibinfo {author} {\bibfnamefont {F.}~\bibnamefont {Beutler}},
  \bibinfo {author} {\bibfnamefont {A.~J.}\ \bibnamefont {Ross}}, \ and\
  \bibinfo {author} {\bibfnamefont {S.}~\bibnamefont {Saito}},\ }\href
  {\doibase 10.1093/mnras/stw1138} {\bibfield  {journal} {\bibinfo  {journal}
  {Mon. Not. Roy. Astron. Soc.}\ }\textbf {\bibinfo {volume} {460}},\ \bibinfo
  {pages} {2453} (\bibinfo {year} {2016})},\ \Eprint
  {http://arxiv.org/abs/1511.00663} {arXiv:1511.00663 [astro-ph.CO]}
  \BibitemShut {NoStop}%
\bibitem [{\citenamefont {White}(2015)}]{White:2015eaa}%
  \BibitemOpen
  \bibfield  {author} {\bibinfo {author} {\bibfnamefont {M.}~\bibnamefont
  {White}},\ }\href {\doibase 10.1093/mnras/stv842} {\bibfield  {journal}
  {\bibinfo  {journal} {Mon. Not. Roy. Astron. Soc.}\ }\textbf {\bibinfo
  {volume} {450}},\ \bibinfo {pages} {3822} (\bibinfo {year} {2015})},\ \Eprint
  {http://arxiv.org/abs/1504.03677} {arXiv:1504.03677 [astro-ph.CO]}
  \BibitemShut {NoStop}%
\bibitem [{\citenamefont {Chen}\ \emph {et~al.}(2019)\citenamefont {Chen},
  \citenamefont {Vlah},\ and\ \citenamefont {White}}]{Chen:2019lpf}%
  \BibitemOpen
  \bibfield  {author} {\bibinfo {author} {\bibfnamefont {S.-F.}\ \bibnamefont
  {Chen}}, \bibinfo {author} {\bibfnamefont {Z.}~\bibnamefont {Vlah}}, \ and\
  \bibinfo {author} {\bibfnamefont {M.}~\bibnamefont {White}},\ }\href
  {\doibase 10.1088/1475-7516/2019/09/017} {\bibfield  {journal} {\bibinfo
  {journal} {JCAP}\ }\textbf {\bibinfo {volume} {09}},\ \bibinfo {pages} {017}
  (\bibinfo {year} {2019})},\ \Eprint {http://arxiv.org/abs/1907.00043}
  {arXiv:1907.00043 [astro-ph.CO]} \BibitemShut {NoStop}%
\bibitem [{\citenamefont {Ota}\ \emph {et~al.}(2021)\citenamefont {Ota},
  \citenamefont {Seo}, \citenamefont {Saito},\ and\ \citenamefont
  {Beutler}}]{Ota:2021caz}%
  \BibitemOpen
  \bibfield  {author} {\bibinfo {author} {\bibfnamefont {A.}~\bibnamefont
  {Ota}}, \bibinfo {author} {\bibfnamefont {H.-J.}\ \bibnamefont {Seo}},
  \bibinfo {author} {\bibfnamefont {S.}~\bibnamefont {Saito}}, \ and\ \bibinfo
  {author} {\bibfnamefont {F.}~\bibnamefont {Beutler}},\ }\href {\doibase
  10.1103/PhysRevD.104.123508} {\bibfield  {journal} {\bibinfo  {journal}
  {Phys. Rev. D}\ }\textbf {\bibinfo {volume} {104}},\ \bibinfo {pages}
  {123508} (\bibinfo {year} {2021})},\ \Eprint
  {http://arxiv.org/abs/2106.00146} {arXiv:2106.00146 [astro-ph.CO]}
  \BibitemShut {NoStop}%
\bibitem [{\citenamefont {Ota}\ \emph {et~al.}(2023)\citenamefont {Ota},
  \citenamefont {Seo}, \citenamefont {Saito},\ and\ \citenamefont
  {Beutler}}]{Ota:2022him}%
  \BibitemOpen
  \bibfield  {author} {\bibinfo {author} {\bibfnamefont {A.}~\bibnamefont
  {Ota}}, \bibinfo {author} {\bibfnamefont {H.-J.}\ \bibnamefont {Seo}},
  \bibinfo {author} {\bibfnamefont {S.}~\bibnamefont {Saito}}, \ and\ \bibinfo
  {author} {\bibfnamefont {F.}~\bibnamefont {Beutler}},\ }\href {\doibase
  10.1103/PhysRevD.107.123523} {\bibfield  {journal} {\bibinfo  {journal}
  {Phys. Rev. D}\ }\textbf {\bibinfo {volume} {107}},\ \bibinfo {pages}
  {123523} (\bibinfo {year} {2023})},\ \Eprint
  {http://arxiv.org/abs/2211.07960} {arXiv:2211.07960 [astro-ph.CO]}
  \BibitemShut {NoStop}%
\bibitem [{\citenamefont {Zhu}\ \emph {et~al.}(2017)\citenamefont {Zhu},
  \citenamefont {Yu}, \citenamefont {Pen}, \citenamefont {Chen},\ and\
  \citenamefont {Yu}}]{Zhu:2016sjc}%
  \BibitemOpen
  \bibfield  {author} {\bibinfo {author} {\bibfnamefont {H.-M.}\ \bibnamefont
  {Zhu}}, \bibinfo {author} {\bibfnamefont {Y.}~\bibnamefont {Yu}}, \bibinfo
  {author} {\bibfnamefont {U.-L.}\ \bibnamefont {Pen}}, \bibinfo {author}
  {\bibfnamefont {X.}~\bibnamefont {Chen}}, \ and\ \bibinfo {author}
  {\bibfnamefont {H.-R.}\ \bibnamefont {Yu}},\ }\href {\doibase
  10.1103/PhysRevD.96.123502} {\bibfield  {journal} {\bibinfo  {journal} {Phys.
  Rev. D}\ }\textbf {\bibinfo {volume} {96}},\ \bibinfo {pages} {123502}
  (\bibinfo {year} {2017})},\ \Eprint {http://arxiv.org/abs/1611.09638}
  {arXiv:1611.09638 [astro-ph.CO]} \BibitemShut {NoStop}%
\bibitem [{\citenamefont {Zhu}\ \emph {et~al.}(2018)\citenamefont {Zhu},
  \citenamefont {Yu},\ and\ \citenamefont {Pen}}]{Zhu:2017vtj}%
  \BibitemOpen
  \bibfield  {author} {\bibinfo {author} {\bibfnamefont {H.-M.}\ \bibnamefont
  {Zhu}}, \bibinfo {author} {\bibfnamefont {Y.}~\bibnamefont {Yu}}, \ and\
  \bibinfo {author} {\bibfnamefont {U.-L.}\ \bibnamefont {Pen}},\ }\href
  {\doibase 10.1103/PhysRevD.97.043502} {\bibfield  {journal} {\bibinfo
  {journal} {Phys. Rev. D}\ }\textbf {\bibinfo {volume} {97}},\ \bibinfo
  {pages} {043502} (\bibinfo {year} {2018})},\ \Eprint
  {http://arxiv.org/abs/1711.03218} {arXiv:1711.03218 [astro-ph.CO]}
  \BibitemShut {NoStop}%
\bibitem [{\citenamefont {Yu}\ \emph {et~al.}(2017)\citenamefont {Yu},
  \citenamefont {Zhu},\ and\ \citenamefont {Pen}}]{Yu:2017tpa}%
  \BibitemOpen
  \bibfield  {author} {\bibinfo {author} {\bibfnamefont {Y.}~\bibnamefont
  {Yu}}, \bibinfo {author} {\bibfnamefont {H.-M.}\ \bibnamefont {Zhu}}, \ and\
  \bibinfo {author} {\bibfnamefont {U.-L.}\ \bibnamefont {Pen}},\ }\href
  {\doibase 10.3847/1538-4357/aa89e7} {\bibfield  {journal} {\bibinfo
  {journal} {Astrophys. J.}\ }\textbf {\bibinfo {volume} {847}},\ \bibinfo
  {pages} {110} (\bibinfo {year} {2017})},\ \Eprint
  {http://arxiv.org/abs/1703.08301} {arXiv:1703.08301 [astro-ph.CO]}
  \BibitemShut {NoStop}%
\bibitem [{\citenamefont {Wang}\ \emph {et~al.}(2017)\citenamefont {Wang},
  \citenamefont {Yu}, \citenamefont {Zhu}, \citenamefont {Yu}, \citenamefont
  {Pan},\ and\ \citenamefont {Pen}}]{Wang:2017jeq}%
  \BibitemOpen
  \bibfield  {author} {\bibinfo {author} {\bibfnamefont {X.}~\bibnamefont
  {Wang}}, \bibinfo {author} {\bibfnamefont {H.-R.}\ \bibnamefont {Yu}},
  \bibinfo {author} {\bibfnamefont {H.-M.}\ \bibnamefont {Zhu}}, \bibinfo
  {author} {\bibfnamefont {Y.}~\bibnamefont {Yu}}, \bibinfo {author}
  {\bibfnamefont {Q.}~\bibnamefont {Pan}}, \ and\ \bibinfo {author}
  {\bibfnamefont {U.-L.}\ \bibnamefont {Pen}},\ }\href {\doibase
  10.3847/2041-8213/aa738c} {\bibfield  {journal} {\bibinfo  {journal}
  {Astrophys. J. Lett.}\ }\textbf {\bibinfo {volume} {841}},\ \bibinfo {pages}
  {L29} (\bibinfo {year} {2017})},\ \Eprint {http://arxiv.org/abs/1703.09742}
  {arXiv:1703.09742 [astro-ph.CO]} \BibitemShut {NoStop}%
\bibitem [{\citenamefont {Schmittfull}\ \emph {et~al.}(2017)\citenamefont
  {Schmittfull}, \citenamefont {Baldauf},\ and\ \citenamefont
  {Zaldarriaga}}]{Schmittfull:2017uhh}%
  \BibitemOpen
  \bibfield  {author} {\bibinfo {author} {\bibfnamefont {M.}~\bibnamefont
  {Schmittfull}}, \bibinfo {author} {\bibfnamefont {T.}~\bibnamefont
  {Baldauf}}, \ and\ \bibinfo {author} {\bibfnamefont {M.}~\bibnamefont
  {Zaldarriaga}},\ }\href {\doibase 10.1103/PhysRevD.96.023505} {\bibfield
  {journal} {\bibinfo  {journal} {Phys. Rev. D}\ }\textbf {\bibinfo {volume}
  {96}},\ \bibinfo {pages} {023505} (\bibinfo {year} {2017})},\ \Eprint
  {http://arxiv.org/abs/1704.06634} {arXiv:1704.06634 [astro-ph.CO]}
  \BibitemShut {NoStop}%
\bibitem [{\citenamefont {Hada}\ and\ \citenamefont
  {Eisenstein}(2018)}]{Hada:2018fde}%
  \BibitemOpen
  \bibfield  {author} {\bibinfo {author} {\bibfnamefont {R.}~\bibnamefont
  {Hada}}\ and\ \bibinfo {author} {\bibfnamefont {D.~J.}\ \bibnamefont
  {Eisenstein}},\ }\href {\doibase 10.1093/mnras/sty1203} {\bibfield  {journal}
  {\bibinfo  {journal} {Mon. Not. Roy. Astron. Soc.}\ }\textbf {\bibinfo
  {volume} {478}},\ \bibinfo {pages} {1866} (\bibinfo {year} {2018})},\ \Eprint
  {http://arxiv.org/abs/1804.04738} {arXiv:1804.04738 [astro-ph.CO]}
  \BibitemShut {NoStop}%
\bibitem [{\citenamefont {Hada}\ and\ \citenamefont
  {Eisenstein}(2019)}]{Hada:2018ziy}%
  \BibitemOpen
  \bibfield  {author} {\bibinfo {author} {\bibfnamefont {R.}~\bibnamefont
  {Hada}}\ and\ \bibinfo {author} {\bibfnamefont {D.~J.}\ \bibnamefont
  {Eisenstein}},\ }\href {\doibase 10.1093/mnras/sty3137} {\bibfield  {journal}
  {\bibinfo  {journal} {Mon. Not. Roy. Astron. Soc.}\ }\textbf {\bibinfo
  {volume} {482}},\ \bibinfo {pages} {5685} (\bibinfo {year} {2019})},\ \Eprint
  {http://arxiv.org/abs/1810.05026} {arXiv:1810.05026 [astro-ph.CO]}
  \BibitemShut {NoStop}%
\bibitem [{\citenamefont {Seo}\ \emph {et~al.}(2022)\citenamefont {Seo},
  \citenamefont {Ota}, \citenamefont {Schmittfull}, \citenamefont {Saito},\
  and\ \citenamefont {Beutler}}]{Seo:2021nev}%
  \BibitemOpen
  \bibfield  {author} {\bibinfo {author} {\bibfnamefont {H.-J.}\ \bibnamefont
  {Seo}}, \bibinfo {author} {\bibfnamefont {A.}~\bibnamefont {Ota}}, \bibinfo
  {author} {\bibfnamefont {M.}~\bibnamefont {Schmittfull}}, \bibinfo {author}
  {\bibfnamefont {S.}~\bibnamefont {Saito}}, \ and\ \bibinfo {author}
  {\bibfnamefont {F.}~\bibnamefont {Beutler}},\ }\href {\doibase
  10.1093/mnras/stac082} {\bibfield  {journal} {\bibinfo  {journal} {Mon. Not.
  Roy. Astron. Soc.}\ }\textbf {\bibinfo {volume} {511}},\ \bibinfo {pages}
  {1557} (\bibinfo {year} {2022})},\ \Eprint {http://arxiv.org/abs/2106.00530}
  {arXiv:2106.00530 [astro-ph.CO]} \BibitemShut {NoStop}%
\bibitem [{\citenamefont {Jain}\ and\ \citenamefont
  {Bertschinger}(1996)}]{Jain:1995kx}%
  \BibitemOpen
  \bibfield  {author} {\bibinfo {author} {\bibfnamefont {B.}~\bibnamefont
  {Jain}}\ and\ \bibinfo {author} {\bibfnamefont {E.}~\bibnamefont
  {Bertschinger}},\ }\href {\doibase 10.1086/176625} {\bibfield  {journal}
  {\bibinfo  {journal} {Astrophys. J.}\ }\textbf {\bibinfo {volume} {456}},\
  \bibinfo {pages} {43} (\bibinfo {year} {1996})},\ \Eprint
  {http://arxiv.org/abs/astro-ph/9503025} {arXiv:astro-ph/9503025} \BibitemShut
  {NoStop}%
\bibitem [{\citenamefont {Scoccimarro}\ and\ \citenamefont
  {Frieman}(1996)}]{Scoccimarro:1995if}%
  \BibitemOpen
  \bibfield  {author} {\bibinfo {author} {\bibfnamefont {R.}~\bibnamefont
  {Scoccimarro}}\ and\ \bibinfo {author} {\bibfnamefont {J.}~\bibnamefont
  {Frieman}},\ }\href {\doibase 10.1086/192306} {\bibfield  {journal} {\bibinfo
   {journal} {Astrophys. J. Suppl.}\ }\textbf {\bibinfo {volume} {105}},\
  \bibinfo {pages} {37} (\bibinfo {year} {1996})},\ \Eprint
  {http://arxiv.org/abs/astro-ph/9509047} {arXiv:astro-ph/9509047} \BibitemShut
  {NoStop}%
\bibitem [{\citenamefont {Kehagias}\ and\ \citenamefont
  {Riotto}(2013)}]{Kehagias:2013yd}%
  \BibitemOpen
  \bibfield  {author} {\bibinfo {author} {\bibfnamefont {A.}~\bibnamefont
  {Kehagias}}\ and\ \bibinfo {author} {\bibfnamefont {A.}~\bibnamefont
  {Riotto}},\ }\href {\doibase 10.1016/j.nuclphysb.2013.05.009} {\bibfield
  {journal} {\bibinfo  {journal} {Nucl. Phys. B}\ }\textbf {\bibinfo {volume}
  {873}},\ \bibinfo {pages} {514} (\bibinfo {year} {2013})},\ \Eprint
  {http://arxiv.org/abs/1302.0130} {arXiv:1302.0130 [astro-ph.CO]} \BibitemShut
  {NoStop}%
\bibitem [{\citenamefont {Peloso}\ and\ \citenamefont
  {Pietroni}(2013)}]{Peloso:2013zw}%
  \BibitemOpen
  \bibfield  {author} {\bibinfo {author} {\bibfnamefont {M.}~\bibnamefont
  {Peloso}}\ and\ \bibinfo {author} {\bibfnamefont {M.}~\bibnamefont
  {Pietroni}},\ }\href {\doibase 10.1088/1475-7516/2013/05/031} {\bibfield
  {journal} {\bibinfo  {journal} {JCAP}\ }\textbf {\bibinfo {volume} {05}},\
  \bibinfo {pages} {031} (\bibinfo {year} {2013})},\ \Eprint
  {http://arxiv.org/abs/1302.0223} {arXiv:1302.0223 [astro-ph.CO]} \BibitemShut
  {NoStop}%
\bibitem [{\citenamefont {Sugiyama}\ and\ \citenamefont
  {Futamase}(2013)}]{Sugiyama:2013pwa}%
  \BibitemOpen
  \bibfield  {author} {\bibinfo {author} {\bibfnamefont {N.~S.}\ \bibnamefont
  {Sugiyama}}\ and\ \bibinfo {author} {\bibfnamefont {T.}~\bibnamefont
  {Futamase}},\ }\href {\doibase 10.1088/0004-637X/769/2/106} {\bibfield
  {journal} {\bibinfo  {journal} {Astrophys. J.}\ }\textbf {\bibinfo {volume}
  {769}},\ \bibinfo {pages} {106} (\bibinfo {year} {2013})},\ \Eprint
  {http://arxiv.org/abs/1303.2748} {arXiv:1303.2748 [astro-ph.CO]} \BibitemShut
  {NoStop}%
\bibitem [{\citenamefont {Sugiyama}\ and\ \citenamefont
  {Spergel}(2014)}]{Sugiyama:2013gza}%
  \BibitemOpen
  \bibfield  {author} {\bibinfo {author} {\bibfnamefont {N.~S.}\ \bibnamefont
  {Sugiyama}}\ and\ \bibinfo {author} {\bibfnamefont {D.~N.}\ \bibnamefont
  {Spergel}},\ }\href {\doibase 10.1088/1475-7516/2014/02/042} {\bibfield
  {journal} {\bibinfo  {journal} {JCAP}\ }\textbf {\bibinfo {volume} {02}},\
  \bibinfo {pages} {042} (\bibinfo {year} {2014})},\ \Eprint
  {http://arxiv.org/abs/1306.6660} {arXiv:1306.6660 [astro-ph.CO]} \BibitemShut
  {NoStop}%
\bibitem [{\citenamefont {Blas}\ \emph {et~al.}(2013)\citenamefont {Blas},
  \citenamefont {Garny},\ and\ \citenamefont {Konstandin}}]{Blas:2013bpa}%
  \BibitemOpen
  \bibfield  {author} {\bibinfo {author} {\bibfnamefont {D.}~\bibnamefont
  {Blas}}, \bibinfo {author} {\bibfnamefont {M.}~\bibnamefont {Garny}}, \ and\
  \bibinfo {author} {\bibfnamefont {T.}~\bibnamefont {Konstandin}},\ }\href
  {\doibase 10.1088/1475-7516/2013/09/024} {\bibfield  {journal} {\bibinfo
  {journal} {JCAP}\ }\textbf {\bibinfo {volume} {09}},\ \bibinfo {pages} {024}
  (\bibinfo {year} {2013})},\ \Eprint {http://arxiv.org/abs/1304.1546}
  {arXiv:1304.1546 [astro-ph.CO]} \BibitemShut {NoStop}%
\bibitem [{\citenamefont {Blas}\ \emph
  {et~al.}(2016{\natexlab{a}})\citenamefont {Blas}, \citenamefont {Garny},
  \citenamefont {Ivanov},\ and\ \citenamefont {Sibiryakov}}]{Blas:2015qsi}%
  \BibitemOpen
  \bibfield  {author} {\bibinfo {author} {\bibfnamefont {D.}~\bibnamefont
  {Blas}}, \bibinfo {author} {\bibfnamefont {M.}~\bibnamefont {Garny}},
  \bibinfo {author} {\bibfnamefont {M.~M.}\ \bibnamefont {Ivanov}}, \ and\
  \bibinfo {author} {\bibfnamefont {S.}~\bibnamefont {Sibiryakov}},\ }\href
  {\doibase 10.1088/1475-7516/2016/07/052} {\bibfield  {journal} {\bibinfo
  {journal} {JCAP}\ }\textbf {\bibinfo {volume} {07}},\ \bibinfo {pages} {052}
  (\bibinfo {year} {2016}{\natexlab{a}})},\ \Eprint
  {http://arxiv.org/abs/1512.05807} {arXiv:1512.05807 [astro-ph.CO]}
  \BibitemShut {NoStop}%
\bibitem [{\citenamefont {Lewandowski}\ and\ \citenamefont
  {Senatore}(2017)}]{Lewandowski:2017kes}%
  \BibitemOpen
  \bibfield  {author} {\bibinfo {author} {\bibfnamefont {M.}~\bibnamefont
  {Lewandowski}}\ and\ \bibinfo {author} {\bibfnamefont {L.}~\bibnamefont
  {Senatore}},\ }\href {\doibase 10.1088/1475-7516/2017/08/037} {\bibfield
  {journal} {\bibinfo  {journal} {JCAP}\ }\textbf {\bibinfo {volume} {08}},\
  \bibinfo {pages} {037} (\bibinfo {year} {2017})},\ \Eprint
  {http://arxiv.org/abs/1701.07012} {arXiv:1701.07012 [astro-ph.CO]}
  \BibitemShut {NoStop}%
\bibitem [{\citenamefont {Creminelli}\ \emph {et~al.}(2013)\citenamefont
  {Creminelli}, \citenamefont {Nore\~na}, \citenamefont {Simonovi\'c},\ and\
  \citenamefont {Vernizzi}}]{Creminelli:2013mca}%
  \BibitemOpen
  \bibfield  {author} {\bibinfo {author} {\bibfnamefont {P.}~\bibnamefont
  {Creminelli}}, \bibinfo {author} {\bibfnamefont {J.}~\bibnamefont
  {Nore\~na}}, \bibinfo {author} {\bibfnamefont {M.}~\bibnamefont
  {Simonovi\'c}}, \ and\ \bibinfo {author} {\bibfnamefont {F.}~\bibnamefont
  {Vernizzi}},\ }\href {\doibase 10.1088/1475-7516/2013/12/025} {\bibfield
  {journal} {\bibinfo  {journal} {JCAP}\ }\textbf {\bibinfo {volume} {12}},\
  \bibinfo {pages} {025} (\bibinfo {year} {2013})},\ \Eprint
  {http://arxiv.org/abs/1309.3557} {arXiv:1309.3557 [astro-ph.CO]} \BibitemShut
  {NoStop}%
\bibitem [{\citenamefont {Zeldovich}(1970)}]{Zeldovich:1969sb}%
  \BibitemOpen
  \bibfield  {author} {\bibinfo {author} {\bibfnamefont {Y.~B.}\ \bibnamefont
  {Zeldovich}},\ }\href@noop {} {\bibfield  {journal} {\bibinfo  {journal}
  {Astron. Astrophys.}\ }\textbf {\bibinfo {volume} {5}},\ \bibinfo {pages}
  {84} (\bibinfo {year} {1970})}\BibitemShut {NoStop}%
\bibitem [{\citenamefont {Eisenstein}\ \emph
  {et~al.}(2007{\natexlab{b}})\citenamefont {Eisenstein}, \citenamefont {Seo},\
  and\ \citenamefont {White}}]{Eisenstein:2006nj}%
  \BibitemOpen
  \bibfield  {author} {\bibinfo {author} {\bibfnamefont {D.~J.}\ \bibnamefont
  {Eisenstein}}, \bibinfo {author} {\bibfnamefont {H.-j.}\ \bibnamefont {Seo}},
  \ and\ \bibinfo {author} {\bibfnamefont {M.~J.}\ \bibnamefont {White}},\
  }\href {\doibase 10.1086/518755} {\bibfield  {journal} {\bibinfo  {journal}
  {Astrophys. J.}\ }\textbf {\bibinfo {volume} {664}},\ \bibinfo {pages} {660}
  (\bibinfo {year} {2007}{\natexlab{b}})},\ \Eprint
  {http://arxiv.org/abs/astro-ph/0604361} {arXiv:astro-ph/0604361} \BibitemShut
  {NoStop}%
\bibitem [{\citenamefont {Crocce}\ and\ \citenamefont
  {Scoccimarro}(2008)}]{Crocce:2007dt}%
  \BibitemOpen
  \bibfield  {author} {\bibinfo {author} {\bibfnamefont {M.}~\bibnamefont
  {Crocce}}\ and\ \bibinfo {author} {\bibfnamefont {R.}~\bibnamefont
  {Scoccimarro}},\ }\href {\doibase 10.1103/PhysRevD.77.023533} {\bibfield
  {journal} {\bibinfo  {journal} {Phys. Rev. D}\ }\textbf {\bibinfo {volume}
  {77}},\ \bibinfo {pages} {023533} (\bibinfo {year} {2008})},\ \Eprint
  {http://arxiv.org/abs/0704.2783} {arXiv:0704.2783 [astro-ph]} \BibitemShut
  {NoStop}%
\bibitem [{\citenamefont {Matsubara}(2008{\natexlab{a}})}]{Matsubara:2007wj}%
  \BibitemOpen
  \bibfield  {author} {\bibinfo {author} {\bibfnamefont {T.}~\bibnamefont
  {Matsubara}},\ }\href {\doibase 10.1103/PhysRevD.77.063530} {\bibfield
  {journal} {\bibinfo  {journal} {Phys. Rev.}\ }\textbf {\bibinfo {volume}
  {D77}},\ \bibinfo {pages} {063530} (\bibinfo {year} {2008}{\natexlab{a}})},\
  \Eprint {http://arxiv.org/abs/0711.2521} {arXiv:0711.2521 [astro-ph]}
  \BibitemShut {NoStop}%
%%CITATION = ARXIV:0711.2521;%%
\bibitem [{\citenamefont {Senatore}\ and\ \citenamefont
  {Zaldarriaga}(2015)}]{Senatore:2014via}%
  \BibitemOpen
  \bibfield  {author} {\bibinfo {author} {\bibfnamefont {L.}~\bibnamefont
  {Senatore}}\ and\ \bibinfo {author} {\bibfnamefont {M.}~\bibnamefont
  {Zaldarriaga}},\ }\href {\doibase 10.1088/1475-7516/2015/02/013} {\bibfield
  {journal} {\bibinfo  {journal} {JCAP}\ }\textbf {\bibinfo {volume} {02}},\
  \bibinfo {pages} {013} (\bibinfo {year} {2015})},\ \Eprint
  {http://arxiv.org/abs/1404.5954} {arXiv:1404.5954 [astro-ph.CO]} \BibitemShut
  {NoStop}%
\bibitem [{\citenamefont {Baldauf}\ \emph {et~al.}(2015)\citenamefont
  {Baldauf}, \citenamefont {Mirbabayi}, \citenamefont {Simonovi\'c},\ and\
  \citenamefont {Zaldarriaga}}]{Baldauf:2015xfa}%
  \BibitemOpen
  \bibfield  {author} {\bibinfo {author} {\bibfnamefont {T.}~\bibnamefont
  {Baldauf}}, \bibinfo {author} {\bibfnamefont {M.}~\bibnamefont {Mirbabayi}},
  \bibinfo {author} {\bibfnamefont {M.}~\bibnamefont {Simonovi\'c}}, \ and\
  \bibinfo {author} {\bibfnamefont {M.}~\bibnamefont {Zaldarriaga}},\ }\href
  {\doibase 10.1103/PhysRevD.92.043514} {\bibfield  {journal} {\bibinfo
  {journal} {Phys. Rev. D}\ }\textbf {\bibinfo {volume} {92}},\ \bibinfo
  {pages} {043514} (\bibinfo {year} {2015})},\ \Eprint
  {http://arxiv.org/abs/1504.04366} {arXiv:1504.04366 [astro-ph.CO]}
  \BibitemShut {NoStop}%
\bibitem [{\citenamefont {Blas}\ \emph
  {et~al.}(2016{\natexlab{b}})\citenamefont {Blas}, \citenamefont {Garny},
  \citenamefont {Ivanov},\ and\ \citenamefont {Sibiryakov}}]{Blas:2016sfa}%
  \BibitemOpen
  \bibfield  {author} {\bibinfo {author} {\bibfnamefont {D.}~\bibnamefont
  {Blas}}, \bibinfo {author} {\bibfnamefont {M.}~\bibnamefont {Garny}},
  \bibinfo {author} {\bibfnamefont {M.~M.}\ \bibnamefont {Ivanov}}, \ and\
  \bibinfo {author} {\bibfnamefont {S.}~\bibnamefont {Sibiryakov}},\ }\href
  {\doibase 10.1088/1475-7516/2016/07/028} {\bibfield  {journal} {\bibinfo
  {journal} {JCAP}\ }\textbf {\bibinfo {volume} {07}},\ \bibinfo {pages} {028}
  (\bibinfo {year} {2016}{\natexlab{b}})},\ \Eprint
  {http://arxiv.org/abs/1605.02149} {arXiv:1605.02149 [astro-ph.CO]}
  \BibitemShut {NoStop}%
\bibitem [{\citenamefont {Senatore}\ and\ \citenamefont
  {Trevisan}(2018)}]{Senatore:2017pbn}%
  \BibitemOpen
  \bibfield  {author} {\bibinfo {author} {\bibfnamefont {L.}~\bibnamefont
  {Senatore}}\ and\ \bibinfo {author} {\bibfnamefont {G.}~\bibnamefont
  {Trevisan}},\ }\href {\doibase 10.1088/1475-7516/2018/05/019} {\bibfield
  {journal} {\bibinfo  {journal} {JCAP}\ }\textbf {\bibinfo {volume} {05}},\
  \bibinfo {pages} {019} (\bibinfo {year} {2018})},\ \Eprint
  {http://arxiv.org/abs/1710.02178} {arXiv:1710.02178 [astro-ph.CO]}
  \BibitemShut {NoStop}%
\bibitem [{\citenamefont {Ivanov}\ and\ \citenamefont
  {Sibiryakov}(2018)}]{Ivanov:2018gjr}%
  \BibitemOpen
  \bibfield  {author} {\bibinfo {author} {\bibfnamefont {M.~M.}\ \bibnamefont
  {Ivanov}}\ and\ \bibinfo {author} {\bibfnamefont {S.}~\bibnamefont
  {Sibiryakov}},\ }\href {\doibase 10.1088/1475-7516/2018/07/053} {\bibfield
  {journal} {\bibinfo  {journal} {JCAP}\ }\textbf {\bibinfo {volume} {07}},\
  \bibinfo {pages} {053} (\bibinfo {year} {2018})},\ \Eprint
  {http://arxiv.org/abs/1804.05080} {arXiv:1804.05080 [astro-ph.CO]}
  \BibitemShut {NoStop}%
\bibitem [{\citenamefont {Lewandowski}\ and\ \citenamefont
  {Senatore}(2020)}]{Lewandowski:2018ywf}%
  \BibitemOpen
  \bibfield  {author} {\bibinfo {author} {\bibfnamefont {M.}~\bibnamefont
  {Lewandowski}}\ and\ \bibinfo {author} {\bibfnamefont {L.}~\bibnamefont
  {Senatore}},\ }\href {\doibase 10.1088/1475-7516/2020/03/018} {\bibfield
  {journal} {\bibinfo  {journal} {JCAP}\ }\textbf {\bibinfo {volume} {03}},\
  \bibinfo {pages} {018} (\bibinfo {year} {2020})},\ \Eprint
  {http://arxiv.org/abs/1810.11855} {arXiv:1810.11855 [astro-ph.CO]}
  \BibitemShut {NoStop}%
\bibitem [{\citenamefont {Sugiyama}\ \emph {et~al.}(2021)\citenamefont
  {Sugiyama}, \citenamefont {Saito}, \citenamefont {Beutler},\ and\
  \citenamefont {Seo}}]{Sugiyama:2020uil}%
  \BibitemOpen
  \bibfield  {author} {\bibinfo {author} {\bibfnamefont {N.~S.}\ \bibnamefont
  {Sugiyama}}, \bibinfo {author} {\bibfnamefont {S.}~\bibnamefont {Saito}},
  \bibinfo {author} {\bibfnamefont {F.}~\bibnamefont {Beutler}}, \ and\
  \bibinfo {author} {\bibfnamefont {H.-J.}\ \bibnamefont {Seo}},\ }\href
  {\doibase 10.1093/mnras/staa3725} {\bibfield  {journal} {\bibinfo  {journal}
  {Mon. Not. Roy. Astron. Soc.}\ }\textbf {\bibinfo {volume} {501}},\ \bibinfo
  {pages} {2862} (\bibinfo {year} {2021})},\ \Eprint
  {http://arxiv.org/abs/2010.06179} {arXiv:2010.06179 [astro-ph.CO]}
  \BibitemShut {NoStop}%
\bibitem [{\citenamefont {Kaiser}(1987)}]{Kaiser:1987qv}%
  \BibitemOpen
  \bibfield  {author} {\bibinfo {author} {\bibfnamefont {N.}~\bibnamefont
  {Kaiser}},\ }\href@noop {} {\bibfield  {journal} {\bibinfo  {journal} {Mon.
  Not. Roy. Astron. Soc.}\ }\textbf {\bibinfo {volume} {227}},\ \bibinfo
  {pages} {1} (\bibinfo {year} {1987})}\BibitemShut {NoStop}%
\bibitem [{\citenamefont {Desjacques}\ \emph {et~al.}(2018)\citenamefont
  {Desjacques}, \citenamefont {Jeong},\ and\ \citenamefont
  {Schmidt}}]{Desjacques:2016bnm}%
  \BibitemOpen
  \bibfield  {author} {\bibinfo {author} {\bibfnamefont {V.}~\bibnamefont
  {Desjacques}}, \bibinfo {author} {\bibfnamefont {D.}~\bibnamefont {Jeong}}, \
  and\ \bibinfo {author} {\bibfnamefont {F.}~\bibnamefont {Schmidt}},\ }\href
  {\doibase 10.1016/j.physrep.2017.12.002} {\bibfield  {journal} {\bibinfo
  {journal} {Phys. Rept.}\ }\textbf {\bibinfo {volume} {733}},\ \bibinfo
  {pages} {1} (\bibinfo {year} {2018})},\ \Eprint
  {http://arxiv.org/abs/1611.09787} {arXiv:1611.09787 [astro-ph.CO]}
  \BibitemShut {NoStop}%
\bibitem [{\citenamefont {Aghanim}\ \emph {et~al.}(2020)\citenamefont {Aghanim}
  \emph {et~al.}}]{Aghanim:2018eyx}%
  \BibitemOpen
  \bibfield  {author} {\bibinfo {author} {\bibfnamefont {N.}~\bibnamefont
  {Aghanim}} \emph {et~al.} (\bibinfo {collaboration} {Planck}),\ }\href
  {\doibase 10.1051/0004-6361/201833910} {\bibfield  {journal} {\bibinfo
  {journal} {Astron. Astrophys.}\ }\textbf {\bibinfo {volume} {641}},\ \bibinfo
  {pages} {A6} (\bibinfo {year} {2020})},\ \Eprint
  {http://arxiv.org/abs/1807.06209} {arXiv:1807.06209 [astro-ph.CO]}
  \BibitemShut {NoStop}%
\bibitem [{\citenamefont {Vlah}\ \emph {et~al.}(2016)\citenamefont {Vlah},
  \citenamefont {Seljak}, \citenamefont {Chu},\ and\ \citenamefont
  {Feng}}]{Vlah:2015zda}%
  \BibitemOpen
  \bibfield  {author} {\bibinfo {author} {\bibfnamefont {Z.}~\bibnamefont
  {Vlah}}, \bibinfo {author} {\bibfnamefont {U.}~\bibnamefont {Seljak}},
  \bibinfo {author} {\bibfnamefont {M.~Y.}\ \bibnamefont {Chu}}, \ and\
  \bibinfo {author} {\bibfnamefont {Y.}~\bibnamefont {Feng}},\ }\href {\doibase
  10.1088/1475-7516/2016/03/057} {\bibfield  {journal} {\bibinfo  {journal}
  {JCAP}\ }\textbf {\bibinfo {volume} {03}},\ \bibinfo {pages} {057} (\bibinfo
  {year} {2016})},\ \Eprint {http://arxiv.org/abs/1509.02120} {arXiv:1509.02120
  [astro-ph.CO]} \BibitemShut {NoStop}%
\bibitem [{\citenamefont {Schmittfull}\ \emph
  {et~al.}(2015{\natexlab{b}})\citenamefont {Schmittfull}, \citenamefont
  {Baldauf},\ and\ \citenamefont {Seljak}}]{Schmittfull:2014tca}%
  \BibitemOpen
  \bibfield  {author} {\bibinfo {author} {\bibfnamefont {M.}~\bibnamefont
  {Schmittfull}}, \bibinfo {author} {\bibfnamefont {T.}~\bibnamefont
  {Baldauf}}, \ and\ \bibinfo {author} {\bibfnamefont {U.}~\bibnamefont
  {Seljak}},\ }\href {\doibase 10.1103/PhysRevD.91.043530} {\bibfield
  {journal} {\bibinfo  {journal} {Phys. Rev. D}\ }\textbf {\bibinfo {volume}
  {91}},\ \bibinfo {pages} {043530} (\bibinfo {year} {2015}{\natexlab{b}})},\
  \Eprint {http://arxiv.org/abs/1411.6595} {arXiv:1411.6595 [astro-ph.CO]}
  \BibitemShut {NoStop}%
\bibitem [{\citenamefont {Bernardeau}\ \emph {et~al.}(2008)\citenamefont
  {Bernardeau}, \citenamefont {Crocce},\ and\ \citenamefont
  {Scoccimarro}}]{Bernardeau:2008fa}%
  \BibitemOpen
  \bibfield  {author} {\bibinfo {author} {\bibfnamefont {F.}~\bibnamefont
  {Bernardeau}}, \bibinfo {author} {\bibfnamefont {M.}~\bibnamefont {Crocce}},
  \ and\ \bibinfo {author} {\bibfnamefont {R.}~\bibnamefont {Scoccimarro}},\
  }\href {\doibase 10.1103/PhysRevD.78.103521} {\bibfield  {journal} {\bibinfo
  {journal} {Phys. Rev.}\ }\textbf {\bibinfo {volume} {D78}},\ \bibinfo {pages}
  {103521} (\bibinfo {year} {2008})},\ \Eprint {http://arxiv.org/abs/0806.2334}
  {arXiv:0806.2334 [astro-ph]} \BibitemShut {NoStop}%
%%CITATION = ARXIV:0806.2334;%%
\bibitem [{\citenamefont {Matsubara}(2008{\natexlab{b}})}]{Matsubara:2008wx}%
  \BibitemOpen
  \bibfield  {author} {\bibinfo {author} {\bibfnamefont {T.}~\bibnamefont
  {Matsubara}},\ }\href {\doibase 10.1103/PhysRevD.78.109901,
  10.1103/PhysRevD.78.083519} {\bibfield  {journal} {\bibinfo  {journal} {Phys.
  Rev.}\ }\textbf {\bibinfo {volume} {D78}},\ \bibinfo {pages} {083519}
  (\bibinfo {year} {2008}{\natexlab{b}})},\ \bibinfo {note} {[Erratum: Phys.
  Rev.D78,109901(2008)]},\ \Eprint {http://arxiv.org/abs/0807.1733}
  {arXiv:0807.1733 [astro-ph]} \BibitemShut {NoStop}%
%%CITATION = ARXIV:0807.1733;%%
\bibitem [{\citenamefont {Langlois}(2019)}]{Langlois:2018dxi}%
  \BibitemOpen
  \bibfield  {author} {\bibinfo {author} {\bibfnamefont {D.}~\bibnamefont
  {Langlois}},\ }\href {\doibase 10.1142/S0218271819420069} {\bibfield
  {journal} {\bibinfo  {journal} {Int. J. Mod. Phys. D}\ }\textbf {\bibinfo
  {volume} {28}},\ \bibinfo {pages} {1942006} (\bibinfo {year} {2019})},\
  \Eprint {http://arxiv.org/abs/1811.06271} {arXiv:1811.06271 [gr-qc]}
  \BibitemShut {NoStop}%
\bibitem [{\citenamefont {Kobayashi}(2019)}]{Kobayashi:2019hrl}%
  \BibitemOpen
  \bibfield  {author} {\bibinfo {author} {\bibfnamefont {T.}~\bibnamefont
  {Kobayashi}},\ }\href {\doibase 10.1088/1361-6633/ab2429} {\bibfield
  {journal} {\bibinfo  {journal} {Rept. Prog. Phys.}\ }\textbf {\bibinfo
  {volume} {82}},\ \bibinfo {pages} {086901} (\bibinfo {year} {2019})},\
  \Eprint {http://arxiv.org/abs/1901.07183} {arXiv:1901.07183 [gr-qc]}
  \BibitemShut {NoStop}%
\bibitem [{\citenamefont {{Sugiyama}}\ \emph {et~al.}(2023)\citenamefont
  {{Sugiyama}}, \citenamefont {{Yamauchi}}, \citenamefont {{Kobayashi}},
  \citenamefont {{Fujita}}, \citenamefont {{Arai}}, \citenamefont {{Hirano}},
  \citenamefont {{Saito}}, \citenamefont {{Beutler}},\ and\ \citenamefont
  {{Seo}}}]{Sugiyama:2023tes}%
  \BibitemOpen
  \bibfield  {author} {\bibinfo {author} {\bibfnamefont {N.~S.}\ \bibnamefont
  {{Sugiyama}}}, \bibinfo {author} {\bibfnamefont {D.}~\bibnamefont
  {{Yamauchi}}}, \bibinfo {author} {\bibfnamefont {T.}~\bibnamefont
  {{Kobayashi}}}, \bibinfo {author} {\bibfnamefont {T.}~\bibnamefont
  {{Fujita}}}, \bibinfo {author} {\bibfnamefont {S.}~\bibnamefont {{Arai}}},
  \bibinfo {author} {\bibfnamefont {S.}~\bibnamefont {{Hirano}}}, \bibinfo
  {author} {\bibfnamefont {S.}~\bibnamefont {{Saito}}}, \bibinfo {author}
  {\bibfnamefont {F.}~\bibnamefont {{Beutler}}}, \ and\ \bibinfo {author}
  {\bibfnamefont {H.-J.}\ \bibnamefont {{Seo}}},\ }\href {\doibase
  10.1093/mnras/stad1505} {\bibfield  {journal} {\bibinfo  {journal} {Mon. Not.
  Roy. Astron. Soc.}\ }\textbf {\bibinfo {volume} {523}},\ \bibinfo {pages}
  {3133} (\bibinfo {year} {2023})},\ \Eprint {http://arxiv.org/abs/2302.06808}
  {arXiv:2302.06808 [astro-ph.CO]} \BibitemShut {NoStop}%
\bibitem [{\citenamefont {Sugiyama}\ \emph {et~al.}(2023)\citenamefont
  {Sugiyama}, \citenamefont {Yamauchi}, \citenamefont {Kobayashi},
  \citenamefont {Fujita}, \citenamefont {Arai}, \citenamefont {Hirano},
  \citenamefont {Saito}, \citenamefont {Beutler},\ and\ \citenamefont
  {Seo}}]{Sugiyama:2023zvd}%
  \BibitemOpen
  \bibfield  {author} {\bibinfo {author} {\bibfnamefont {N.~S.}\ \bibnamefont
  {Sugiyama}}, \bibinfo {author} {\bibfnamefont {D.}~\bibnamefont {Yamauchi}},
  \bibinfo {author} {\bibfnamefont {T.}~\bibnamefont {Kobayashi}}, \bibinfo
  {author} {\bibfnamefont {T.}~\bibnamefont {Fujita}}, \bibinfo {author}
  {\bibfnamefont {S.}~\bibnamefont {Arai}}, \bibinfo {author} {\bibfnamefont
  {S.}~\bibnamefont {Hirano}}, \bibinfo {author} {\bibfnamefont
  {S.}~\bibnamefont {Saito}}, \bibinfo {author} {\bibfnamefont
  {F.}~\bibnamefont {Beutler}}, \ and\ \bibinfo {author} {\bibfnamefont
  {H.-J.}\ \bibnamefont {Seo}},\ }\href {\doibase 10.1093/mnras/stad1935}
  {\bibfield  {journal} {\bibinfo  {journal} {Mon. Not. Roy. Astron. Soc.}\
  }\textbf {\bibinfo {volume} {524}},\ \bibinfo {pages} {1651} (\bibinfo {year}
  {2023})},\ \Eprint {http://arxiv.org/abs/2305.01142} {arXiv:2305.01142
  [astro-ph.CO]} \BibitemShut {NoStop}%
\bibitem [{\citenamefont {Yoo}\ \emph {et~al.}(2011)\citenamefont {Yoo},
  \citenamefont {Dalal},\ and\ \citenamefont {Seljak}}]{Yoo:2011tq}%
  \BibitemOpen
  \bibfield  {author} {\bibinfo {author} {\bibfnamefont {J.}~\bibnamefont
  {Yoo}}, \bibinfo {author} {\bibfnamefont {N.}~\bibnamefont {Dalal}}, \ and\
  \bibinfo {author} {\bibfnamefont {U.}~\bibnamefont {Seljak}},\ }\href
  {\doibase 10.1088/1475-7516/2011/07/018} {\bibfield  {journal} {\bibinfo
  {journal} {JCAP}\ }\textbf {\bibinfo {volume} {07}},\ \bibinfo {pages} {018}
  (\bibinfo {year} {2011})},\ \Eprint {http://arxiv.org/abs/1105.3732}
  {arXiv:1105.3732 [astro-ph.CO]} \BibitemShut {NoStop}%
\bibitem [{\citenamefont {Chisari}\ and\ \citenamefont
  {Pontzen}(2019)}]{Chisari:2019tig}%
  \BibitemOpen
  \bibfield  {author} {\bibinfo {author} {\bibfnamefont {N.~E.}\ \bibnamefont
  {Chisari}}\ and\ \bibinfo {author} {\bibfnamefont {A.}~\bibnamefont
  {Pontzen}},\ }\href {\doibase 10.1103/PhysRevD.100.023543} {\bibfield
  {journal} {\bibinfo  {journal} {Phys. Rev. D}\ }\textbf {\bibinfo {volume}
  {100}},\ \bibinfo {pages} {023543} (\bibinfo {year} {2019})},\ \Eprint
  {http://arxiv.org/abs/1905.02078} {arXiv:1905.02078 [astro-ph.CO]}
  \BibitemShut {NoStop}%
\bibitem [{\citenamefont {Eisenstein}\ and\ \citenamefont
  {Hu}(1998)}]{Eisenstein:1997ik}%
  \BibitemOpen
  \bibfield  {author} {\bibinfo {author} {\bibfnamefont {D.~J.}\ \bibnamefont
  {Eisenstein}}\ and\ \bibinfo {author} {\bibfnamefont {W.}~\bibnamefont
  {Hu}},\ }\href {\doibase 10.1086/305424} {\bibfield  {journal} {\bibinfo
  {journal} {Astrophys. J.}\ }\textbf {\bibinfo {volume} {496}},\ \bibinfo
  {pages} {605} (\bibinfo {year} {1998})},\ \Eprint
  {http://arxiv.org/abs/astro-ph/9709112} {arXiv:astro-ph/9709112 [astro-ph]}
  \BibitemShut {NoStop}%
%%CITATION = ASTRO-PH/9709112;%%
\bibitem [{\citenamefont {White}(2010)}]{White:2010qd}%
  \BibitemOpen
  \bibfield  {author} {\bibinfo {author} {\bibfnamefont {M.}~\bibnamefont
  {White}},\ }\href@noop {} {\  (\bibinfo {year} {2010})},\ \Eprint
  {http://arxiv.org/abs/1004.0250} {arXiv:1004.0250 [astro-ph.CO]} \BibitemShut
  {NoStop}%
\bibitem [{\citenamefont {Alam}\ \emph {et~al.}(2017)\citenamefont {Alam} \emph
  {et~al.}}]{BOSS:2016wmc}%
  \BibitemOpen
  \bibfield  {author} {\bibinfo {author} {\bibfnamefont {S.}~\bibnamefont
  {Alam}} \emph {et~al.} (\bibinfo {collaboration} {BOSS}),\ }\href {\doibase
  10.1093/mnras/stx721} {\bibfield  {journal} {\bibinfo  {journal} {Mon. Not.
  Roy. Astron. Soc.}\ }\textbf {\bibinfo {volume} {470}},\ \bibinfo {pages}
  {2617} (\bibinfo {year} {2017})},\ \Eprint {http://arxiv.org/abs/1607.03155}
  {arXiv:1607.03155 [astro-ph.CO]} \BibitemShut {NoStop}%
\bibitem [{\citenamefont {Hinton}\ \emph {et~al.}(2017)\citenamefont {Hinton}
  \emph {et~al.}}]{Hinton:2016atz}%
  \BibitemOpen
  \bibfield  {author} {\bibinfo {author} {\bibfnamefont {S.~R.}\ \bibnamefont
  {Hinton}} \emph {et~al.},\ }\href {\doibase 10.1093/mnras/stw2725} {\bibfield
   {journal} {\bibinfo  {journal} {Mon. Not. Roy. Astron. Soc.}\ }\textbf
  {\bibinfo {volume} {464}},\ \bibinfo {pages} {4807} (\bibinfo {year}
  {2017})},\ \Eprint {http://arxiv.org/abs/1611.08040} {arXiv:1611.08040
  [astro-ph.CO]} \BibitemShut {NoStop}%
\bibitem [{\citenamefont {Alam}\ \emph {et~al.}(2021)\citenamefont {Alam} \emph
  {et~al.}}]{eBOSS:2020yzd}%
  \BibitemOpen
  \bibfield  {author} {\bibinfo {author} {\bibfnamefont {S.}~\bibnamefont
  {Alam}} \emph {et~al.} (\bibinfo {collaboration} {eBOSS}),\ }\href {\doibase
  10.1103/PhysRevD.103.083533} {\bibfield  {journal} {\bibinfo  {journal}
  {Phys. Rev. D}\ }\textbf {\bibinfo {volume} {103}},\ \bibinfo {pages}
  {083533} (\bibinfo {year} {2021})},\ \Eprint
  {http://arxiv.org/abs/2007.08991} {arXiv:2007.08991 [astro-ph.CO]}
  \BibitemShut {NoStop}%
\bibitem [{\citenamefont {Moon}\ \emph {et~al.}(2023)\citenamefont {Moon} \emph
  {et~al.}}]{DESI:2023bgx}%
  \BibitemOpen
  \bibfield  {author} {\bibinfo {author} {\bibfnamefont {J.}~\bibnamefont
  {Moon}} \emph {et~al.} (\bibinfo {collaboration} {DESI}),\ }\href@noop {} {\
  (\bibinfo {year} {2023})},\ \Eprint {http://arxiv.org/abs/2304.08427}
  {arXiv:2304.08427 [astro-ph.CO]} \BibitemShut {NoStop}%
\bibitem [{\citenamefont {Chen}\ \emph {et~al.}(2024)\citenamefont {Chen} \emph
  {et~al.}}]{Chen:2024tfp}%
  \BibitemOpen
  \bibfield  {author} {\bibinfo {author} {\bibfnamefont {S.-F.}\ \bibnamefont
  {Chen}} \emph {et~al.},\ }\href@noop {} {\  (\bibinfo {year} {2024})},\
  \Eprint {http://arxiv.org/abs/2402.14070} {arXiv:2402.14070 [astro-ph.CO]}
  \BibitemShut {NoStop}%
\bibitem [{\citenamefont {Blas}\ \emph {et~al.}(2011)\citenamefont {Blas},
  \citenamefont {Lesgourgues},\ and\ \citenamefont {Tram}}]{Blas:2011rf}%
  \BibitemOpen
  \bibfield  {author} {\bibinfo {author} {\bibfnamefont {D.}~\bibnamefont
  {Blas}}, \bibinfo {author} {\bibfnamefont {J.}~\bibnamefont {Lesgourgues}}, \
  and\ \bibinfo {author} {\bibfnamefont {T.}~\bibnamefont {Tram}},\ }\href
  {\doibase 10.1088/1475-7516/2011/07/034} {\bibfield  {journal} {\bibinfo
  {journal} {JCAP}\ }\textbf {\bibinfo {volume} {07}},\ \bibinfo {pages} {034}
  (\bibinfo {year} {2011})},\ \Eprint {http://arxiv.org/abs/1104.2933}
  {arXiv:1104.2933 [astro-ph.CO]} \BibitemShut {NoStop}%
\bibitem [{\citenamefont {{Hamilton}}(2000)}]{Hamilton:2000MNRAS.312..257H}%
  \BibitemOpen
  \bibfield  {author} {\bibinfo {author} {\bibfnamefont {A.~J.~S.}\
  \bibnamefont {{Hamilton}}},\ }\href {\doibase
  10.1046/j.1365-8711.2000.03071.x} {\bibfield  {journal} {\bibinfo  {journal}
  {Mon. Not. Roy. Astron. Soc.}\ }\textbf {\bibinfo {volume} {312}},\ \bibinfo
  {pages} {257} (\bibinfo {year} {2000})},\ \Eprint
  {http://arxiv.org/abs/astro-ph/9905191} {arXiv:astro-ph/9905191 [astro-ph]}
  \BibitemShut {NoStop}%
\bibitem [{\citenamefont {Fang}\ \emph {et~al.}(2020)\citenamefont {Fang},
  \citenamefont {Krause}, \citenamefont {Eifler},\ and\ \citenamefont
  {MacCrann}}]{Fang:2019xat}%
  \BibitemOpen
  \bibfield  {author} {\bibinfo {author} {\bibfnamefont {X.}~\bibnamefont
  {Fang}}, \bibinfo {author} {\bibfnamefont {E.}~\bibnamefont {Krause}},
  \bibinfo {author} {\bibfnamefont {T.}~\bibnamefont {Eifler}}, \ and\ \bibinfo
  {author} {\bibfnamefont {N.}~\bibnamefont {MacCrann}},\ }\href {\doibase
  10.1088/1475-7516/2020/05/010} {\bibfield  {journal} {\bibinfo  {journal}
  {JCAP}\ }\textbf {\bibinfo {volume} {05}},\ \bibinfo {pages} {010} (\bibinfo
  {year} {2020})},\ \Eprint {http://arxiv.org/abs/1911.11947} {arXiv:1911.11947
  [astro-ph.CO]} \BibitemShut {NoStop}%
\bibitem [{\citenamefont {Hirano}\ \emph {et~al.}(2018)\citenamefont {Hirano},
  \citenamefont {Kobayashi}, \citenamefont {Tashiro},\ and\ \citenamefont
  {Yokoyama}}]{Hirano:2018uar}%
  \BibitemOpen
  \bibfield  {author} {\bibinfo {author} {\bibfnamefont {S.}~\bibnamefont
  {Hirano}}, \bibinfo {author} {\bibfnamefont {T.}~\bibnamefont {Kobayashi}},
  \bibinfo {author} {\bibfnamefont {H.}~\bibnamefont {Tashiro}}, \ and\
  \bibinfo {author} {\bibfnamefont {S.}~\bibnamefont {Yokoyama}},\ }\href
  {\doibase 10.1103/PhysRevD.97.103517} {\bibfield  {journal} {\bibinfo
  {journal} {Phys. Rev. D}\ }\textbf {\bibinfo {volume} {97}},\ \bibinfo
  {pages} {103517} (\bibinfo {year} {2018})},\ \Eprint
  {http://arxiv.org/abs/1801.07885} {arXiv:1801.07885 [astro-ph.CO]}
  \BibitemShut {NoStop}%
\bibitem [{\citenamefont {Hirano}\ \emph {et~al.}(2020)\citenamefont {Hirano},
  \citenamefont {Kobayashi}, \citenamefont {Yamauchi},\ and\ \citenamefont
  {Yokoyama}}]{Hirano:2020dom}%
  \BibitemOpen
  \bibfield  {author} {\bibinfo {author} {\bibfnamefont {S.}~\bibnamefont
  {Hirano}}, \bibinfo {author} {\bibfnamefont {T.}~\bibnamefont {Kobayashi}},
  \bibinfo {author} {\bibfnamefont {D.}~\bibnamefont {Yamauchi}}, \ and\
  \bibinfo {author} {\bibfnamefont {S.}~\bibnamefont {Yokoyama}},\ }\href
  {\doibase 10.1103/PhysRevD.102.103505} {\bibfield  {journal} {\bibinfo
  {journal} {Phys. Rev. D}\ }\textbf {\bibinfo {volume} {102}},\ \bibinfo
  {pages} {103505} (\bibinfo {year} {2020})},\ \Eprint
  {http://arxiv.org/abs/2008.02798} {arXiv:2008.02798 [gr-qc]} \BibitemShut
  {NoStop}%
\bibitem [{\citenamefont {Crisostomi}\ \emph {et~al.}(2020)\citenamefont
  {Crisostomi}, \citenamefont {Lewandowski},\ and\ \citenamefont
  {Vernizzi}}]{Crisostomi:2019vhj}%
  \BibitemOpen
  \bibfield  {author} {\bibinfo {author} {\bibfnamefont {M.}~\bibnamefont
  {Crisostomi}}, \bibinfo {author} {\bibfnamefont {M.}~\bibnamefont
  {Lewandowski}}, \ and\ \bibinfo {author} {\bibfnamefont {F.}~\bibnamefont
  {Vernizzi}},\ }\href {\doibase 10.1103/PhysRevD.101.123501} {\bibfield
  {journal} {\bibinfo  {journal} {Phys. Rev. D}\ }\textbf {\bibinfo {volume}
  {101}},\ \bibinfo {pages} {123501} (\bibinfo {year} {2020})},\ \Eprint
  {http://arxiv.org/abs/1909.07366} {arXiv:1909.07366 [astro-ph.CO]}
  \BibitemShut {NoStop}%
\bibitem [{\citenamefont {Lewandowski}(2020)}]{Lewandowski:2019txi}%
  \BibitemOpen
  \bibfield  {author} {\bibinfo {author} {\bibfnamefont {M.}~\bibnamefont
  {Lewandowski}},\ }\href {\doibase 10.1088/1475-7516/2020/08/044} {\bibfield
  {journal} {\bibinfo  {journal} {JCAP}\ }\textbf {\bibinfo {volume} {08}},\
  \bibinfo {pages} {044} (\bibinfo {year} {2020})},\ \Eprint
  {http://arxiv.org/abs/1912.12292} {arXiv:1912.12292 [astro-ph.CO]}
  \BibitemShut {NoStop}%
\bibitem [{\citenamefont {Scoccimarro}\ \emph {et~al.}(1999)\citenamefont
  {Scoccimarro}, \citenamefont {Couchman},\ and\ \citenamefont
  {Frieman}}]{Scoccimarro:1999ed}%
  \BibitemOpen
  \bibfield  {author} {\bibinfo {author} {\bibfnamefont {R.}~\bibnamefont
  {Scoccimarro}}, \bibinfo {author} {\bibfnamefont {H.~M.~P.}\ \bibnamefont
  {Couchman}}, \ and\ \bibinfo {author} {\bibfnamefont {J.~A.}\ \bibnamefont
  {Frieman}},\ }\href {\doibase 10.1086/307220} {\bibfield  {journal} {\bibinfo
   {journal} {Astrophys. J.}\ }\textbf {\bibinfo {volume} {517}},\ \bibinfo
  {pages} {531} (\bibinfo {year} {1999})},\ \Eprint
  {http://arxiv.org/abs/astro-ph/9808305} {arXiv:astro-ph/9808305 [astro-ph]}
  \BibitemShut {NoStop}%
%%CITATION = ASTRO-PH/9808305;%%
\end{thebibliography}%

\end{document}